\documentclass[a4paper,12pt]{article}
\pdfoutput=1

\usepackage[T1]{fontenc}
\usepackage[utf8]{inputenc}
\usepackage[english]{babel}
\usepackage{authblk}
\usepackage[labelfont=bf]{caption}
\usepackage{hyperref}
\usepackage{parskip}
\usepackage{graphicx}
\usepackage{subcaption}
\usepackage{multirow}
\usepackage{fancyvrb}
\usepackage{amsmath, amsthm, amssymb}
\usepackage{braket}
\usepackage{slashed}
\usepackage{bbold}
\usepackage[makeroom]{cancel}
\usepackage{cleveref}

\newcommand{\mytablename}{Table}
\newcommand{\myfigurename}{Figure}
\newcommand{\mysectionname}{Sect.}
\newcommand{\myappendixname}{Appendix}
\newcommand{\myrefname}{Ref.}
\newcommand{\myrefsname}{Refs.}
\newcommand{\eq}{Eq.}
\newcommand{\eqs}{Eqs.}

\DeclareMathOperator{\Tr}{\mathrm{Tr}}
\DeclareMathOperator{\sun}{{\mathrm{SU}}}
\DeclareMathOperator{\re}{\operatorname{Re}}

\DeclareMathOperator{\poa}{\Pi_\mathfrak{g}}

\newcommand{\CCel}{C\nolinebreak\hspace{-.05em}\raisebox{.4ex}{\tiny\bf +}\nolinebreak\hspace{-.10em}\raisebox{.4ex}{\tiny\bf +}}
\def\CCel{{C\nolinebreak[4]\hspace{-.05em}\raisebox{.4ex}{\tiny\bf ++}11 }}

\let\oldfootnote\footnote
\def\footnote{\ifhmode\unskip\fi\oldfootnote}

\newcommand*{\mailto}[1]{\href{mailto:#1}{{#1}}}
\title{\textbf{Large-order NSPT for lattice gauge theories with fermions: the plaquette in massless QCD}}
\author[1]{L. Del Debbio\thanks{\mailto{luigi.del.debbio@ed.ac.uk}}}
\author[2]{F. Di Renzo\thanks{\mailto{francesco.direnzo@unipr.it}}}
\author[1]{G. Filaci\thanks{\mailto{g.filaci@ed.ac.uk}}}
\affil[1]{Higgs Centre for Theoretical Physics, School of Physics \& Astronomy, University of Edinburgh, EH9 3FD, U.K.}
\affil[2]{Dipartimento di Scienze Matematiche, Fisiche e Informatiche, Universit\`{a} di Parma and INFN, Gruppo Collegato di Parma, I-43100 Parma, Italy}

\date{}

\begin{document}

\maketitle

\begin{abstract}
	\noindent
	Numerical Stochastic Perturbation Theory (NSPT) allows for perturbative
	computations in quantum field theory. We present an implementation of NSPT
	that yields results for high orders in the perturbative expansion of lattice
	gauge theories coupled to fermions. The zero-momentum mode is removed by
	imposing twisted boundary conditions; in turn, twisted boundary conditions
	require us to introduce a smell degree of freedom in order to include
	fermions in the fundamental representation. As a first application, we
	compute the critical mass of two flavours of Wilson fermions up to order
	$O(\beta^{-7})$ in a $\sun(3)$ gauge theory. We also implement, for the
	first time, staggered fermions in NSPT. The residual chiral symmetry of
	staggered fermions protects the theory from an additive mass
	renormalisation. We compute the	perturbative expansion of the plaquette with
	two flavours of massless staggered fermions up to order $O(\beta^{-35})$ in
	a $\sun(3)$ gauge theory, and investigate the renormalon behaviour of such
	series. We are able to subtract the power divergence in the Operator Product
	Expansion (OPE) for the plaquette and estimate the gluon condensate in
	massless QCD. Our results confirm that NSPT provides a viable way to probe
	systematically the asymptotic behaviour of perturbative series in QCD and,
	eventually, gauge theories with fermions in higher representations.
\end{abstract}

\clearpage
\tableofcontents
\clearpage
\flushbottom

\section{Introduction}
\label{sec:introduction}
The success of perturbation theory in High Energy Physics (HEP) can hardly be
denied. In particular, in asymptotically free theories, field correlators at
short distances are reliably approximated by perturbative expansions in the
running coupling at a large momentum scale. At the same time, even in these
(lucky) cases, it is mandatory to have some control on nonperturbative effects,
i.e. contributions that scale like powers of the QCD scale
$\Lambda_\mathrm{QCD}$. We will often refer to these as {\em power corrections}.
A tool to take the latter into account was suggested back in the late seventies.
This goes under the name of QCD sum rules, or Shifman-Vainshtein-Zakharov (SVZ)
sum rules~\cite{Shifman:1978bx,Shifman:1978by}. One of the authors defined the
method as ``an expansion of the correlation functions in the vacuum
condensates''~\cite{Shifman:1998rb}. These condensates are the vacuum
expectation value of the operators that emerge in the Operator Product Expansion
(OPE) for the relevant correlation function. In the OPE formalism the
condensates are fundamental quantities, which are in principle supposed to
parametrise power corrections in a universal way. By determining
the value of a condensate in one context, one gains insight into different
physical processes; this has in turn motivated several approaches to the
determination of condensates. Having said all this, the sad news is that not all
the condensates have actually the same status. In particular not all the
condensates can be defined in a neat way, which ultimately means disentangled
from perturbation theory. While this is the case for the chiral condensate, the
same cannot be said for the gluon condensate, which is the one we will be
concerned with in this work. 

Based on a separation of scales, the OPE makes pretty clear what can/must be
computed in perturbation theory, i.e. the Wilson coefficients. Still, this does
not automatically imply that perturbative and nonperturbative contributions are
separated in a clear-cut way. The key issue is that perturbative expansions in
HEP are expected to be asymptotic ones on very general grounds. In particular,
the series in asymptotically free theories are plagued by ambiguities which are
due to so-called infrared renormalons~\cite{tHooft:1977xjm,Beneke:1998ui}.
From a technical point of view, renormalons show up as singularities which are
encountered if one tries to Borel resum the perturbative series. All in all,
there is a power-like ambiguity in any procedure one can devise in order to sum
the series, and this ambiguity unavoidably reshuffles perturbative and
nonperturbative contributions in the structure of the OPE. Being the Wilson
coefficients affected by ambiguities that are power corrections, the general
strategy is to reabsorb the latter in the definition of the condensates. This
amounts to a prescription to give a precise meaning both to the perturbative
series and to the condensates that appear in the OPE.

The idea of determining the gluon condensate from nonperturbative (Monte Carlo)
measurements in lattice gauge theories dates back to the eighties and early
nineties~\cite{DiGiacomo:1981lcx,Alles:1993dn,Campostrini:1989uj, Alles:1992pq}. Based on
symmetry grounds and dimensional counting, the two leading contributions in the
OPE for the basic plaquette are given by the identity operator and the gluon
condensate. Both operators appear multiplied by Wilson coefficients that can be
computed in perturbation theory, and in particular the coefficient that
multiplies the identity operator is simply the perturbative expansion of the
plaquette. Other operators that appear in the OPE are of higher dimension, and
their contributions are therefore suppressed by powers of $a
\Lambda_\mathrm{QCD}$.
Subtracting from a nonperturbative (Monte Carlo) measurement of the plaquette
the sum of the perturbative series, and repeating the procedure at different
values of the coupling, the signature of asymptotic scaling, i.e. the signature
of a quantity of (mass) dimension four, should become visible. With renormalons
attracting more and more attention, it eventually became clear that such a
procedure must be deeply affected by the ambiguities we discussed above,
suggesting that a precise definition of the resummed perturbative expansion is
necessary. 

In the meantime Numerical Stochastic Perturbation Theory
(NSPT)~\cite{DiRenzo:1994sy} was developed as a new tool for computing high
orders in lattice perturbation theory. NSPT paved the way to the evaluation of
many more terms in the perturbative expansion of the plaquette, and in turn made
it at least conceivable that the behaviour of the series could be understood at
the level of pinning down the correct order of magnitude of the ambiguity
involved. Results of early investigations~\cite{DiRenzo:1995qc} were
interesting: for the first time, it was clear that very high order contributions
can be computed in perturbative series for lattice gauge theories. Unfortunately
the pioneering NSPT studies of that time were far away from computing the series
up to the orders at which the renormalon growth actually shows up in its full
glory. With limited computing power available, a way out was sought in the form
of a change of scheme (i.e. a scheme in which the renormalon behaviour is best
recognised, possibly at lower orders than in the lattice scheme). Still, the
numerical results were in the end puzzling as for consequences, since trying to
sum the series from the information available even suggested the idea that an
unexpected contribution from a dimension-$2$ operator was
present~\cite{Burgio:1997hc}. Other attempts were made~\cite{Horsley:2001uy},
but it eventually took roughly twenty years before the renormalon behaviour was
actually captured~\cite{Bauer:2011ws,Bali:2013pla,Bali:2014fea,Bali:2014sja}, needless to
say, via NSPT~\footnote{One should note that one of the reason why the renormalon
	growth was correctly reproduced and the OPE correctly reconstructed is the
	adoption of twisted boundary conditions: in this way zero modes are absent and
	the theoretical picture is clear.}. In $\sun(3)$ Yang-Mills theory the IR
renormalon was indeed directly inspected, and the finite-size effects that are
unavoidable on finite lattices assessed. The bottom line is that the victory is
twofold. On one side, the renormalon growth is indeed proved to be present as
conjectured (ironically, in a scheme - the lattice - which one would have
regarded as the very worst to perform the computations). Given this, one has a
prescription to sum the series and perform the subtraction (if sufficiently high
orders are available, one can look for the inversion point in the series, where
contributions start to grow and a minimum indetermination in summing the series
can be attained).

The present work is a first attempt at performing the determination of the gluon
condensate from the plaquette in full QCD, i.e. with fermionic contributions
taken into account. The main focus here is in developing the NSPT technology,
and present a first set of results, which allow a definition of the gluon
condensate. In particular for the first exploration, we use existing Monte Carlo
simulations for the plaquette in full QCD, as detailed below. Having ascertained
that the procedure is viable, a precise determination of the condensate in
full QCD will require a dedicated Monte Carlo simulation, with a careful choice
of the fermionic action. On top of being interesting {\em per se}, the
methodology presented here opens the way to other applications, in which
different colour groups and different matter contents can be investigated. The
final goal would be to inspect whether in a theory that has an IR fixed point,
the renormalon growth is tamed, as one would expect in theories where the
condensates vanish. We defer these questions to future investigations, hoping to
gain extra insight into the task of identifying the boundaries of the conformal
window. 

The paper is organised as follows.
In \mysectionname~\ref{sec:nspt} we review briefly how NSPT can be applied to lattice gauge theories.
In \mysectionname~\ref{sec:tbc} twisted boundary conditions for fermions in the fundamental representation are introduced.
In \mysectionname~\ref{sec:fermions} we discuss how to take into account fermions with smell in NSPT.
We present our results for the expansion of the critical mass of Wilson fermions in \mysectionname~\ref{sec:criticalmass},
and for the expansion of the plaquette with staggered fermions in \mysectionname~\ref{sec:plaquette}.
In \mysectionname~\ref{sec:gluonc} we investigate the asymptotic behaviour of the expansion of the plaquette and extract the gluon condensate in massless QCD.
In \mysectionname~\ref{sec:conclusions} we draw our conclusions and present some possible future steps.

\section{Lattice gauge theories in NSPT}
\label{sec:nspt}
Let us here summarise the main steps in defining NSPT for lattice gauge
theories. Rather than trying to give a comprehensive review of the method, we
aim here to introduce a consistent notation that will allow us to discuss the
new developments in the rest of the paper. For a more detailed discussion of the
NSPT formulation, the interested reader can consult e.g.
\myrefname~\cite{DiRenzo:2004hhl}, whose notation we shall try to follow
consistently~\footnote{For convenience, we summarise our group theory conventions in
	\myappendixname~\ref{sec:group-theory-conv}.}. In particular, we assume to work with a hypercubic lattice
with volume $L^4=a^4N^4$ and assume the lattice spacing $a$ to be $1$, unless
where stated otherwise. We use $x,y,z$ for position indices,
$\mu,\nu,\rho=1,\ldots,4$ for Lorentz indices and
$\alpha,\beta,\gamma=1,\ldots,4$ for Dirac indices.

The original formulation of NSPT is based on the Stochastic Quantization formulation of lattice
field theories, in the case at hand lattice gauge theories. For the purposes of this study, we focus on gauge
theories that are defined by the Euclidean Wilson action for the gauge
group $\sun(N_c)$:
\begin{equation}
	\label{eq:WilsGaugeActionDef}
	S_G\left[U\right] = -\frac{\beta}{2 N_c} 
	\sum_\Box \mathrm{Tr} \left( U_\Box + {U_\Box}^\dagger \right)\, ,
\end{equation}
where $U_\Box$ is the product of the link variables, denoted $U_{\mu}(x)$,
around the $1\times 1$ plaquette $\Box$, and the sum extends over all the
plaquettes in the lattice. Introducing a stochastic time $t$, a field
$U_{\mu}(x;t)$ can be defined that satisfies the Langevin equation
\begin{equation}
	\label{eq:LangevinEq}
	\frac{\partial}{\partial t} U_{\mu}(x;t) = 
	i \Big[
	-\nabla_{x\mu} S_G[U_{\mu}(x;t)] - \eta_{\mu}(x;t)
	\Big] U_{\mu}(x;t)\, .
\end{equation}
As detailed in \myappendixname~\ref{sec:group-theory-conv}, we have denoted by
$\nabla_{x\mu}$ the left derivative in the group; $\eta$ is a
stochastic variable defined in the algebra of the group, 
\begin{equation}
	\label{eq:EtaDef}
	\eta_{\mu}(x;t) = \sum_a T^a \eta_{\mu}^a(x;t) \, ,
\end{equation}
where $T^a$ are the generators of the group, and $\eta_{\mu}^a(x;t)$
are Gaussian variables such that
\begin{equation}
	\label{eq:EtaMeanVar}
	\langle \eta_{\mu}^a(x;t) \rangle = 0\, , \quad 
	\langle \eta_{\mu}^a(x;t)\, \eta^{b}_{\nu}(y;t')\rangle =
	2 \delta^{ab} \delta_{\mu\nu} \delta_{xy} \delta(t-t')\, .
\end{equation}
The key point of Stochastic Quantization is that the large-$t$
distribution of observables built from the solution of the Langevin
equation above corresponds to the distribution that defines the path
integral of the quantum theory~\cite{Parisi:1980ys,Batrouni:1985jn}:
\begin{equation}
	\label{eq:StochQuantExp}
	\lim_{t\to\infty} \langle O[U(t)]\rangle = 
	\frac{1}{Z} \int \mathcal{D}U\, e^{-S_G[U]} O[U]\, .
\end{equation}
In order to develop NSPT, the dynamical variables $U_{\mu}(x;t)$
can be expanded in powers of the coupling constant $g$, which is
given in the lattice formulation by $\beta^{-1/2}$:
\begin{equation}
	\label{eq:NSPTExp}
	U_{\mu}(x;t) \mapsto 
	1 + \sum_{k=1} \beta^{-k/2} U_{\mu}^{(k)}(x;t)\, .
\end{equation}
Solving the Langevin equation, \eq~\eqref{eq:LangevinEq}, order by
order in $\beta^{-1/2}$ yields a system of coupled equations for the
{\em perturbative components} of the link variables
$U_{\mu}^{(k)}(x;t)$.\\
Expanding the solution of Langevin equation in powers of the coupling is a standard
approach to proving the equivalence of stochastic and canonical
quantisation, i.e. \eq~\eqref{eq:StochQuantExp} \cite{Floratos:1982xj}, and was the starting
point for stochastic perturbation theory: with this respect NSPT is just the numerical implementation
of the latter on a computer. The idea of studying the
convergence properties of a stochastic process order by order after an expansion in the
coupling is actually quite general. In this spirit different NSPT schemes can be
set up, also based on stochastic differential equations different from
Langevin \cite{DallaBrida:2017tru,DallaBrida:2017pex}.

\paragraph{Euler integrator}

Discretising the stochastic time in steps of size $\epsilon$ allows a
numerical integration of the Langevin equation,
\begin{equation}
	\label{eq:EulerInt}
	U_{\mu}(x;t+\epsilon) =
	e^{-F_{\mu}(x;t)}\, U_{\mu}(x;t)\, ,
\end{equation}
where the force driving the evolution is 
\begin{align}
	\label{eq:StochForce}
	F_{\mu}(x;t) &= i\left[\epsilon \nabla_{x\mu} S_G[U(t)] + \sqrt{\epsilon}\,
	\eta_{\mu}(x;t)\right] \notag\\
	&= \epsilon\, \frac{\beta}{2 N_c} \sum_{U_\Box \supset U_{\mu}(x)}
	\poa(U_\Box)
	+ \sqrt{\epsilon} \,\eta_{\mu}(x;t)\, 
\end{align}
and the operator $\poa$ projects on the
algebra (see \myappendixname~\ref{sec:group-theory-conv}).
Note that \eq~\eqref{eq:StochForce} does not lend itself to a
perturbative solution in powers of $\beta^{-1/2}$, since there is a
mismatch between the deterministic drift term, which starts at order
$\beta^{1/2}$, and the stochastic noise, which is of order
$\beta^0$. This is easily resolved by rescaling the integration step
by a factor of $\beta$, so that both contributions start at order
$\beta^{-1/2}$. Denoting the new time step $\tau = \epsilon \beta$,
the force term becomes
\begin{equation}
	\label{eq:StochForceTwo}
	F_{\mu}(x;t) = \frac{\tau}{\beta}  \nabla_{x\mu} S_G[U(t)] +
	\sqrt{\frac{\tau}{\beta}} \,
	\eta_{\mu}(x;t) \, .
\end{equation}
Expanding $F$ in powers of $\beta^{-1/2}$,
\begin{equation}
	\label{eq:FExp}
	F_{\mu}(x;t) = \sum_{k=1} \beta^{-k/2} F^{(k)}_{\mu}(x;t)\, ,
\end{equation}
leads to a system of coupled equations for the evolution of the
coefficients of the perturbative expansion of $U$. Omitting Lorentz and position indices, we get
\begin{subequations}
	\label{eq:eqmotionNSPT}
	\begin{align}
		U^{(1)}(t+\tau) &= U^{(1)}(t) - F^{(1)}(t) \\
		U^{(2)}(t+\tau) &= U^{(2)}(t) - F^{(2)}(t) + \frac12 F^{(1)}(t)^2 
		- F^{(1)}(t) U^{(1)}(t) \\
		&\ldots\notag
	\end{align}
\end{subequations}
where $\eta$ only contributes to the $F^{(1)}$ term.

\paragraph{Stochastic gauge fixing}

The zero modes of the gauge action do not generate a deterministic
drift term in the Langevin equation, and therefore their evolution in
stochastic time is entirely driven by the stochastic noise, which
gives rise to diverging fluctuations. This phenomenon is well known
since the early days of NSPT, see e.g. \myrefname~\cite{DiRenzo:1994av}, and
is cured by the so-called stochastic gauge fixing
procedure~\cite{Zwanziger:1981kg} applied to the theory formulated on
the lattice. The procedure implemented in this work alternates an
integration step as described above with a gauge
transformation:
\begin{equation}
	\label{eq:StochGaugeFix}
	U_{\mu}(x) \mapsto e^{w(x)} U_{\mu}(x) e^{-w(x+\hat\mu)}\, ,
\end{equation}
where the field $w(x)$ is defined in the algebra of the group,
\begin{equation}
	\label{eq:StochGaugeTransf}
	w(x) = - \alpha \poa\left(\sum_\mu \nabla^*_\mu U_{\mu}(x)\right)\,.
\end{equation}
$\alpha$ is a free parameter, which we choose equal to $0.1$ and $\nabla^*_\mu$ is the backward derivative in direction $\mu$.
Note that there is nothing compelling in the choice for $w(x)$. 
In this work we make the same choice as in \myrefname~\cite{DiRenzo:1994av},
which is slightly different from the one adopted
in \myrefname~\cite{DiRenzo:2004hhl}: the corresponding gauge transformation does 
not lead, if iterated, to the Landau gauge. In NSPT the gauge transformation is
expanded in powers of the 
coupling, 
\begin{equation}
	\label{eq:StochGaugePert}
	w(x) = \sum_{k=1} \beta^{-k/2} w^{(k)}(x)\, ,
\end{equation}
and the transformation in \eq~\eqref{eq:StochGaugeFix} is implemented
order by order in perturbation theory. 

The combined step for the integrator adopted in this work can be
summarised as
\begin{subequations}
	\begin{align}
		U_{\mu}(x)' &= e^{-F_{\mu}(x;t)}\, U_{\mu}(x;t)\, , \\
		U_{\mu}(x;t+\tau) &= e^{w[U'](x)} U_{\mu}(x)' e^{-w[U'](x+\hat\mu)}\, ,
	\end{align}
\end{subequations}
where all the terms are expanded in powers of $\beta^{-1/2}$, and the
perturbative components are updated.

\paragraph{Runge-Kutta integrator}

Higher order integrators, in particular Runge-Kutta schemes, have been
used for the lattice version of the Langevin equation since the early
days~\cite{Batrouni:1985jn}. A new, very effective second-order
integration scheme 
for NSPT in lattice gauge theories has been introduced
in \myrefname~\cite{Bali:2013pla}. While we have tested Runge-Kutta schemes
ourselves for pure gauge NSPT simulations, in this work we adhere to
the simpler Euler scheme: when making use of the (standard) stochastic
evaluation of the fermionic equations of motion (see \mysectionname~\ref{sec:fermions}), Runge-Kutta schemes
are actually more demanding (extra terms are needed \cite{Batrouni:1985ye,Kronfeld:1986dv}).

\section{Twisted boundary conditions and smell}
\label{sec:tbc}
When a theory is defined in finite volume, the fields can be required
to satisfy any boundary conditions that are compatible with the
symmetries of the action. We adopt twisted boundary conditions
(TBC)~\cite{tHooft:1979rtg} in order to remove the zero-mode of the
gauge field, and have an unambiguous perturbative expansion, which is
not plagued by toron vacua~\cite{GonzalezArroyo:1981vw}. The gauge
fields undergo a constant gauge transformation when translated by a
multiple of the lattice size; therefore twisted boundary conditions in
direction $\hat\nu$ are 
\begin{equation}
	\label{eq:gaugetwist} 
	U_\mu(x+L\hat\nu)=\Omega_\nu
	U_\mu(x)\Omega_\nu^\dag\,, 
\end{equation} 
where $\Omega_\mu \in \sun(N_c)$ are a set of constant matrices satisfying
\begin{equation} 
	\label{eq:twistalgebra} 
	\Omega_\nu\Omega_\mu =
	z_{\mu\nu} \Omega_\mu\Omega_\nu\,,\qquad z_{\mu\nu}\in Z_{N_c}\,.
\end{equation}

Fermions in the adjoint representation
can be introduced in a straightforward manner; the boundary conditions
with the fermionic field in the matrix representation read
\begin{equation} 
	\label{eq:AdjFermionBC}
	\psi(x+L\hat\nu)=\Omega_\nu \psi(x)\Omega_\nu^\dag\,.
\end{equation}

The inclusion of fermions in the fundamental representation is not
straightforward; indeed, the gauge transformation for the fermions
when translated by a multiple of the lattice size reads
\begin{equation}
	\label{eq:fundfermwrong}
	\psi(x+L\hat\nu)=\Omega_\nu \psi(x)\, ,
\end{equation}
leading to an ambiguous definition of
$\psi(x+L\hat\mu+L\hat\nu)$.  An idea to overcome this
problem, proposed in \myrefname~\cite{Parisi:1984cy} and implemented e.g. in
\myrefname~\cite{Hao:2007iz}, is to introduce a new quantum number so that
fermions exist in different copies, or \emph{smells}, which transform
into each other according to the antifundamental representation of
$\sun(N_c)$.  The theory has a new global symmetry, but physical
observables are singlets under the smell group. Thus, configurations
related by a smell transformations are equivalent, and in finite volume
we are free to substitute \eq~\eqref{eq:fundfermwrong} with
\begin{equation}
	\label{eq:smellfundamentaltransformation}
	\psi(x+L\hat\nu)_{ir}=\sum_{j,s}\big(\Omega_\nu\big)_{ij}
	\psi(x)_{js}\big(\Lambda_\nu^\dag\big)_{s
		r}\, , 
\end{equation}
where $\Lambda_\nu\in \sun(N_c)$. It is useful to think of the fermion
field as a matrix in colour-smell space.  If the transformation
matrices in smell space satisfy the same relations as in
\eq~\eqref{eq:twistalgebra} (in particular we choose them to be equal
to the $\Omega$s), then twisted boundary conditions are
well-defined.

It is worth pointing out that, through a change of variable in the path integral~\cite{GonzalezArroyo:1982hz,Luscher:1985wf}, twisted boundary conditions could be equivalently implemented by multiplying particular sets of plaquettes in the action by suitable elements of $Z_{N_c}$ and considering the fields to be periodic.
This change of variable works only in the pure gauge or fermions in the adjoint representation cases. Thus, the explicit transformation of
\eq~\eqref{eq:smellfundamentaltransformation} is required when fermions in the fundamental representation with smell are considered.

\section{Fermions in NSPT}
\label{sec:fermions}
If $S_F=\sum_{x,y}\bar\psi(x) M[U] \psi(y)$ is the action of a single fermion, then dynamical fermions in NSPT can be included thanks to a new term in the drift, as shown in \myrefsname~\cite{Batrouni:1985jn,DiRenzo:2000qe}: the determinant arising from $N_f$ degenerate fermions can be rewritten as
\begin{equation}
	\det(M)^{N_f} = \exp\left(N_f\Tr\ln M\right)
\end{equation}
and can be taken into account by adding $-{N_f}\Tr\ln M$ to the gauge action. From the Lie derivative of the additional term and recalling that a rescaled time step $\tau=\epsilon/\beta$ is used in the Euler update, we obtain the new contribution
\begin{equation}
	\label{eq:exactdrift}
	F^{f}_\mu(x)= -i\,\frac{\tau N_f}{\beta}\sum_a T^a \Tr(\nabla^a_{x\mu} M) M^{-1}
\end{equation}
to be added to the pure gauge drift.
It is important to note that the coefficient of $iT^a$ is purely real because the Wilson operator is $\gamma_5$-Hermitian and the staggered operator is antihermitian: this is consistent with the drift being an element of the algebra.
The trace can be evaluated stochastically: \eq~\eqref{eq:exactdrift} is replaced by
\begin{equation}
	F^{f}_\mu(x) =-i\frac{\tau N_f}{\beta}\sum_a T^a \re \xi^*(\nabla^a_{x\mu} M) M^{-1}\xi
\end{equation}
thanks to the introduction of a new complex Gaussian noise $\xi$ satisfying~\footnote{
	Obviously $\xi$ does not have any Dirac structure in the staggered case. The noise can be built from the independent generation of real and imaginary part with zero mean and variance $1/2$.}
\begin{equation}
	\braket{\xi^*(y)_{\beta i r}\xi(z)_{\gamma js}} = \delta_{yz}	\delta_{\beta\gamma}\delta_{ij}\delta_{rs}\,.
\end{equation}
The real part must be enforced, otherwise the dynamics would lead the links out of the group since the drift would be guaranteed to be in the algebra only on average. In NSPT, the Dirac operator inherits a formal perturbative expansion from the links, $M=\sum_{n=0}^\infty \beta^{-n} M^{(n)}$, so the inverse $\psi=M^{-1}\xi$ can be computed efficiently from the knowledge of the inverse free operator via the recursive formula
\begin{subequations}
	\begin{align}
		\psi^{(0)}&={M^{(0)}}^{-1}\xi\\
		\psi^{(n)}&=-{M^{(0)}}^{-1}\sum_{j=0}^{n-1}M^{(n-j)}\psi^{(j)}\,.
	\end{align}
\end{subequations}
The inverse of the free operator is conveniently applied in Fourier space.

If fermions have smell, then the rescaling $N_f\to N_f/N_c$ is required in order to have $N_f$ flavours in the infinite-volume limit. In other words, this is the same as considering the $N_c$th root of the determinant of the fermion operator. In principle such rooted determinant could come from a nonlocal action, because twisted boundary conditions break the invariance under smell transformations. Nevertheless, this rooting procedure is sound since we know in advance that in the infinite-volume limit all the dependence on boundary conditions will be lost and the determinant will factorise as the fermion determinant of a single smell times the identity in smell space. It is also possible to show with arguments similar to those presented in \myrefname~\cite{Sharpe:2006re} that, if the theory without smell is renormalisable, this operation leads to a perturbatively renormalisable theory as well.
Below we describe in detail Wilson and staggered fermions in the fundamental representation, so we explicitly rescale $N_f\to N_f/N_c$.  It is also important to remember that the fermion field, seen as a matrix in colour-smell space, is not required to be traceless, thus its Fourier zero-mode does not vanish: we require antiperiodic boundary conditions in time direction not to hit the pole of the free propagator in the massless case.
We avoid twisted boundary conditions in time direction because in the massless case it might happen for the free fermion propagator to develop a pole at some particular momenta.

\subsection{Wilson fermions}
The Wilson Dirac operator and its Lie derivative are
\begin{subequations}
	\begin{align}
		M_{y\beta i r, z\gamma j s} &=(m+4)\delta_{rs}\delta_{yz} \delta_{\beta\gamma}\delta_{ij}+\sum_\mu\left[D(\mu)+\gamma_5 D(\mu)^\dag\gamma_5\right]_{y\beta i r, z\gamma j s}\\
		\nabla^a_{x,\mu}M_{y\beta i r, z\gamma j s} &= i\delta_{xy}[T^aD(\mu)]_{y\beta i r, z\gamma j s}-i\delta_{xz}[\gamma_5D(\mu)^\dag\gamma_5T^a]_{y\beta i r, z\gamma j s}\,,
	\end{align}
\end{subequations}
where the non-diagonal term has been expressed through
\begin{equation}
	D(\mu)_{y\beta i r, z\gamma j s}=-\frac{1}{2}\delta_{rs}\delta_{y,z-\hat\mu}(1-\gamma_\mu)_{\beta\gamma}U_\mu(y)_{ij}\,.
\end{equation}
We must give a perturbative structure to the mass $m=\sum_{n=0}^\infty \beta^{-n} m^{(n)}$ to account for an additive mass renormalisation, see \mysectionname~\ref{sec:criticalmass}.
The stochastic evaluation of the trace leads to
\begin{equation}
	\label{eq:stoctrace}
	\xi^*(\nabla^a_{x\mu} M) M^{-1}\xi=
	i\Tr T^a \sum_\beta\left( \varphi^{(\mu)}(x)_\beta\,\xi(x)^\dag_\beta-\psi(x)_\beta\,\tilde\varphi^{(\mu)}(x)_\beta^\dag\right)\,,
\end{equation}
where $\varphi^{(\mu)}=D(\mu)\psi$, $\tilde\varphi^{(\mu)}=\gamma_5D(\mu)\gamma_5\xi$ and the fermion fields have been represented as matrices in colour-smell space. After taking the real part, the fermion drift can be finally written as
\begin{align}
	\label{eq:wilsondrift}
	F^{f}_\mu(x)_{ij}&=\frac{1}{2}\frac{N_f}{N_c}\frac{\tau}{\beta}\sum_a T^a_{ij}
	\Tr T^a \sum_\beta\left[\left( \varphi^{(\mu)}(x)_\beta\,\xi(x)^\dag_\beta-\psi(x)_\beta\,\tilde\varphi^{(\mu)}(x)_\beta^\dag\right)-\text{h.c.}\right]=\notag\\
	&=\frac{1}{2}\frac{N_f}{N_c}\frac{\tau}{\beta}\poa\left[\sum_\beta\left( \varphi^{(\mu)}(x)_\beta\,\xi(x)^\dag_\beta+\tilde\varphi^{(\mu)}(x)_\beta\,\psi(x)_\beta^\dag\right)\right]_{ij}\,.
\end{align}
In \myappendixname~\ref{sec:driftoptimisation} the actual implementation of the fermion drift is described (only one of the two terms in \eq~\eqref{eq:wilsondrift} is actually needed).

With the Fourier transform described in \myappendixname~\ref{sec:appendixft}, the inverse free Wilson operator with twisted boundary conditions is diagonal in momentum space and can be expressed as
\begin{equation}
	\label{eq:wilsonpropagator}
	{M^{(0)}}^{-1}_{k,p}=\delta_{k_\parallel p_\parallel}\delta_{k_\perp p_\perp}\frac{2\sum_{\mu}\sin^2\frac{k_\mu}{2}+m^{(0)}-i\sum_{\mu}\gamma_\mu \sin k_\mu}{\left(2\sum_{\mu}\sin^2\frac{k_\mu}{2}+m^{(0)}\right)^2+\sum_{\mu} \sin^2 k_\mu}\,.
\end{equation}

\subsection{Staggered fermions}
We implemented for the first time staggered fermions in NSPT.
The staggered field has no Dirac structure and describes four physical fermions in the continuum limit. Therefore, we rescale $N_f\to N_f/4$ and the staggered operator is understood to be rooted when the number of flavour is not a multiple of four.
The staggered Dirac operator and its Lie derivative are
\begin{subequations}
	\begin{align}
		M_{y i r, z j s} &=m\delta_{rs}\delta_{yz}	\delta_{ij}+\sum_\mu\left[D(\mu)-D(\mu)^\dag\right]_{y i r, z j s}\\
		\nabla^a_{x,\mu}M_{y i r, z j s} &= i\delta_{xy}[T^aD(\mu)]_{y i r, z j s}+i\delta_{xz}[D(\mu)^\dag T^a]_{y i r, z j s}\,,
	\end{align}
\end{subequations}
where the non-diagonal term has been expressed through
\begin{equation}
	D(\mu)_{y i r, z j s}=\frac{1}{2}\alpha_\mu(y)\delta_{rs}\delta_{y,z-\hat\mu}U_\mu(y)_{ij}
\end{equation}
and $\alpha_\mu(x)=(-1)^{\sum_{\nu=1}^{\mu-1}x_\nu}$ is the staggered phase.
The stochastic evaluation of the trace is analogous to the Wilson fermion case and \eq~\eqref{eq:stoctrace} becomes
\begin{equation}
	\label{eq:staggereddrift}
	\xi^*(\nabla^a_{x\mu} M) M^{-1}\xi=
	i\Tr T^a \left( \varphi^{(\mu)}(x)\,\xi(x)^\dag-\psi(x)\,\tilde\varphi^{(\mu)}(x)^\dag\right)\,,
\end{equation}
with $\varphi^{(\mu)}=D(\mu)\psi$ and $\tilde\varphi^{(\mu)}=-D(\mu)\xi$, leading to the final expression
\begin{equation}
	F^{f}_\mu(x)_{ij}=\frac{1}{2}\frac{N_f}{4N_c}\frac{\tau}{\beta}\poa\left( \varphi^{(\mu)}(x)\,\xi(x)^\dag+\tilde\varphi^{(\mu)}(x)\,\psi(x)^\dag\right)_{ij}\,.
\end{equation}
Again, the actual implementation of the staggered drift is shown in \myappendixname~\ref{sec:driftoptimisation}.

With the Fourier transform described in \myappendixname~\ref{sec:appendixft}, the inverse free staggered operator with twisted boundary conditions is found to be
\begin{equation}
	\label{eq:staggeredpropagator}
	{M^{(0)}}^{-1}_{k,p}=\delta_{k_\perp p_\perp}\frac{m\delta_{k_\parallel p_\parallel}-i\sum_\mu\sin k_\mu\,\bar\delta(k_\parallel+\pi\bar\mu-p_\parallel)}{\sum_\mu \sin^2 k_\mu+m^2}\,,
\end{equation}
where $\bar 1=0$, $\overline{\mu+1}=\bar\mu+\hat\mu$ and $\bar\delta$ is the periodic Kronecker delta, with support in $0\mod2\pi$.
The propagator is not diagonal in momentum space because the action depends explicitly on the position through $\alpha_\mu(x)$, but it is simple enough to avoid a complete matrix multiplication over all the degrees of freedom. If we aim to compute ${M^{(0)}}^{-1}v$ for some field $v$ in momentum space, it is useful to represent $v(p_\parallel)_{p_\perp}$ as matrices $N_c\times N_c$ with indices ${\tilde n_1,\tilde n_2}$ defined at each $p_\parallel$ site $(n_1,n_2,n_3,n_4)$ (see again \myappendixname~\ref{sec:appendixft}). Then the non-diagonal terms become diagonal when shifting iteratively $v$ by $L/2$ in the $p_\parallel$ space. Incidentally, we must consider $L$ to be even so that at the same time $L/2$ is well defined and (in the massless case) no spurious pole is hit when \eq~\eqref{eq:staggeredpropagator} is evaluated in finite volume: this stems from the fact that the staggered action is only invariant under translation of two lattice spacings, therefore twisted boundary conditions would be inconsistent for $L$ odd.

\section{The critical mass of Wilson fermions}
\label{sec:criticalmass}
The inverse of the Wilson fermion propagator in momentum space can be expressed
as
\begin{align}
	a\Gamma(ap,am,\beta^{-1}) &= aS(ap,am,\beta^{-1})^{-1} = \nonumber \\
	\label{eq:GammaDef}	
	&= i\sum_\mu\gamma_\mu 
	\overline{(ap_\mu)}+\frac{1}{2}\widehat{(ap)}^2+am-
	a\Sigma(ap,am,\beta^{-1})\,,
\end{align}
where $\bar v_\mu = \sin v_\mu$, $\hat v_\mu = 2\sin(\frac{v_\mu}{2})$ and
$\Sigma(ap,am,\beta^{-1})$ is the self energy. In this section the lattice
spacing $a$ is written explicitly.
Wilson fermions are not equipped with chiral symmetry when the bare mass
$m$ vanishes: the self
energy at zero momentum is affected by a power divergence $a^{-1}$, which has to
be cured by an additive renormalisation.
In an on-shell renormalisation scheme, the
critical value of the bare mass, $m_c$, for which the lattice theory describes
massless fermions, is given by the solution of
\begin{equation}
	\label{eq:onshellscheme}
	am_c - a\Sigma(ap=0,am_c,\beta^{-1}) = 0\,.
\end{equation}
As observed in \myrefname~\cite{Bochicchio:1985xa}, this prescription matches the one
obtained by requiring the chiral Ward identity to hold in the continuum limit.
Expanding \eq~\eqref{eq:onshellscheme} defines the critical mass order by
order in perturbation theory. The perturbative expansion of the inverse
propagator is 
\begin{equation}
	\label{eq:GammaExpansion}
	a\Gamma(ap,am,\beta^{-1}) = \sum_{n=0} \Gamma^{(n)}\left(ap, am \right) \beta^{-n} \, ,
\end{equation}	
where we have indicated explicitly the dependence of the coefficients on the bare mass $am$.
The functions $\Gamma^{(n)}(ap,am)$ are matrices in Dirac space; since we are interested in the small momentum region and $\Gamma^{(n)}(0,am)$ is proportional to the identity, we consider $\Gamma^{(n)}(ap,am)$ as scalar functions: when $ap\neq0$ a projection onto the identity is understood.
Plugging the perturbative expansion of the critical mass
\begin{equation}
	\label{eq:CritMassExpansion}
	am_c = \sum_{n=1} m_c^{(n)} \beta^{-n}
\end{equation}
into \eq~\eqref{eq:GammaExpansion} results in
\begin{equation}
	\label{eq:gammaDefinition}
	a\Gamma(ap,am_c,\beta^{-1}) = \sum_{n=0} \gamma^{(n)}\left(ap\right) \beta^{-n} = \sum_{n=0} \left[m_c^{(n)}+\chi^{(n)}\left(ap\right) \right] \beta^{-n} \, ,
\end{equation}
where the dependence of $\gamma^{(n)}$ on $m_c^{(n)}$ has been made explicit and $\chi^{(n)}$ depends only on $m_c^{(0)},\dots,m_c^{(n-1)}$. Therefore, the renormalisation condition in \eq~\eqref{eq:onshellscheme} becomes order by order
\begin{equation}
	\label{eq:renormalisationconditiongamma}
	\gamma^{(n)}(0)=0
	\qquad\text{or}\qquad
	m_c^{(n)}=-\chi^{(n)}(0)\,.
\end{equation}
For illustration, we can compute the recursive solution of
\eq~\eqref{eq:onshellscheme} at first- and second-order in the expansion in
powers of $\beta^{-1}$, which yields
\begin{subequations}
	\label{eq:RecSolnFirstOrder}
	\begin{align}
		& \gamma^{(1)}(0)= \Gamma^{(1)}(0,0) + m_c^{(1)} = 0 \, , \\
		\label{eq:RecSolnSecondOrder}
		& \gamma^{(2)}(0) = m^{(1)}_c \,\frac{\partial \Gamma^{(1)}}{\partial(am)}\bigg|_{ap=0,am=0}+ \Gamma^{(2)}(0,0) + m_c^{(2)}= 0\, .
	\end{align}
\end{subequations}
Both results are familiar from analytical calculations of the critical mass. The
first equation encodes the fact that the mass counterterm at first order in
perturbation theory is given by the one-loop diagrams computed at zero bare
mass. The second equation states that the second-order correction is given by
summing two-loop diagrams evaluated at vanishing bare mass, and one-loop
diagrams with the insertion of the $O\left(\beta^{-1}\right)$ counterterm, see
e.g. \myrefname~\cite{Follana:2000mn}.

It should also be noted that, when working in finite volume, momenta are
quantised. Unless periodic boundary conditions are used, $p=0$ is not an allowed
value for the momentum of the states in a box. Therefore, condition
\eqref{eq:onshellscheme} can only be imposed after extrapolating the value of
$\Sigma$ to vanishing momentum. The detailed implementation is discussed below
in \mysectionname~\ref{sec:cmprocedure}.

Critical masses have been computed analytically up to two
loops~\cite{Follana:2000mn,Caracciolo:2001ki}, and in NSPT at
three and four loops~\cite{DiRenzo:2004ju,DiRenzo:2006qtj}. High-order
perturbation theory with massless Wilson fermions requires the tuning of the
critical mass at the same order in $\beta^{-1}$, and it is possible to
determine this renormalisation using NSPT. Let us illustrate the
strategy in detail. We begin by collecting configurations for different time
steps $\tau$ of the stochastic process; for each configuration the gauge is
fixed to the Landau gauge~\cite{Rossi:1987hv,Davies:1987vs}. The propagator at momentum $p$ is
computed by applying the inverse Dirac operator to a point source in momentum
space,
\begin{equation}
	\label{eq:propagatormeasure}	
	S(p)_{\alpha\beta}= \langle\sum_{q\gamma} 
	M\left[U\right]^{-1}_{pq,\alpha\gamma}\delta_{qp}\delta_{\gamma\beta}\rangle_\text{MC}\,.
\end{equation}
For each simulation at a given value of $\tau$, the error bars are computed as detailed in \myappendixname~\ref{sec:autocorrelation}.
The propagator with periodic boundary conditions is a (diagonal) matrix in colour and momentum space and has a Dirac structure; it is important to stress again that with TBC there is not a colour structure any more and the momentum has a finer quantisation.
The average over all the configurations gives the Monte Carlo estimate of $S(p)$. We
can now extrapolate the stochastic time step to zero and invert the propagator
to obtain $S(p)^{-1}$. Finally, the
inverse propagator is projected onto the identity in Dirac space.
All these operations are performed
order by order in perturbation theory keeping in mind that, after the measure of
the propagator, all perturbative orders $\beta^{-k/2}$ with an odd $k$ are
discarded, since the expansion in powers of $\beta^{-1/2}$ is an artefact of
NSPT. The errors can be estimated by bootstrapping the whole procedure.

The legacy of this process is the measure of the functions $\gamma^{(n)}(ap)$, as it is clear from \eq~\eqref{eq:gammaDefinition}. The renormalisation condition in \eq~\eqref{eq:renormalisationconditiongamma} must then be imposed: this can be done iteratively one order after the other.
When all the coefficients up to some $m_c^{(n)}$ are included in the simulation, all the $\gamma$ functions up to $\gamma^{(n)}(ap)$ extrapolate to zero; on the other hand, from $\gamma^{(n+1)}(0)$ we can read $-m_c^{(n+1)}$.
In order to move on and compute
the following coefficient of the critical mass, a new set of configurations where $m_c^{(n+1)}$ is taken into account must be generated.

The procedure we described is well defined and even theoretically clean, since it enlightens the status of our $m_c$ as a perturbative additive renormalisation: once it is plugged in at a given order, the renormalised mass turns out to be zero at the prescribed order. On the other side, it is not at all the only possible procedure. The prescription of the authors of \myrefname~\cite{DallaBrida:2017pex} is to expand the solution of the stochastic process both in the coupling and in the mass counterterm. 
This is in the same spirit of \myrefname~\cite{DiRenzo:2008en}: the solution of the stochastic process can be expanded in more than one parameter and once a precise power counting is in place, the resulting hierarchy of equations can be exactly truncated at any given order. 
There are pros and contras for both approaches, i.e. the one we followed and the double expansion. The latter can provide a better handle on estimating errors due to the critical mass value; on the other side, it is expected to be numerical more demanding. All in all, we did not push Wilson fermions to very high orders: moving to the staggered formulation was by far the most natural option for the purpose of this work.

\subsection{Zero-momentum extrapolation and valence twist}
\label{sec:cmprocedure}

Since in finite volume it is possible to measure $\Gamma(ap)$ only for
discretised non-zero momenta, the data need to be extrapolated to zero
momentum using a suitable functional form. The strategy adopted in the
literature -- see for example \eqs~(13) and (14) in
\myrefname~\cite{DiRenzo:2006qtj} -- is based on expanding the quantities of
interest in powers of $ap$. In the infinite-volume limit, such an expansion
leads to a hypercubic symmetric Taylor expansion composed of invariants in $ap$,
logarithms of $ap$ and ratios of invariants; an explicit one-loop computation to
order $a^2$ is shown e.g. in \eq~(24) of \myrefname~\cite{Constantinou:2009tr}. The
ratios and the logarithms arise because we are expanding a nonanalytic function
of the lattice spacing: infrared divergences appear when expanding the
integrands in $ap$. On the other hand, working consistently in finite volume
does not cause any infrared divergence: expressions for $\gamma^{(n)}(ap)$ will be just sums of ratios of trigonometric functions,
which we can expand in $ap$ obtaining simply a combination of polynomial lattice
invariants~\footnote{Expanding in $ap$ and sending the lattice size to infinity
	are operations that do not commute; in particular this gives rise to different
	series in the finite- and infinite-volume cases.}.

Still, this is not enough for a reliable extrapolation to vanishing
momenta. In order to understand better the range of momenta that allow a
reliable extrapolation, we computed $\gamma^{(1)}(ap)$ in twisted lattice
perturbation theory (see \myappendixname~\ref{sec:twistedpt}). As a cross-check of our
calculation we verified that $\gamma^{(1)}(0)$ is gauge-invariant (this result
must be true at all orders because of the gauge-invariance of the pole
mass~\cite{Kronfeld:1998di}). It can be seen from the analytic
expansion of $\gamma^{(1)}(ap)$ that even the lowest momentum allowed on our
finite-size lattices, $ap_{1,2,3}=0$, $ap_4 = \pi/L$, is far from the convergence region of this
series. This happens even for reasonably big lattices, $L\lesssim32$. In order
to increase the range of available momenta, we use $\theta$-boundary
conditions~\cite{Bedaque:2004kc} for the valence fermions,
\begin{equation}
	\psi(x+L\hat4)=e^{i\theta}\psi(x)\,,
\end{equation}
thereby reaching momenta $p_4 = \theta/L$ which are within the
convergence radius of the $ap$-expansion. The hypercubic series becomes just a
polynomial in $(ap_4)^2$ by setting all the other components to zero.

\begin{figure}[tbp]
	\centering
	\begin{subfigure}{.49\textwidth}
		\includegraphics[width=\textwidth,trim={0.1cm 0 1.4cm 0},clip]{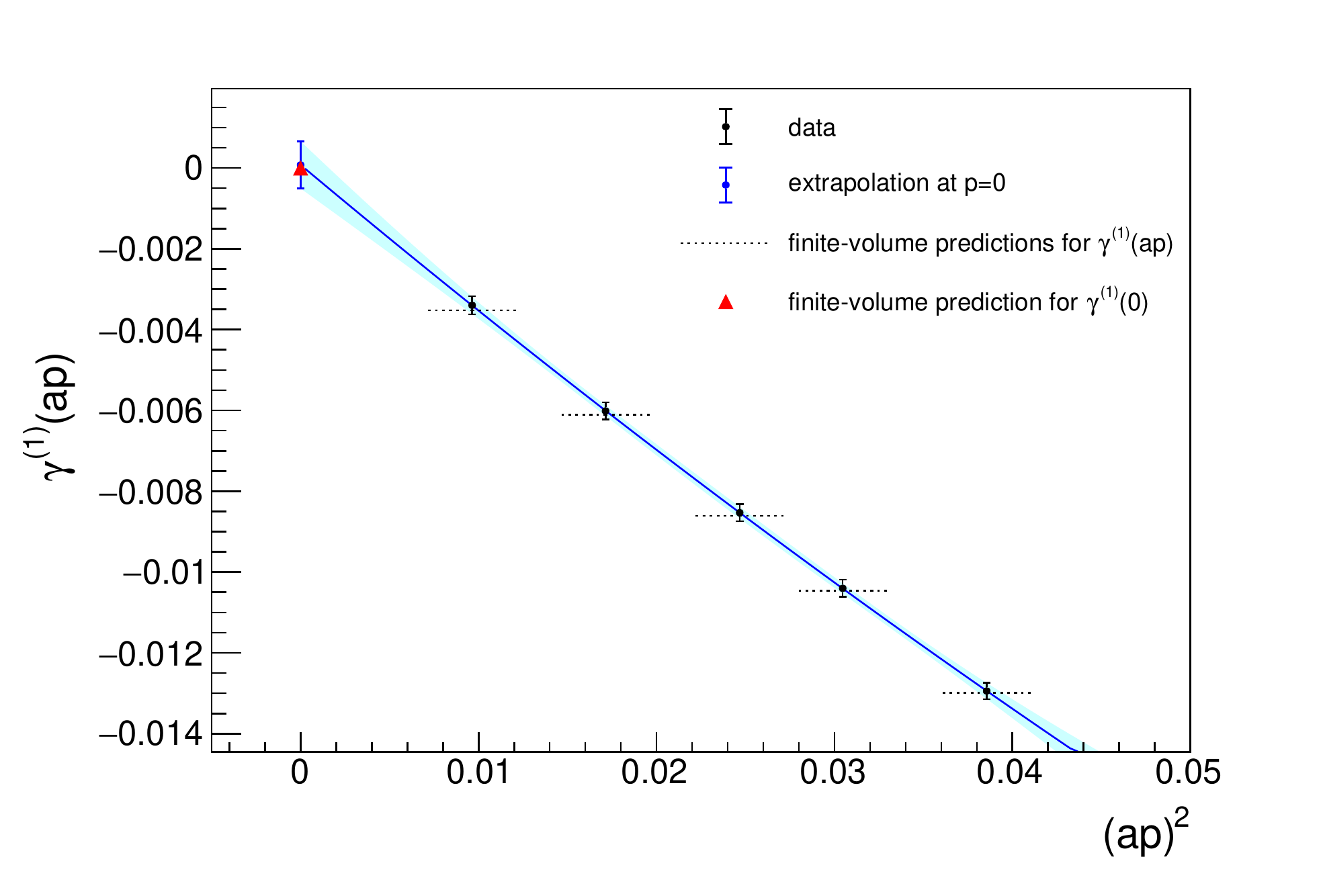}
	\end{subfigure}%
	\begin{subfigure}{.49\textwidth}
		\includegraphics[width=\textwidth,trim={0.1cm 0 1.4cm 0},clip]{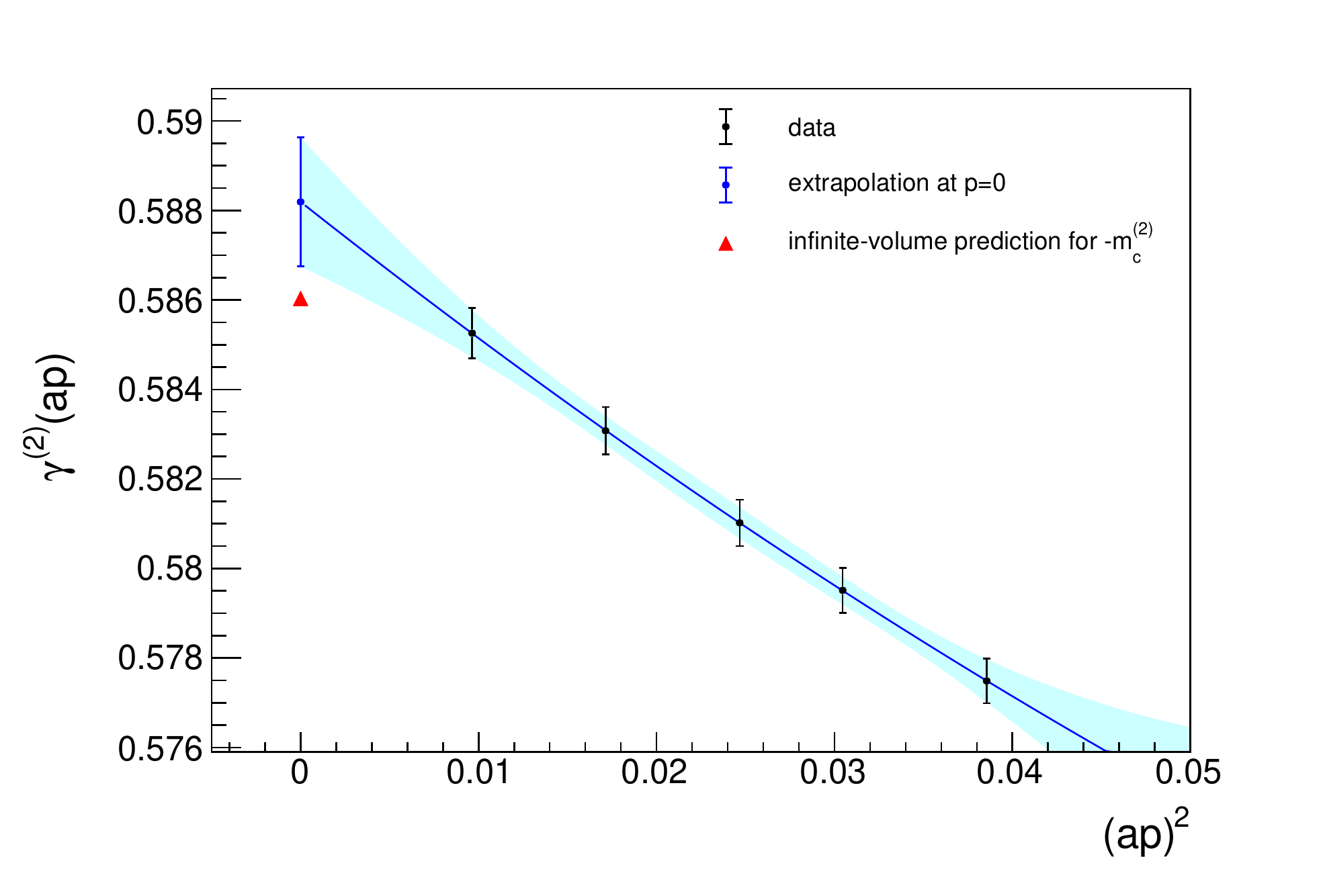}
	\end{subfigure}
	\caption{\label{fig:cmplot}Measure of $\gamma^{(1)}(ap)$ (left panel) and
		$\gamma^{(2)}(ap)$ (right panel) for a $12^4$ lattice with twisted boundary
		conditions on a plane, $N_c = 2$ and $N_f = 2$ Wilson fermions. The
		analytic finite-volume critical mass $m_c^{(1)}$ is included in the
		simulation. A second-order polynomial in $(ap)^2$ is used for fitting. Most
		analytic finite-volume predictions have been drawn as lines to help the eye
		in the comparison. The difference with the prediction in the right panel is
		to be ascribed to the fact that we are able to resolve finite volume
		effects.}
\end{figure}

\begin{figure}[tbp]
	\centering
	\begin{subfigure}{.49\textwidth}
		\includegraphics[width=\textwidth,trim={0.1cm 0 1.4cm 0},clip]{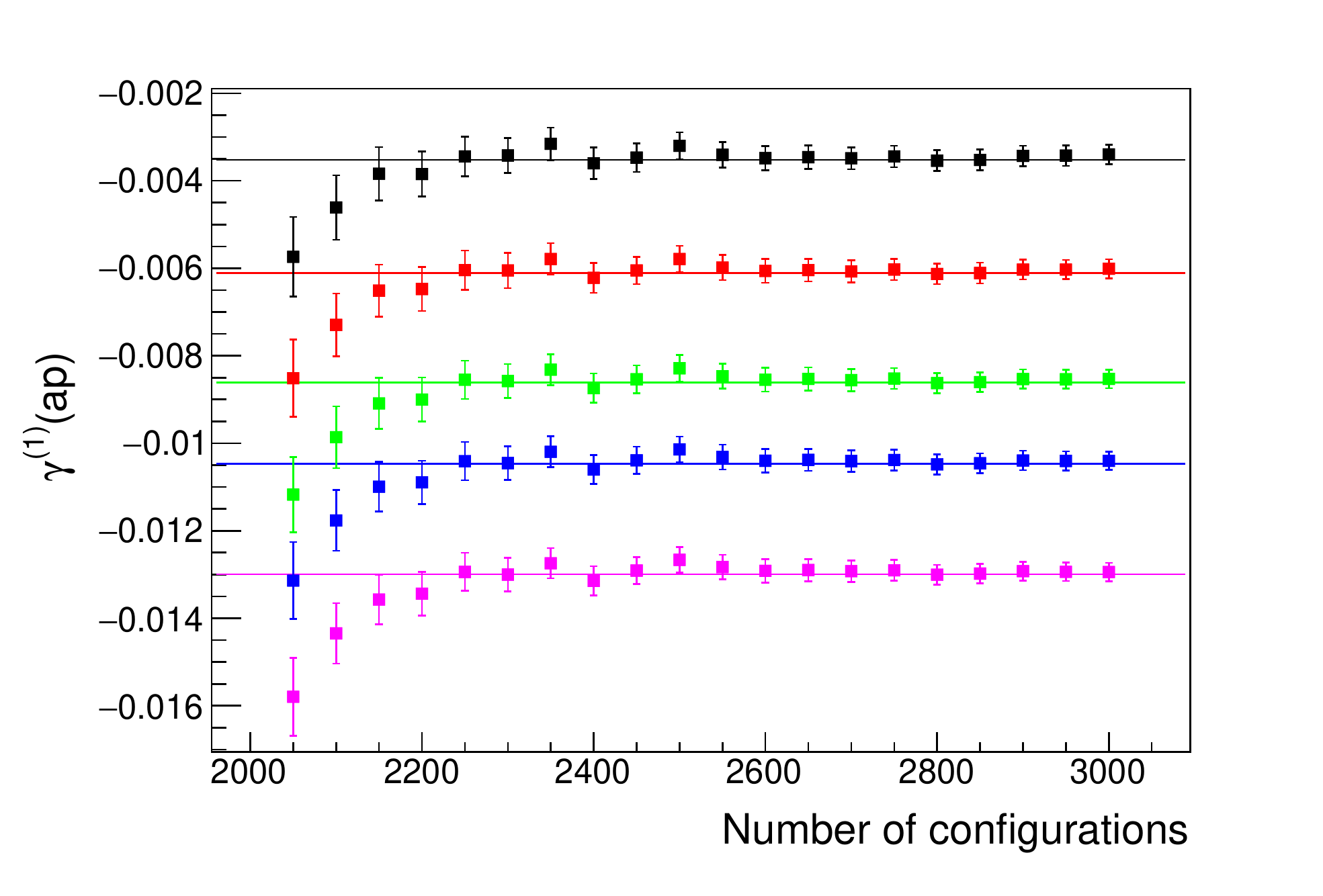}
	\end{subfigure}%
	\begin{subfigure}{.49\textwidth}
		\includegraphics[width= \textwidth,trim={0.1cm 0 1.4cm 0},clip]{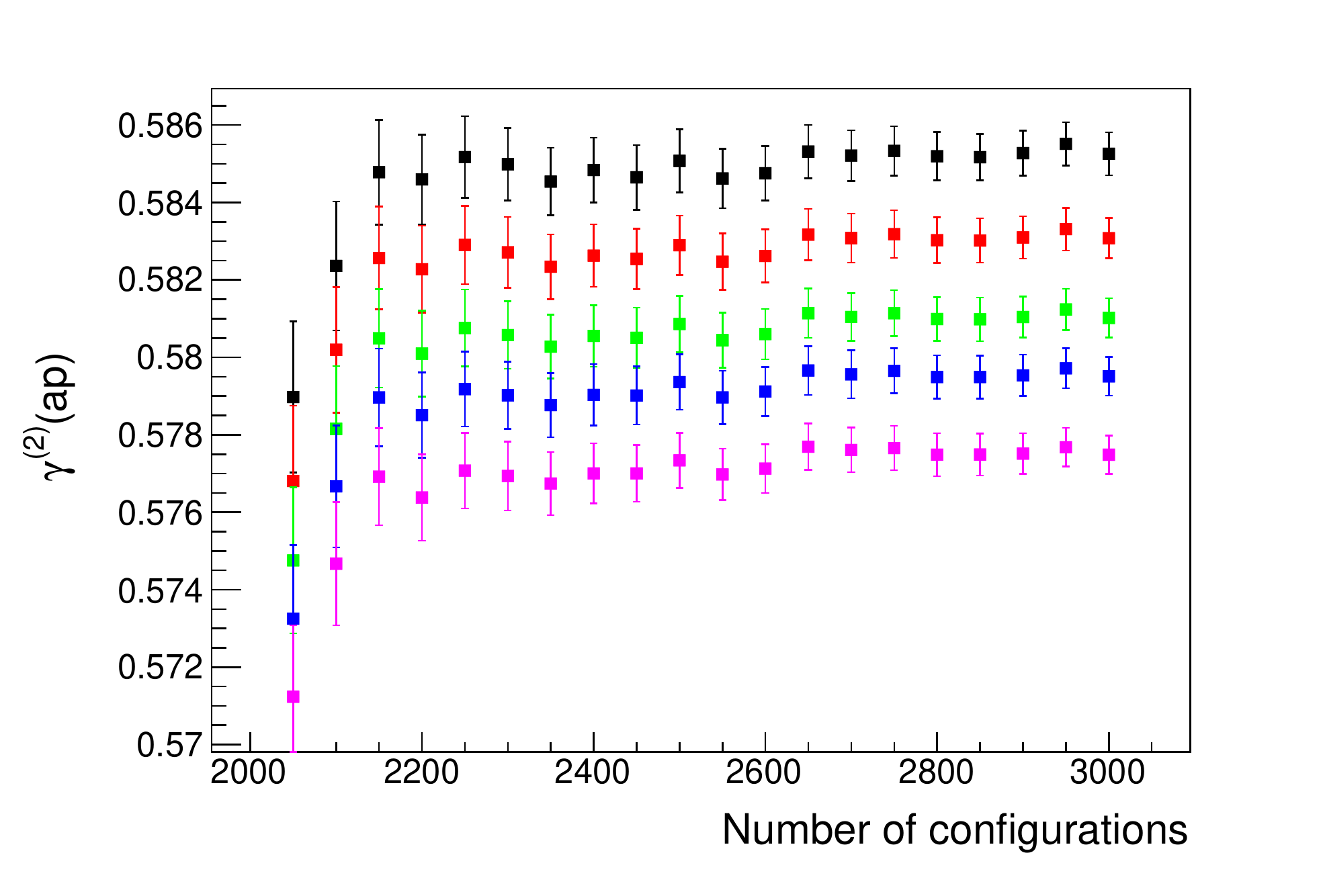}
	\end{subfigure}
	\caption{\label{fig:thermalisation} Same as \myfigurename~\ref{fig:cmplot} with
		data drawn as a function of the number of configurations included in the
		analysis. Each colour corresponds to a different momentum. Horizontal lines
		are the analytical predictions.}
\end{figure}

The agreement between data and the analytic finite-volume calculations can
be seen in \myfigurename~\ref{fig:cmplot}. It is worthwhile to emphasise that measuring
such low momenta requires a careful analysis of the thermalisation. At the
lowest order we can check directly when the measures agree with the theoretical
predictions. At higher orders, it is necessary to wait until the statistical
average has clearly stabilised, as shown in \myfigurename~\ref{fig:thermalisation}.
This kind of analysis is computationally intensive: in the case at hand, we
performed up to $5 \cdot 10^6$ lattice sweeps, saving one propagator every
$10^3$ sweeps. The first $2 \cdot 10^3$ configurations have been discarded in
the analysis.


\subsection{A first attempt for high-order critical mass for SU(3), 
	$N_f = 2$}
\label{sec:FirstAttemptMcrit}

We determined the first $7$ coefficients of the critical mass for $N_c = 3$
and $N_f = 2$ on a $16^4$ lattice with twisted boundary conditions on a plane.
The twist matrices are
\begin{equation}
	\label{eq:twistmatrices}
	\Omega_1=
	\begin{pmatrix}
		e^{-i\frac{2\pi}{3}} & 0 & 0 \\
		0 & 1 & 0 \\
		0 & 0 & e^{i\frac{2\pi}{3}}
	\end{pmatrix}
	\qquad
	\Omega_2=
	\begin{pmatrix}
		0 & 1 & 0 \\
		0 & 0 & 1 \\
		1 & 0 & 0
	\end{pmatrix}\,,
\end{equation}
corresponding to $z_{12}=\exp{\left(i\frac{2\pi}{3}\right)}$. Configurations are
collected at three different time steps, $\tau=0.005$, $0.008$, $0.01$. Because
the volume and the number of colours are large compared to the former test in
\myfigurename~\ref{fig:cmplot}, it is computationally too expensive to replicate the
same statistics at all orders: we settled for $5\cdot10^5$ sweeps at the
smallest $\tau$, measuring the propagator every $r=10^3$ sweeps. At larger time steps, we rescale these numbers to keep the product $r\cdot\tau$ constant.
The propagator is measured at the smallest available momentum, which has
$\theta/L$ in the time component and vanishes elsewhere; we choose three
different values for the phase of the valence twist,
$\theta=\pi/2$, $2\pi/3$, $4\pi/5$. Extrapolations to zero momentum are performed
using a linear fit in $(ap)^2$. The analysis is performed on different subsets of the data~\footnote{
	The different subsets are built by varying the number of initial configurations that are excluded in the analysis and by rejecting data at different rates.
}
to estimate systematic errors. The total error is the sum in quadrature of half the spread around the
central value among the different fits and the largest error from the fits.

The procedure described in \mysectionname~\ref{sec:cmprocedure}, even though
well-defined, is found to be numerically unstable at high orders. The number of
propagators required to reach a clear plateau, like the ones shown in
\myfigurename~\ref{fig:thermalisation}, is beyond what it can be reasonably collected
with the current NSPT implementations. Therefore, we decided to proceed
with a smaller statistics and to add a new systematic uncertainty for the
extrapolated coefficients, as explained below. It has to be emphasised that once a coefficient of the critical mass is determined, only the central value is used as input for the following runs: even if we could collect enough statistics and manage to reduce the error, that is not included in the simulations. This makes the impact of the uncertainty of $m_c^{(n)}$ on
$m_c^{(n+1)}$ and higher hard to assess; also, performing simulations
for several values of each coefficient is not feasible. To be conservative, we
adopted the following strategy. Once a critical mass $m_c^{(n)}$ is determined
and put in the next-order simulation, the corresponding $\gamma^{(n)}(ap)$ should
extrapolate to zero. If it extrapolates to $\epsilon_n$, we take
$|\epsilon_n/m_c^{(n)}|$ as an estimate of the relative systematic error to be
added in quadrature to the determination of all the higher-order critical
masses.

\begin{figure}[tbp]
	\centering
	\begin{subfigure}{.49\textwidth}
		\includegraphics[width=\textwidth,trim={0.1cm 0 1.4cm 0},clip]{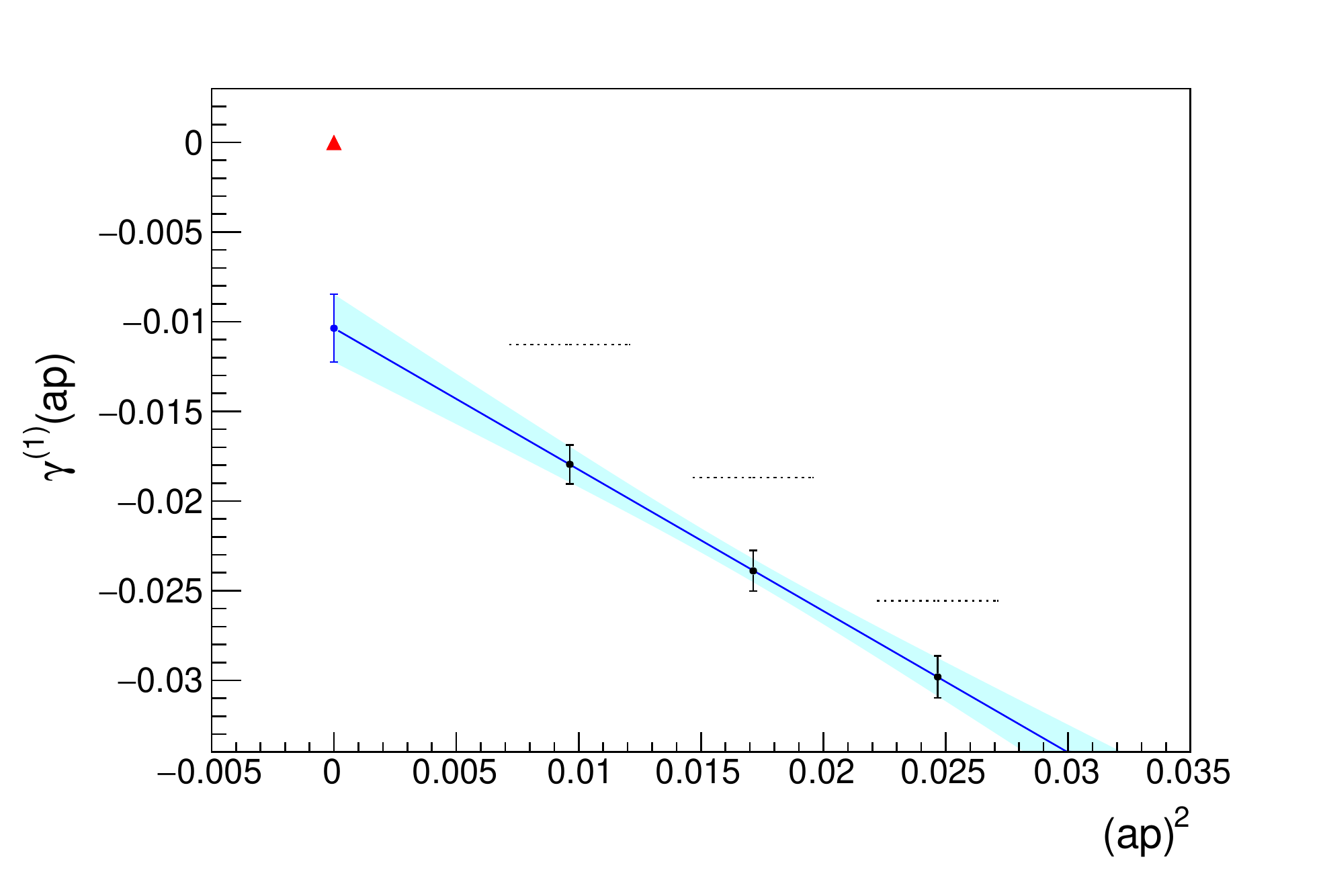}
	\end{subfigure}%
	\begin{subfigure}{.49\textwidth}
		\includegraphics[width=\textwidth,trim={0.1cm 0 1.4cm 0},clip]{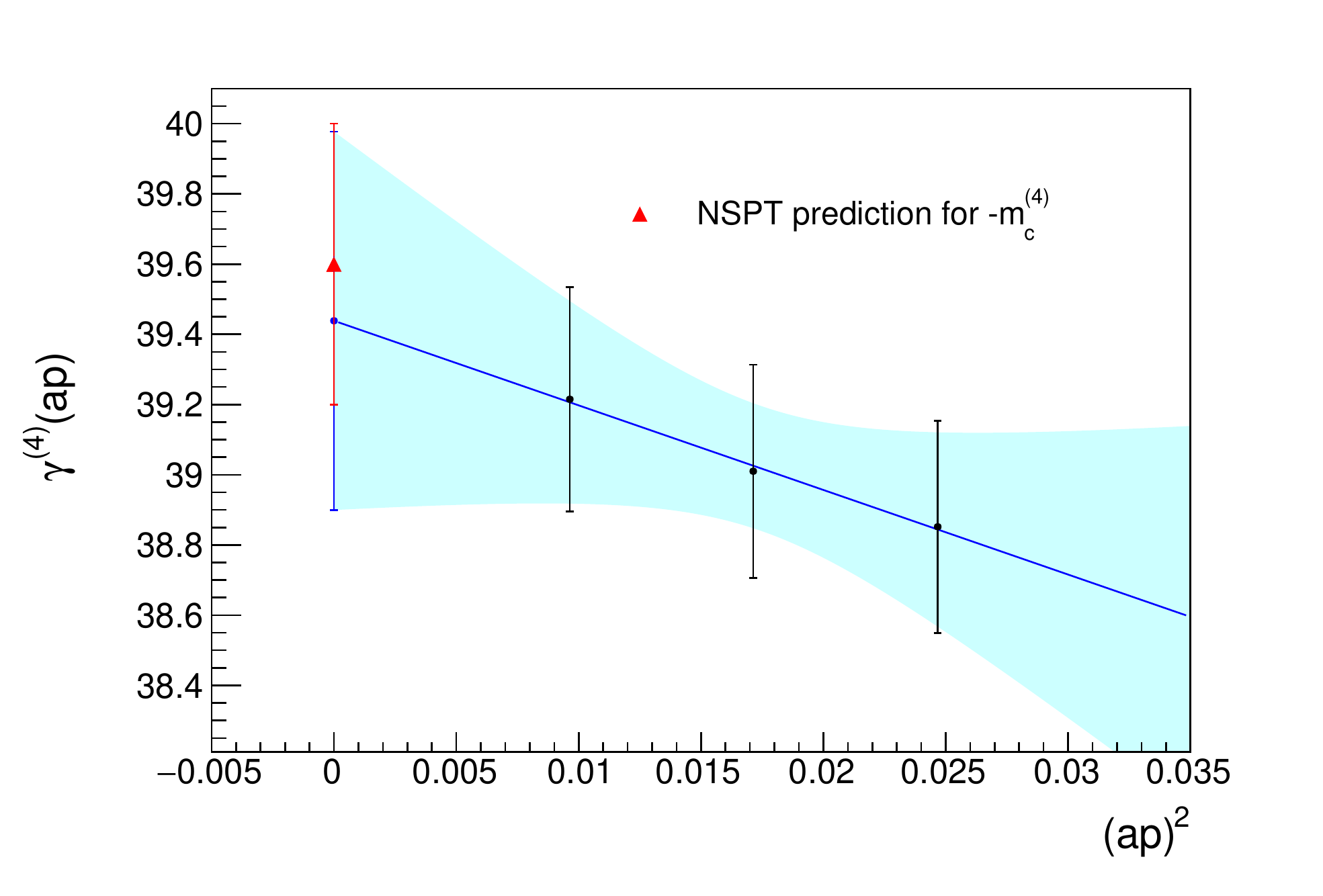}
	\end{subfigure}
	\caption{\label{fig:mcal4} Determination of the coefficient $m_c^{(4)}$.
		Although $\gamma^{(1)}(ap)$ does not extrapolate to zero, the extrapolation
		of $\gamma^{(4)}(ap)$ is compatible with the value known
		from \myrefname~\cite{DiRenzo:2006qtj}. Notation as in \myfigurename~\ref{fig:cmplot}.}
\end{figure}

\begin{figure}[tbp]
	\centering
	\begin{subfigure}{.49\textwidth}
		\includegraphics[width=\textwidth,trim={0.1cm 0 1.4cm 0},clip]{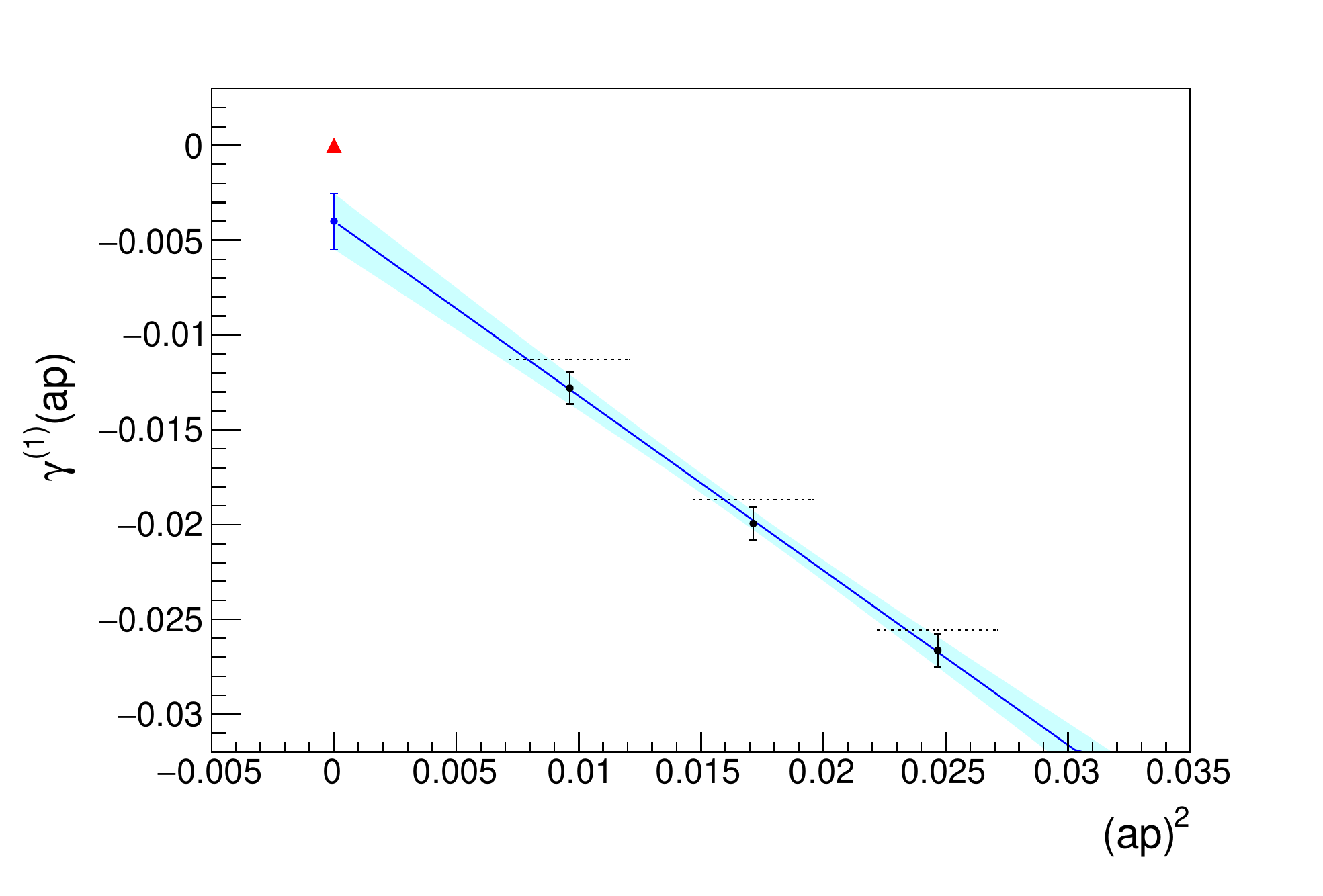}
	\end{subfigure}%
	\begin{subfigure}{.49\textwidth}
		\includegraphics[width=\textwidth,trim={0.1cm 0 1.4cm 0},clip]{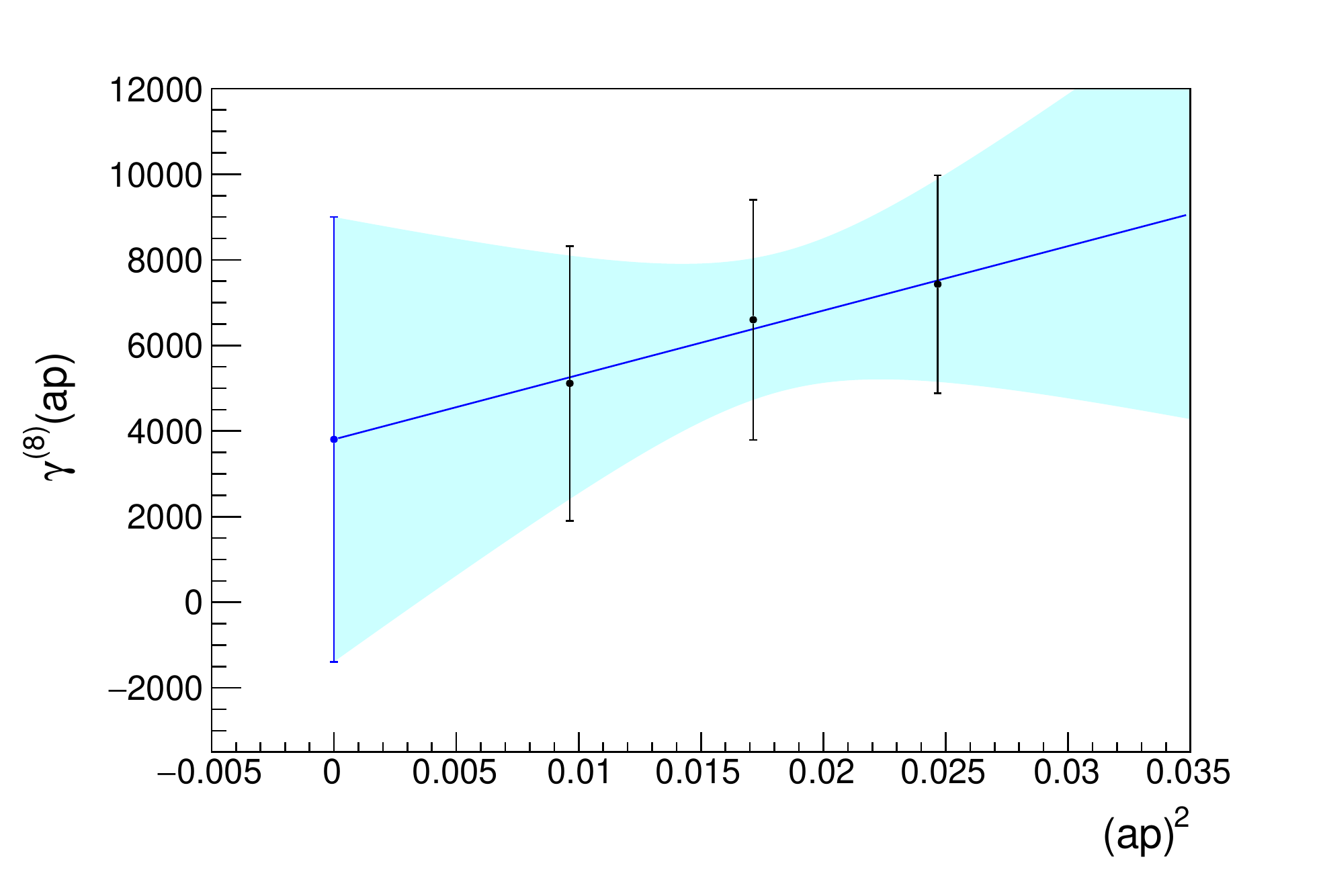}
	\end{subfigure}
	\caption{\label{fig:mcal8}Determination of the coefficient $m_c^{(8)}$. The
		errors overshadow the value of the critical mass, which is compatible with
		zero. Notation as in \myfigurename~\ref{fig:cmplot}.}
\end{figure}

Despite these instabilities, the lower-order results are close to the known
coefficients (keeping in mind that we might resolve finite-volume effects), as
it can be seen for example in \myfigurename~\ref{fig:mcal4}. We stopped the procedure
at $m_c^{(8)}$, when the errors started dominating over the central value of the
coefficient, see \myfigurename~\ref{fig:mcal8}. Our results are summarised in \mytablename~\ref{tab:criticalmasses}.

\begin{table}[tbp]
	\centering
	\begin{tabular}{|c|c|c|}
		\hline
		$n$ & $-m_c^{(n)}$ on $16^4$ & $-m_c^{(n)}$ in infinite volume\\
		\hline
		$1$ & $2.61083\dots$ & $2.60571\dots$\\
		$2$ & $4.32(3)$ & $4.293(1)$~\cite{Follana:2000mn,Caracciolo:2001ki}\\
		$3$ & $1.21(1)\cdot10^1$ & $1.178(5)\cdot10^1$~\cite{DiRenzo:2004ju,DiRenzo:2006qtj}\\
		$4$ & $3.9(2)\cdot10^1$ & $3.96(4)\cdot10^1$~\cite{DiRenzo:2006qtj}\\
		$5$ & $1.7(2) \cdot10^2$ & -\\
		$6$ & $5(1) \cdot10^2$ & -\\
		$7$ & $2(1) \cdot10^3$ & -\\
		\hline
	\end{tabular}
	\caption{\label{tab:criticalmasses} Critical masses for $N_c=3$,
		$N_f=2$ Wilson fermions determined with NSPT on a $16^4$ lattice with
		twisted boundary condition on a plane, compared with the known values in
		infinite volume. The $n=1$ coefficient has been determined analytically
		in twisted lattice perturbation theory; many digits have been used in the
		actual simulation.}
\end{table}

\section{Perturbative expansion of the plaquette}
\label{sec:plaquette}
Following \myrefname~\cite{Bali:2014fea}, we define the average plaquette
\begin{equation}
	\label{eq:PlaqDef}
	P=\frac{1}{6N_cL^4}\sum_\Box\re\Tr\left(1-U_\Box\right)\, ,
\end{equation}
so that the value of $P$ ranges between 0, when all link variables are equal to the identity, and 1.
The plaquette expectation value has the perturbative expansion
\begin{equation}
	\label{eq:PlaqPertExp}
	\braket{P}_\text{pert}=\sum_{n=0}^\infty p_n\,\beta^{-(n+1)}\, ;
\end{equation}
the coefficients $p_n$ are obtained from the Langevin process.

\subsection{Simulation details}
We run NSPT simulations of an $\sun(3)$ gauge theory with $N_f=2$ massless
staggered fermions in the fundamental representation, measuring the average
plaquette after each Langevin update. Twisted boundary conditions are imposed on
a plane, with twist matrices chosen as in \eq~\eqref{eq:twistmatrices}. These
simulations have been mostly run with the GridNSPT code on KNL and Skylake nodes
provided by the Cambridge Service for Data Driven Discovery (CSD3);
simulations on the smallest lattice have been run on the Skylake nodes
on the Marconi system provided by CINECA in Bologna.
The main features of our code are described in \myappendixname~\ref{sec:codenspt}.
We simulate
$24^4,28^4,32^4,48^4$ volumes up to order $\beta^{-40}$ in the expansion of the
links. We gradually switch on higher orders when the plaquette at lower orders
is thermalised. 
Because of the instabilities discussed in \mysectionname~\ref{sec:NumInst},
results are presented only up to the order shown in \mytablename~\ref{tab:runsummary}.
All simulations are run independently at three different time steps, and we have
at least $5\cdot10^3$ measures for the largest order at the smallest time step.
The length of the runs at larger time steps is rescaled to have approximately
the same Langevin time history for all $\tau$.

\begin{table}[tbp]
	\centering
	\begin{tabular}{|c|c|c|}
		\hline
		$L$ & $\tau$ & $n_\text{max}$\\
		\hline 
		\multirow{3}{*}{$24$} 
		& $0.005$ & $35$\\
		& $0.0075$ & $35$\\
		& $0.01$ & $35$\\
		\hline 
		\multirow{3}{*}{$28$} 
		& $0.005$ & $29$\\
		& $0.008$ & $35$\\
		& $0.01$ & $35$\\
		\hline 
		\multirow{3}{*}{$32$} 
		& $0.005$ & $33$\\
		& $0.008$ & $35$\\
		& $0.01$ & $35$\\
		\hline 
		\multirow{3}{*}{$48$} 
		& $0.005$ & $35$\\
		& $0.008$ & $35$\\
		& $0.01$ & $35$\\
		\hline
	\end{tabular}
	\caption{\label{tab:runsummary} Summary of the ensembles for $N_c=3$ and $N_f=2$ staggered fermions. The order $n_{\text{max}}$ is the highest order at which the plaquette $p_n$ has been measured.}
\end{table}

\subsection{Numerical instabilities}
\label{sec:NumInst}

The study of the NSPT hierarchy of stochastic processes is not
trivial. While there are general results for the convergence of the 
generic correlation function of a finite number of perturbative 
components of the fields~\cite{Alfieri:2000ce,DiRenzo:2004hhl}, the
study of variances is more involved, and many results can only come from
direct inspection of the outcome of numerical simulations. In particular, 
one should keep in mind that in the context of (any formulation of)
NSPT, variances are not an intrinsic property of the theory
under study; in other words, they are not obtained as field correlators of the underlying theory.
Big fluctuations and correspondingly huge variances were observed at (terrifically)
high orders in toy models~\cite{Alfieri:2000ce}: signals are plagued 
by several spikes and it is found by inspection that a fluctuation at a given 
order is reflected and amplified at higher orders. All in all, variances increase with the perturbative order (not
surprisingly, given the recursive nature of the equations of motion). 
Moving to more realistic theories, a robust rule of thumb is that, as
expected on general grounds, the larger the number of degrees of
freedom, the less severe the problems with fluctuations are. In
particular, we have not yet found (nor has anyone else reported) big
problems with fluctuations in the computation of high orders in pure
Yang-Mills theory. 

\begin{figure}[tbp]
	\centering
	\begin{subfigure}{.49\textwidth}
		\includegraphics[width=\textwidth,trim={0.1cm 0 1.4cm 0},clip]{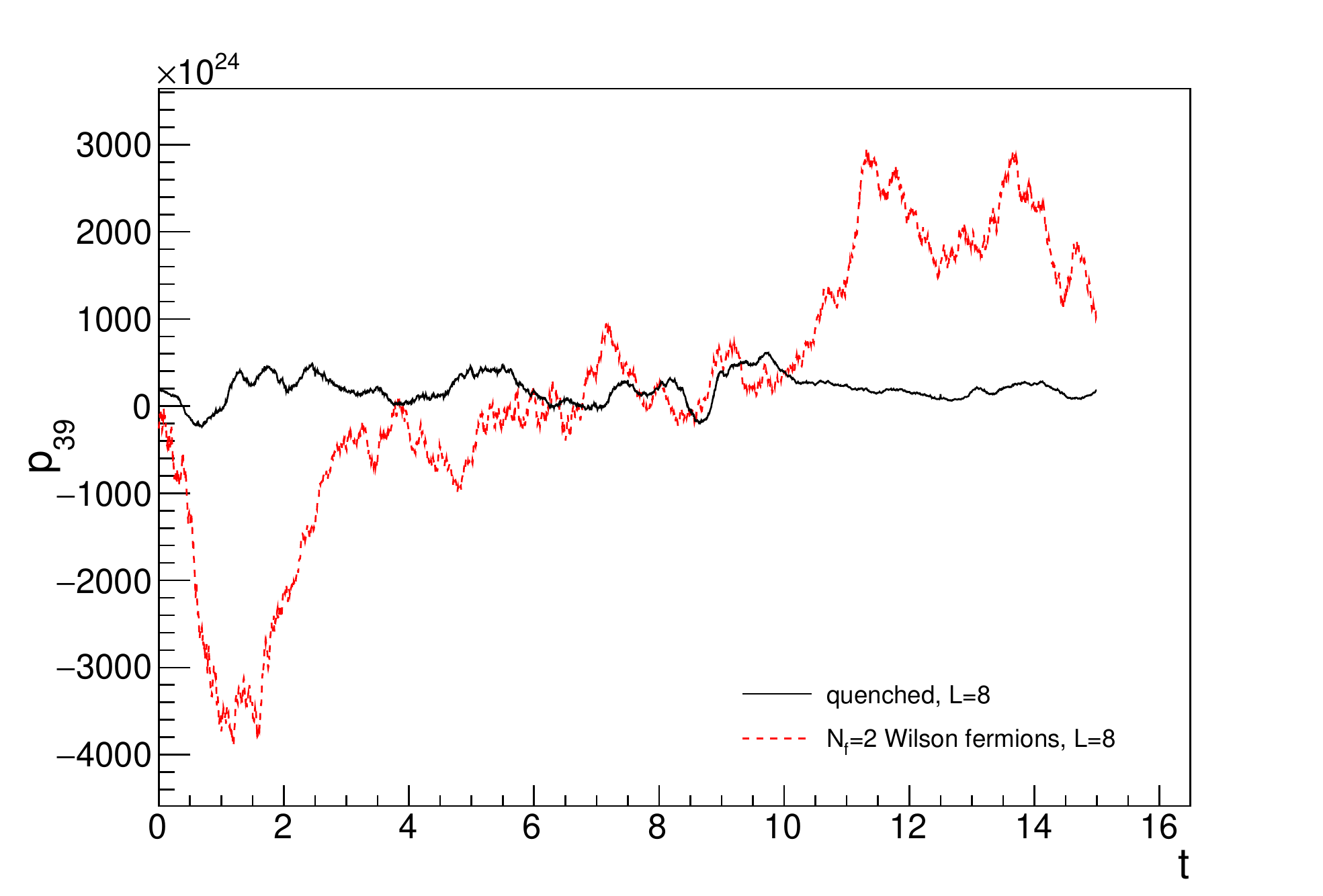}
	\end{subfigure}%
	\begin{subfigure}{.49\textwidth}
		\includegraphics[width=\textwidth,trim={0.1cm 0 1.4cm 0},clip]{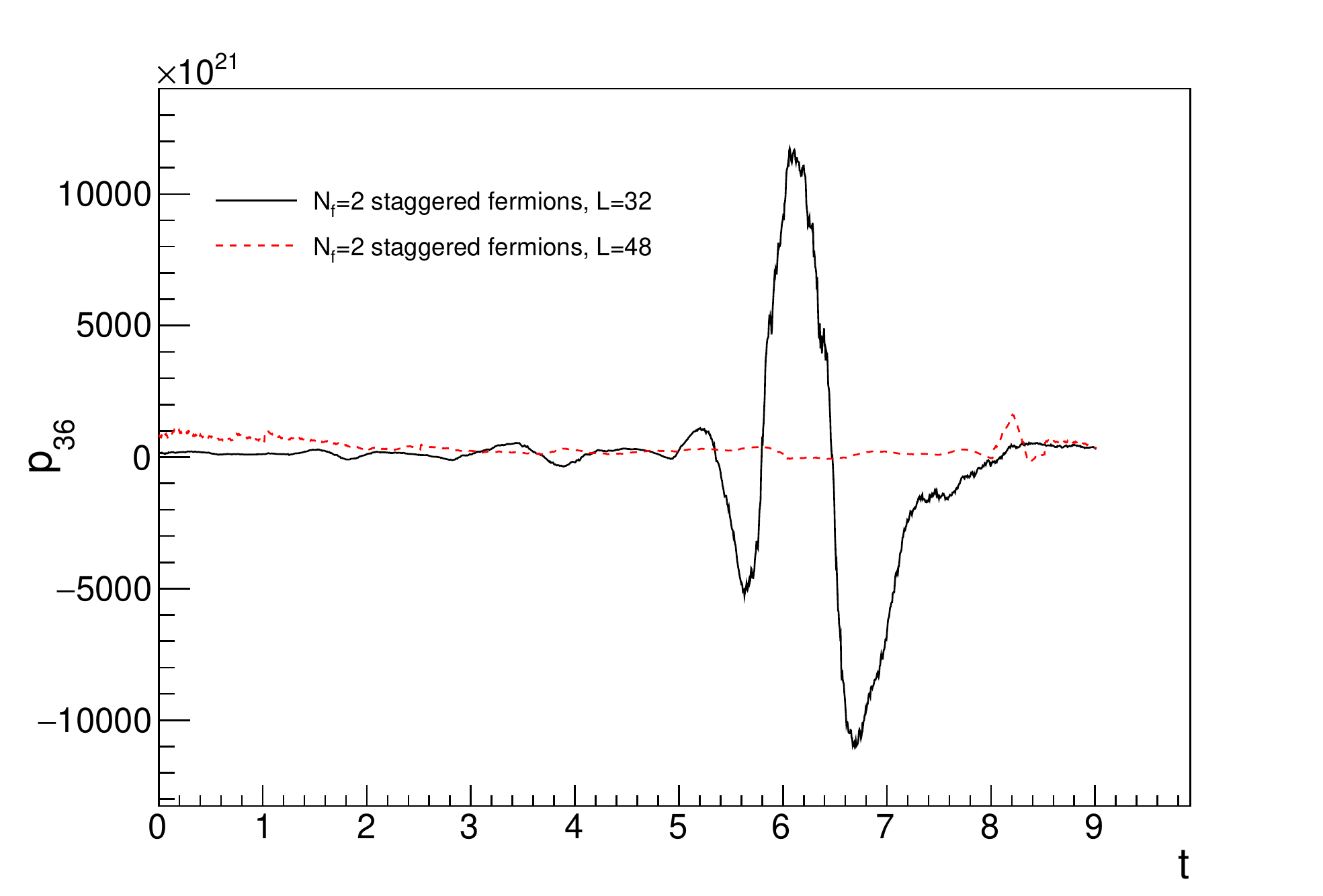}
	\end{subfigure}
	\caption{\label{fig:spikes}In the left panel, signal samples of the coefficient $p_{39}$ taken from a $8^4$ lattice with TBC in three directions. The simulation with Wilson fermions has been performed for illustrative reasons and the bare mass has been set to zero. In the right panel, signal samples of the coefficient $p_{36}$ with TBC on a plane and staggered fermions. In both panels $\tau=0.005$ and the origin of $t$ is set arbitrarily. It is evident that in the quenched case we could extract the plaquette coefficient even from a small volume, while fermions introduce instabilities that can be mitigated by considering bigger lattices. While we have chosen these two particular examples for illustration purposes, the appearance of spikes is a general phenomenon that we observe for orders approximately  $\geq 30$  on the volumes under study.}
\end{figure}

We now found that the introduction of fermions indeed causes
instabilities at orders as high as the ones we are considering
in this work. Once again, this effect can be tamed by working on
increasingly large volumes. Once a fluctuation takes place, the
restoring force would eventually take the signal back around its
average value but in practice this mechanism is not always effective. 
At high orders the instabilities can be so frequent and large that the 
signal is actually lost, and the average
value of the plaquette becomes negligible compared to its standard deviation, as it is illustrated in \myfigurename~\ref{fig:spikes}.
The order at which the signal is lost is pushed to higher values
by increasing the volume, but eventually uncontrolled fluctuations 
will dominate.
Moreover, we find that spikes tend to happen more frequently at
smaller $\tau$. Roughly speaking, this does not come as a surprise,
since at smaller time steps one has to live with a larger number of
sweeps, thereby increasing the
chances of generating large fluctuations when computing the force fields.
In \mytablename~\ref{tab:runsummary} the orders available at each volume and 
time step are shown in detail.

\subsection{Determination of the $p_n$}
\begin{figure}[tbp]
	\centering
	\begin{subfigure}{.49\textwidth}
		\includegraphics[width=\textwidth,trim={0.1cm 0 1.4cm 0},clip]{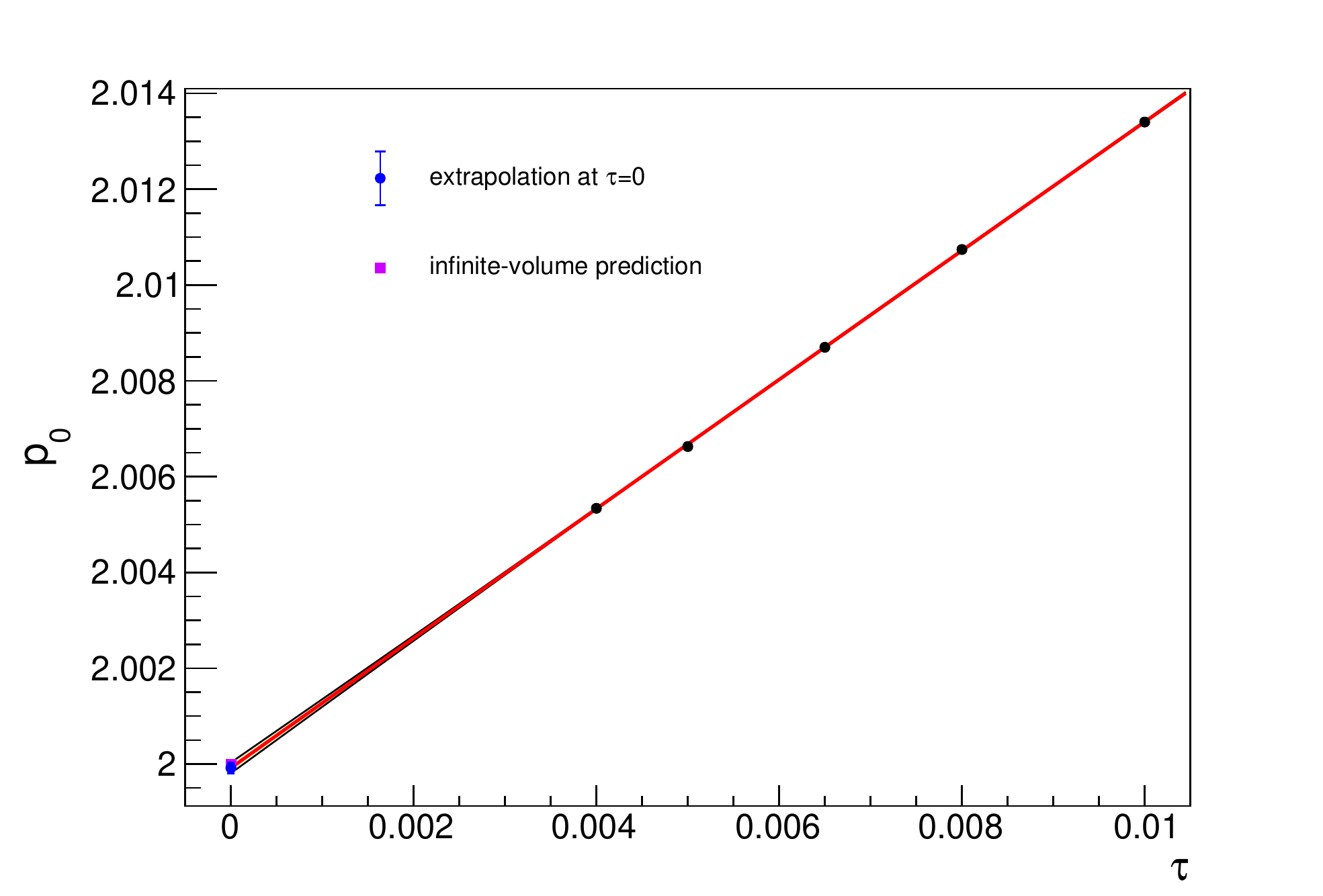}
	\end{subfigure}%
	\begin{subfigure}{.49\textwidth}
		\includegraphics[width=\textwidth,trim={0.1cm 0 1.4cm 0},clip]{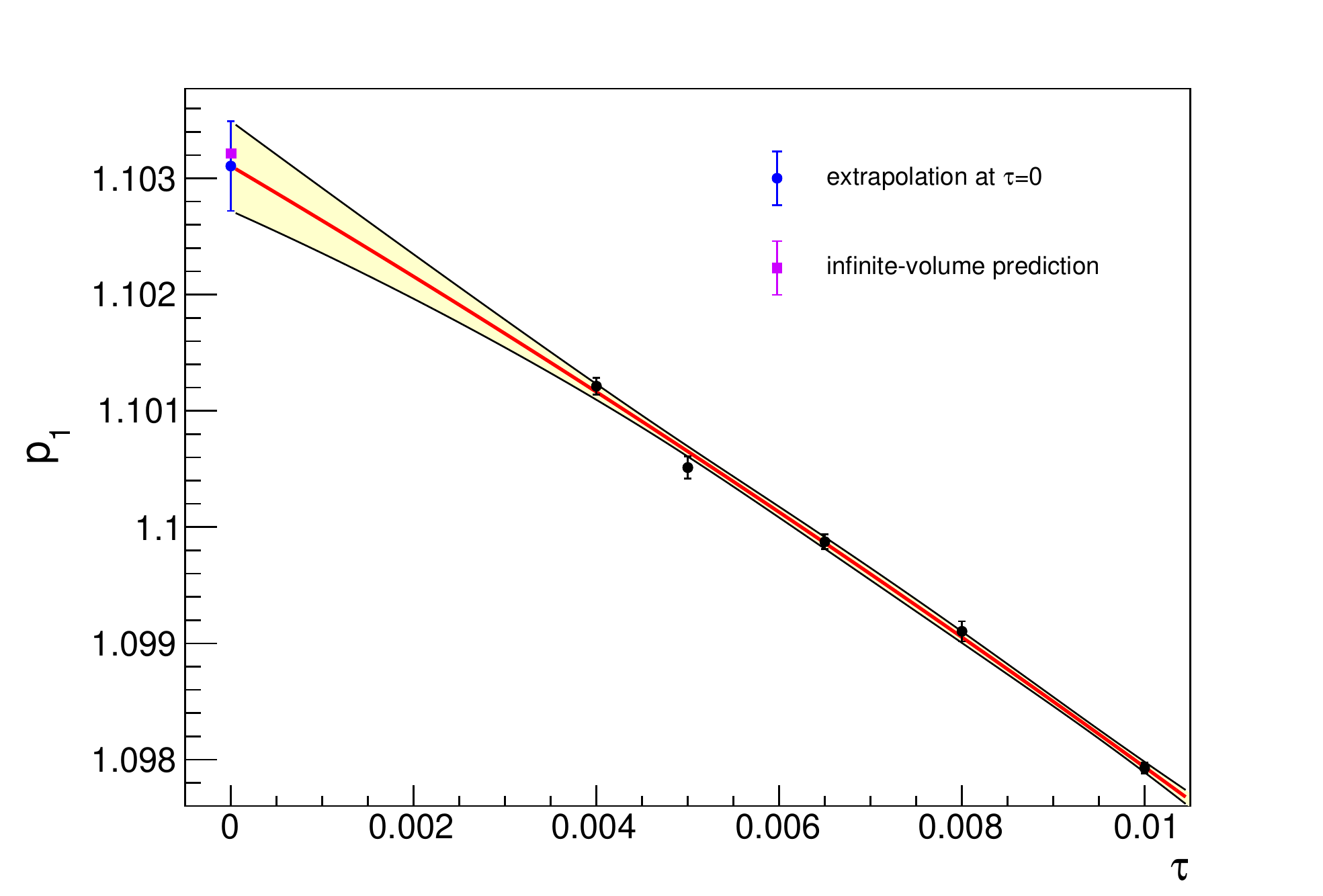}
	\end{subfigure}
	\caption{\label{fig:tinytau}Determination of $p_0$, $p_1$ at $L=48$. Dedicated
		simulations for these two coefficients have been performed at
		$\tau=0.004$ and $\tau=0.0065$. We extrapolate to zero time step with a second order polynomial in $\tau$.
		The extrapolated values are
		$p_0 = 1.9999(1)$ and $p_1 = 1.1031(4)$ with reduced
		$\chi^2$ respectively equal to $1.710$ and $1.477$.}
\end{figure}

The lowest coefficients have already been computed analytically.
In particular, in twisted lattice perturbation theory we have that
\begin{equation}
	p_0=\frac{1}{6}\sum_{\mu>\nu}\frac{N_c}{2}\cancel{\sum_p}\,(1-\delta_{p_\perp,0})\,
	\frac{\hat p^2_\mu+\hat p^2_\nu}{\hat p^2}
	=\frac{N_c^2-1}{4}
\end{equation}
is volume independent~\cite{Perez:2017jyq}. The infinite-volume value of $p_1$ can be obtained adding to the pure gauge contribution~\cite{Alles:1998is},
\begin{equation}
	p_{1,g}=4N_c^2(N_c^2-1)\left(0.0051069297-\frac{1}{128N_c^2}\right)\,,
\end{equation}
the contribution due to staggered fermions~\cite{Bali:2002wf},
\begin{equation}
	\label{eq:P1fBali}
	p_{1,f}= -1.2258(7) \cdot 10^{-3}\,(N_c^2-1)2N_cN_f\,.
\end{equation}
For the specific case $N_c=3,N_f=2$, we find $p_1=1.10312(7)$.  We also
computed the fermion contribution to $p_1$ in twisted lattice
perturbation theory~\footnote{We are grateful to M.~Garc\'ia~P\'erez and
	A.~Gonz\'alez-Arroyo for providing us the gluon contribution in
	finite volume.}. The finite-volume result is $p_1=1.10317022\dots$ at $L=8$, therefore we expect
finite volume effects to be negligible in the lattices we are
employing.  In particular, we improved the determination of $p_{1,f}$ in \eq~\eqref{eq:P1fBali} using the finite volume calculations at
$L=16$ as the central value, and the variation between $L=16$ and $L=14$ as an estimate of its uncertainty, leading to $p_{1,f}=-0.0587909(3)N_f$ for $N_c=3$, and hence
$p_1=1.1032139(6)$ for $N_f=2$.  Trying to extract $p_0$ and $p_1$
from our data at $L=48$, we realise that even $\tau^2$ effects in the
extrapolation must be considered because of the very high precision of
the measurements. For these two coefficients, a dedicated study at
has been performed, which required new simulations at time steps $\tau=0.004$ and $\tau=0.0065$; the agreement
with the analytic calculations is found to be excellent, see
\myfigurename~\ref{fig:tinytau}.

\begin{figure}[tbp]
	\centering
	\begin{subfigure}{.49\textwidth}
		\includegraphics[width=\textwidth,trim={0.1cm 0 1.4cm 0},clip]{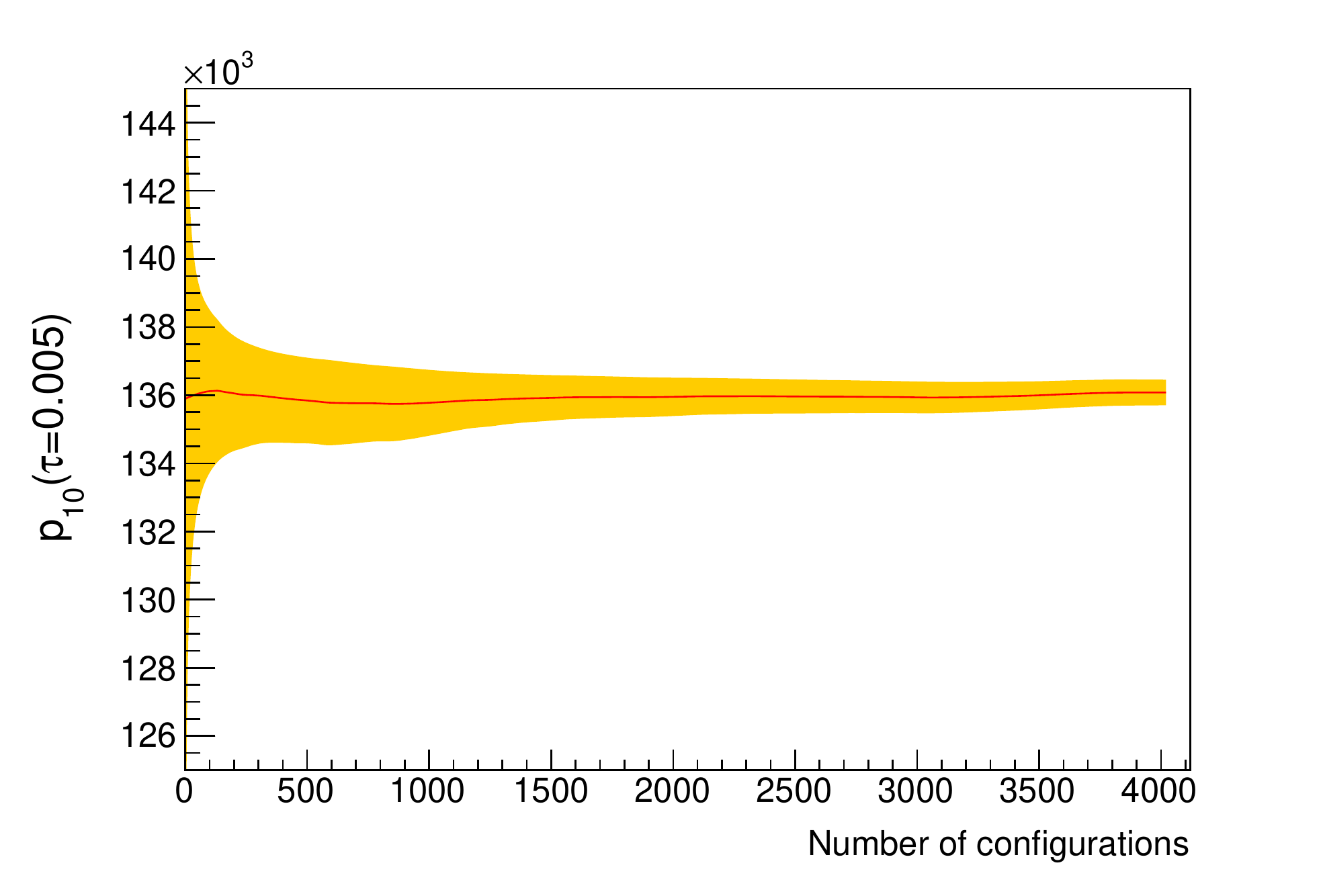}
	\end{subfigure}%
	\begin{subfigure}{.49\textwidth}
		\includegraphics[width=\textwidth,trim={0.1cm 0 1.4cm 0},clip]{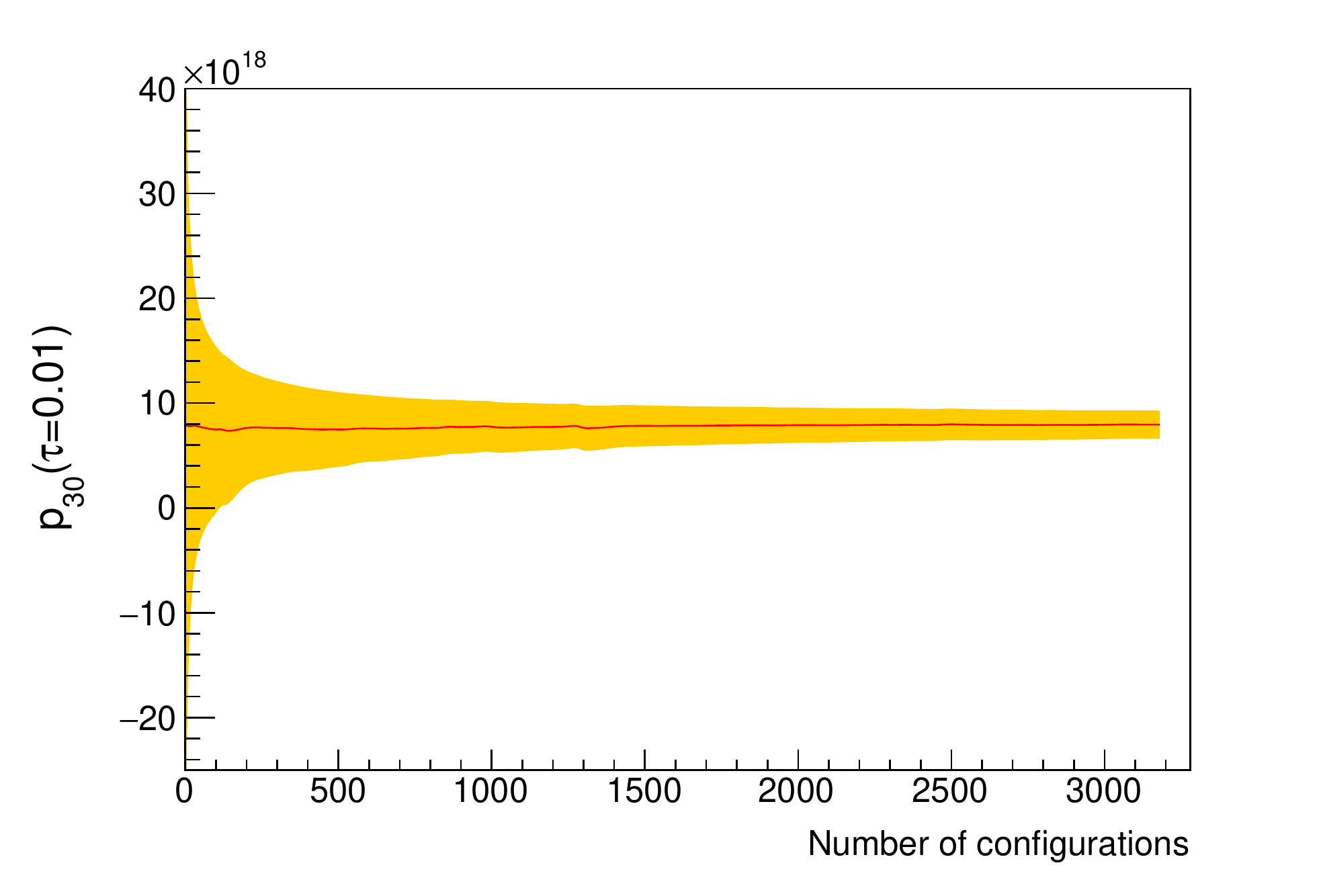}
	\end{subfigure}
	\caption{\label{fig:movingaverage}Average of two plaquette coefficients at $L=48$ as a function of the number of configurations. The error band corresponds to the standard deviation of the sample.}
	\centering
	\begin{subfigure}{.49\textwidth}
		\includegraphics[width=\textwidth,trim={0.8cm 0 1.2cm 0},clip]{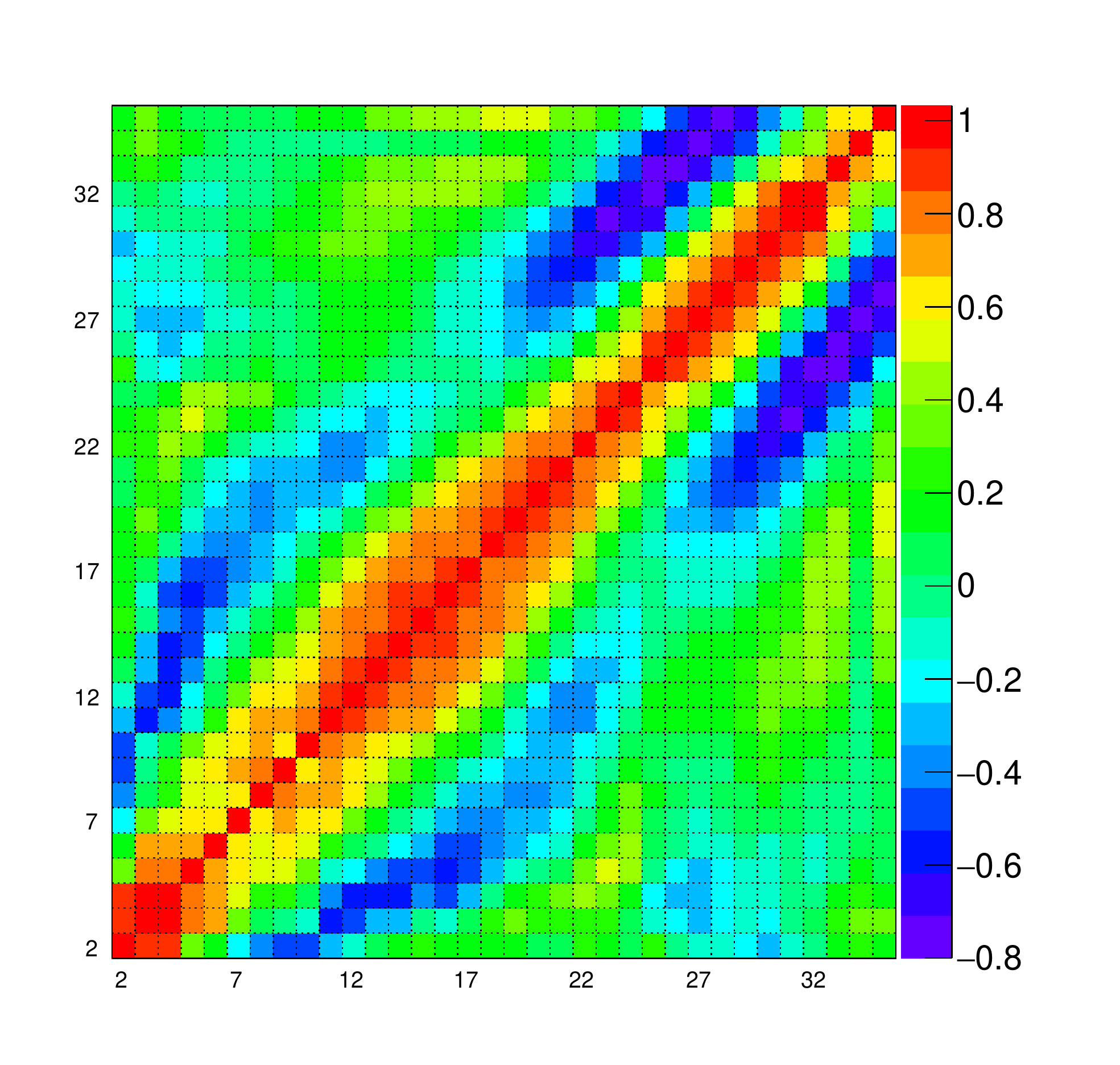}
	\end{subfigure}%
	\begin{subfigure}{.49\textwidth}
		\includegraphics[width=\textwidth,trim={0.8cm 0 1.2cm 0},clip]{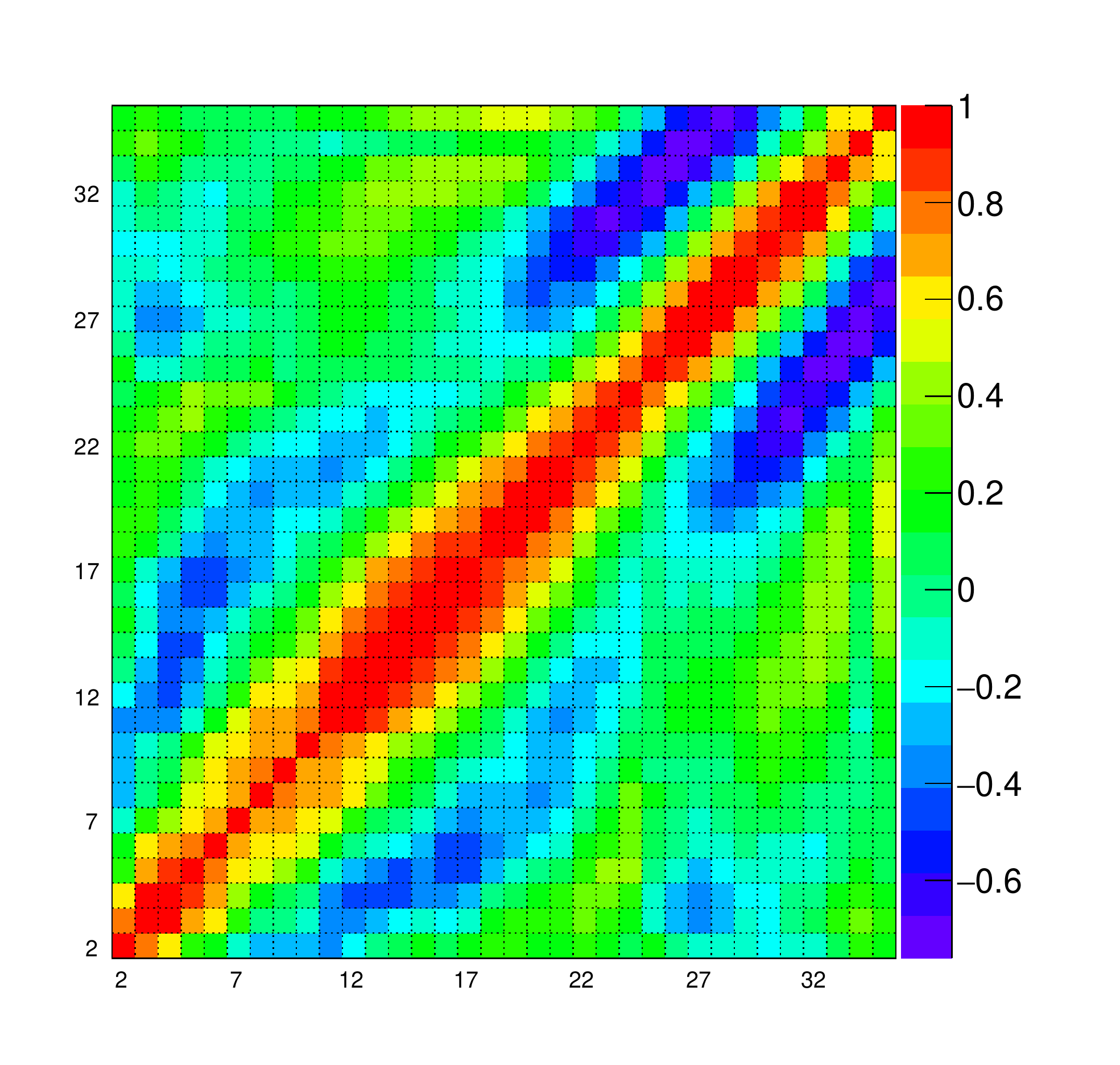}
	\end{subfigure}
	\caption{\label{fig:corrmatrix}In the left panel, correlation matrix between the coefficients $p_2,\ldots,p_{35}$ at $L=48$ extracted from the combined fit procedure. The entrances can be bigger than $1$ because the matrix is not positive definite. In the right panel, the nearest correlation matrix obtained with Higham's algorithm ($\delta=10^{-10}$).}
\end{figure}

Therefore, $p_0$ and $p_1$ are set to their infinite-volume values and excluded from the analysis of the numerical simulations. The remaining orders are obtained from NSPT. The value $p_{n,\tau}$ for the plaquette at order $n$ and time step $\tau$ is computed from the average of the fields generated by the stochastic process, after discarding a number of thermalisation steps.  The moving averages result to be stable, as can be seen in the two examples of \myfigurename~\ref{fig:movingaverage}. In order to exploit all the available data, the thermalisation is set differently at different orders.
The covariance $\text{Cov}(n,m)_\tau$ between $p_{n,\tau}$ and $p_{m,\tau}$ is computed taking into account autocorrelations and cross-correlations, as explained in detail in \myappendixname~\ref{sec:autocorrelation}. Clearly there is no correlation between different $\tau$. In order to estimate the covariance when two orders have different thermalisations, we take into account only the largest set of common values where both are thermalised. This pairwise estimation of the covariance matrix does not guarantee positive definiteness, therefore we rely on Higham's algorithm, which we describe in \myappendixname~\ref{sec:higham}, to find the nearest positive definite covariance matrix; the procedure introduces some dependence on a tolerance $\delta$. 
The extrapolation to vanishing time step is performed by minimising
\begin{equation}
	\chi^2=\sum_{n,m}^{n_{max}}\sum_{\tau} (p_{n,\tau}-a_n\tau-p_n)\, \text{Cov}^{-1}(n,m)_\tau \,(p_{m,\tau}-a_m\tau-p_m)\,,
\end{equation}
where the coefficients $a_n$ are the slopes of the combined linear fits. The interesting fit results are the values of the extrapolated plaquettes $p_n$ and their covariance matrix $\text{Cov}(n,m)$.
In general, because of the available statistics and the intrinsic fluctuations of the observable, the lower-order values are measured more accurately compared to the higher-order ones; the same holds for the estimate of the entries the covariance matrix.
Since, in principle, the plaquette at a certain order could be extracted without any knowledge about its higher-order values, we can get the best estimate for a $p_n$ by implementing the fit iteratively, increasing $n_{max}$ from $0$ to the maximum available order.
At each iteration, we determine the order with the minimum number of measures $N_\text{min}$ and rescale the entries of the covariance matrix so that there is a common normalisation ($N=N_\text{min}$ in \eq~\eqref{eq:covaverage}) for all the matrix elements. In this way, all the data are exploited for the determination of the covariance of the process, and the non-positive definiteness of the covariance of the averages arises only from the presence of autocorrelations and cross-correlations. Higham's algorithm is then applied to $\text{Cov}(n,m)_\tau$ restricted to $n_{max}$ orders. At this stage, minimising the $\chi^2$ allows us to extract $p_{n_{max}}$ with $\text{Cov}(n_{max},m)$ for $m\leq n_{max}$. The tolerance of Higham's algorithm is tuned so that the covariance matrix is able to represent our data, i.e. so that the reduced chi-squared is close to $1$.
The combined fit determines also the plaquettes at orders lower than $n_{max}$, which are always checked and found to be in agreement, within errors, with their previous determination at smaller $n_{max}$. An example of a correlation matrix extracted with this procedure is in \myfigurename~\ref{fig:corrmatrix}, where clear structures of correlated and anticorrelated coefficients are visible. The results of the combined extrapolations are summarised in \mytablename~\ref{tab:pntable}.

\begin{table}[tbp]
	\centering
	\begin{tabular}{|c|c|c|c|}
		\hline
		\multicolumn{4}{|c|}{$L=24$}\\
		\hline
		$n$ & $p_n$ & $\chi^2/\text{dof}$ & $\delta$ \\
		\hline
		$2$ & $2.536(1)$ & $2.178$ & $-$ \\
		$3$ & $7.622(6)$ & $1.079$ & $0.1$ \\
		$4$ & $2.626(3) \cdot 10^{1}$ & $0.735$ & $0.1$ \\
		$5$ & $9.84(1) \cdot 10^{1}$ & $0.615$ & $0.1$ \\
		$6$ & $3.906(6) \cdot 10^{2}$ & $0.828$ & $0.01$ \\
		$7$ & $1.615(3) \cdot 10^{3}$ & $0.529$ & $0.01$ \\
		$8$ & $6.89(2) \cdot 10^{3}$ & $0.581$ & $0.01$ \\
		$9$ & $3.021(9) \cdot 10^{4}$ & $0.421$ & $0.01$ \\
		$10$ & $1.357(5) \cdot 10^{5}$ & $0.861$ & $0.01$ \\
		$11$ & $6.09(3) \cdot 10^{5}$ & $0.940$ & $0.01$ \\
		$12$ & $2.80(2) \cdot 10^{6}$ & $0.753$ & $0.01$ \\
		$13$ & $1.302(9) \cdot 10^{7}$ & $0.690$ & $0.01$ \\
		$14$ & $6.14(4) \cdot 10^{7}$ & $0.570$ & $0.01$ \\
		$15$ & $2.94(2) \cdot 10^{8}$ & $0.652$ & $0.01$ \\
		$16$ & $1.41(1) \cdot 10^{9}$ & $0.797$ & $0.01$ \\
		$17$ & $6.79(6) \cdot 10^{9}$ & $0.758$ & $0.01$ \\
		$18$ & $3.31(3) \cdot 10^{10}$ & $0.730$ & $0.01$ \\
		$19$ & $1.65(2) \cdot 10^{11}$ & $0.678$ & $0.01$ \\
		$20$ & $8.3(1) \cdot 10^{11}$ & $0.732$ & $0.01$ \\
		$21$ & $4.15(7) \cdot 10^{12}$ & $0.755$ & $0.01$ \\
		$22$ & $2.08(5) \cdot 10^{13}$ & $0.590$ & $0.1$ \\
		$23$ & $10.0(4) \cdot 10^{13}$ & $0.569$ & $0.1$ \\
		$24$ & $5.0(2) \cdot 10^{14}$ & $0.543$ & $0.1$ \\
		$25$ & $2.5(1) \cdot 10^{15}$ & $0.485$ & $0.1$ \\
		$26$ & $1.34(4) \cdot 10^{16}$ & $1.140$ & $0.01$ \\
		$27$ & $6.6(2) \cdot 10^{16}$ & $1.054$ & $0.01$ \\
		$28$ & $3.2(2) \cdot 10^{17}$ & $0.479$ & $0.1$ \\
		$29$ & $1.6(1) \cdot 10^{18}$ & $1.124$ & $0.01$ \\
		$30$ & $7.6(7) \cdot 10^{18}$ & $0.836$ & $0.01$ \\
		$31$ & $3.6(6) \cdot 10^{19}$ & $0.456$ & $0.01$ \\
		$32$ & $1.8(4) \cdot 10^{20}$ & $0.443$ & $0.01$ \\
		$33$ & $9(3) \cdot 10^{20}$ & $0.445$ & $0.01$ \\
		$34$ & $5(2) \cdot 10^{21}$ & $0.432$ & $0.01$ \\
		$35$ & $3(1) \cdot 10^{22}$ & $0.425$ & $0.01$ \\
		\hline
	\end{tabular}
	\qquad
	\begin{tabular}{|c|c|c|c|}
		\hline
		\multicolumn{4}{|c|}{$L=28$}\\
		\hline
		$n$ & $p_n$ & $\chi^2/\text{dof}$ & $\delta$ \\
		\hline
		$2$ & $2.537(1)$ & $0.032$ & $-$ \\
		$3$ & $7.639(7)$ & $1.136$ & $0.625$ \\
		$4$ & $2.636(3) \cdot 10^{1}$ & $0.648$ & $0.5$ \\
		$5$ & $9.89(2) \cdot 10^{1}$ & $0.853$ & $0.1$ \\
		$6$ & $3.934(7) \cdot 10^{2}$ & $0.593$ & $0.1$ \\
		$7$ & $1.630(4) \cdot 10^{3}$ & $0.480$ & $0.1$ \\
		$8$ & $6.97(2) \cdot 10^{3}$ & $0.707$ & $0.1$ \\
		$9$ & $3.05(1) \cdot 10^{4}$ & $0.927$ & $0.1$ \\
		$10$ & $1.366(5) \cdot 10^{5}$ & $0.753$ & $0.1$ \\
		$11$ & $6.21(3) \cdot 10^{5}$ & $0.599$ & $0.1$ \\
		$12$ & $2.87(1) \cdot 10^{6}$ & $0.512$ & $0.1$ \\
		$13$ & $1.338(7) \cdot 10^{7}$ & $0.443$ & $0.1$ \\
		$14$ & $6.31(4) \cdot 10^{7}$ & $0.401$ & $0.1$ \\
		$15$ & $3.01(2) \cdot 10^{8}$ & $0.360$ & $0.1$ \\
		$16$ & $1.44(1) \cdot 10^{9}$ & $1.012$ & $0.01$ \\
		$17$ & $6.96(7) \cdot 10^{9}$ & $0.998$ & $0.01$ \\
		$18$ & $3.36(3) \cdot 10^{10}$ & $0.972$ & $0.01$ \\
		$19$ & $1.63(2) \cdot 10^{11}$ & $0.953$ & $0.01$ \\
		$20$ & $8.0(1) \cdot 10^{11}$ & $0.884$ & $0.01$ \\
		$21$ & $3.89(6) \cdot 10^{12}$ & $0.829$ & $0.01$ \\
		$22$ & $1.91(3) \cdot 10^{13}$ & $0.821$ & $0.01$ \\
		$23$ & $9.5(2) \cdot 10^{13}$ & $0.873$ & $0.01$ \\
		$24$ & $4.7(1) \cdot 10^{14}$ & $0.851$ & $0.01$ \\
		$25$ & $2.34(6) \cdot 10^{15}$ & $0.764$ & $0.01$ \\
		$26$ & $1.14(3) \cdot 10^{16}$ & $0.695$ & $0.01$ \\
		$27$ & $5.7(2) \cdot 10^{16}$ & $0.687$ & $0.01$ \\
		$28$ & $2.8(1) \cdot 10^{17}$ & $0.671$ & $0.01$ \\
		$29$ & $1.5(1) \cdot 10^{18}$ & $0.462$ & $0.01$ \\
		$30$ & $7.1(7) \cdot 10^{18}$ & $0.855$ & $0.001$ \\
		$31$ & $4.2(7) \cdot 10^{19}$ & $0.663$ & $0.001$ \\
		$32$ & $2.0(4) \cdot 10^{20}$ & $0.661$ & $0.001$ \\
		$33$ & $10(3) \cdot 10^{20}$ & $0.651$ & $0.001$ \\
		$34$ & $4(2) \cdot 10^{21}$ & $0.516$ & $0.001$ \\
		$35$ & $2(1) \cdot 10^{22}$ & $0.519$ & $0.001$ \\
		\hline
	\end{tabular}
\end{table}
\begin{table}[tbp]
	\centering
	\begin{tabular}{|c|c|c|c|}
		\hline
		\multicolumn{4}{|c|}{$L=32$}\\
		\hline
		$n$ & $p_n$ & $\chi^2/\text{dof}$ & $\delta$ \\
		\hline
		$2$ & $2.5370(8)$ & $0.249$ & $-$ \\
		$3$ & $7.627(4)$ & $1.182$ & $-$ \\
		$4$ & $2.633(2) \cdot 10^{1}$ & $2.412$ & $-$ \\
		$5$ & $9.882(9) \cdot 10^{1}$ & $1.378$ & $0.5$ \\
		$6$ & $3.926(5) \cdot 10^{2}$ & $1.015$ & $0.1$ \\
		$7$ & $1.626(2) \cdot 10^{3}$ & $0.730$ & $0.1$ \\
		$8$ & $6.96(1) \cdot 10^{3}$ & $0.929$ & $0.01$ \\
		$9$ & $3.050(6) \cdot 10^{4}$ & $0.772$ & $0.01$ \\
		$10$ & $1.367(4) \cdot 10^{5}$ & $0.638$ & $0.01$ \\
		$11$ & $6.22(2) \cdot 10^{5}$ & $0.963$ & $0.01$ \\
		$12$ & $2.86(1) \cdot 10^{6}$ & $0.645$ & $0.1$ \\
		$13$ & $1.337(6) \cdot 10^{7}$ & $0.771$ & $0.1$ \\
		$14$ & $6.29(3) \cdot 10^{7}$ & $0.861$ & $0.1$ \\
		$15$ & $3.00(2) \cdot 10^{8}$ & $0.952$ & $0.1$ \\
		$16$ & $1.438(9) \cdot 10^{9}$ & $1.012$ & $0.1$ \\
		$17$ & $6.94(5) \cdot 10^{9}$ & $0.996$ & $0.1$ \\
		$18$ & $3.34(3) \cdot 10^{10}$ & $1.000$ & $0.1$ \\
		$19$ & $1.63(2) \cdot 10^{11}$ & $0.965$ & $0.1$ \\
		$20$ & $7.90(8) \cdot 10^{11}$ & $1.053$ & $0.01$ \\
		$21$ & $3.86(4) \cdot 10^{12}$ & $0.995$ & $0.01$ \\
		$22$ & $1.90(2) \cdot 10^{13}$ & $0.957$ & $0.01$ \\
		$23$ & $9.4(1) \cdot 10^{13}$ & $0.949$ & $0.01$ \\
		$24$ & $4.74(9) \cdot 10^{14}$ & $0.979$ & $0.01$ \\
		$25$ & $2.39(5) \cdot 10^{15}$ & $0.967$ & $0.01$ \\
		$26$ & $1.22(3) \cdot 10^{16}$ & $0.921$ & $0.01$ \\
		$27$ & $6.3(2) \cdot 10^{16}$ & $0.871$ & $0.01$ \\
		$28$ & $3.2(1) \cdot 10^{17}$ & $0.849$ & $0.01$ \\
		$29$ & $1.63(9) \cdot 10^{18}$ & $0.812$ & $0.01$ \\
		$30$ & $8.6(7) \cdot 10^{18}$ & $0.779$ & $0.01$ \\
		$31$ & $4.5(9) \cdot 10^{19}$ & $0.743$ & $0.01$ \\
		$32$ & $1.9(3) \cdot 10^{20}$ & $0.723$ & $0.01$ \\
		$33$ & $9(2) \cdot 10^{20}$ & $0.723$ & $0.01$ \\
		$34$ & $5(1) \cdot 10^{21}$ & $0.702$ & $0.01$ \\
		$35$ & $1(1) \cdot 10^{22}$ & $0.663$ & $0.01$ \\
		\hline
	\end{tabular}
	\qquad
	\begin{tabular}{|c|c|c|c|}
		\hline
		\multicolumn{4}{|c|}{$L=48$}\\
		\hline
		$n$ & $p_n$ & $\chi^2/\text{dof}$ & $\delta$ \\
		\hline
		$2$ & $2.5354(7)$ & $2.745$ & $-$ \\
		$3$ & $7.615(3)$ & $1.454$ & $0.01$ \\
		$4$ & $2.623(1) \cdot 10^{1}$ & $1.428$ & $0.1$ \\
		$5$ & $9.826(6) \cdot 10^{1}$ & $1.673$ & $0.1$ \\
		$6$ & $3.897(3) \cdot 10^{2}$ & $1.653$ & $0.1$ \\
		$7$ & $1.613(2) \cdot 10^{3}$ & $1.338$ & $0.1$ \\
		$8$ & $6.88(1) \cdot 10^{3}$ & $1.194$ & $0.1$ \\
		$9$ & $3.007(6) \cdot 10^{4}$ & $1.079$ & $0.1$ \\
		$10$ & $1.341(3) \cdot 10^{5}$ & $0.998$ & $0.1$ \\
		$11$ & $6.08(1) \cdot 10^{5}$ & $0.925$ & $0.1$ \\
		$12$ & $2.793(6) \cdot 10^{6}$ & $1.108$ & $0.01$ \\
		$13$ & $1.297(3) \cdot 10^{7}$ & $0.978$ & $0.01$ \\
		$14$ & $6.08(2) \cdot 10^{7}$ & $0.883$ & $0.01$ \\
		$15$ & $2.87(1) \cdot 10^{8}$ & $1.067$ & $0.01$ \\
		$16$ & $1.370(5) \cdot 10^{9}$ & $1.013$ & $0.01$ \\
		$17$ & $6.57(3) \cdot 10^{9}$ & $0.951$ & $0.01$ \\
		$18$ & $3.16(1) \cdot 10^{10}$ & $0.930$ & $0.01$ \\
		$19$ & $1.530(6) \cdot 10^{11}$ & $0.938$ & $0.01$ \\
		$20$ & $7.45(3) \cdot 10^{11}$ & $0.890$ & $0.01$ \\
		$21$ & $3.65(1) \cdot 10^{12}$ & $0.824$ & $0.01$ \\
		$22$ & $1.796(9) \cdot 10^{13}$ & $0.748$ & $0.01$ \\
		$23$ & $8.88(5) \cdot 10^{13}$ & $0.691$ & $0.01$ \\
		$24$ & $4.41(3) \cdot 10^{14}$ & $0.636$ & $0.01$ \\
		$25$ & $2.19(2) \cdot 10^{15}$ & $0.575$ & $0.01$ \\
		$26$ & $1.09(1) \cdot 10^{16}$ & $0.548$ & $0.01$ \\
		$27$ & $5.46(9) \cdot 10^{16}$ & $0.538$ & $0.01$ \\
		$28$ & $2.74(6) \cdot 10^{17}$ & $0.523$ & $0.01$ \\
		$29$ & $1.38(4) \cdot 10^{18}$ & $0.511$ & $0.01$ \\
		$30$ & $7.0(3) \cdot 10^{18}$ & $0.492$ & $0.01$ \\
		$31$ & $3.5(2) \cdot 10^{19}$ & $0.494$ & $0.01$ \\
		$32$ & $1.7(1) \cdot 10^{20}$ & $0.503$ & $0.01$ \\
		$33$ & $8.3(7) \cdot 10^{20}$ & $1.062$ & $0.001$ \\
		$34$ & $5.2(6) \cdot 10^{21}$ & $1.090$ & $0.001$ \\
		$35$ & $2.3(6) \cdot 10^{22}$ & $0.486$ & $0.01$ \\
		\hline
	\end{tabular}
	\caption{\label{tab:pntable} Plaquette coefficients from the combined fit for $L=24$, $28$, $32$, $48$. The tolerance $\delta$ is given only when the covariance matrix is found not to be positive definite.}
\end{table}

\section{Gluon condensate}
\label{sec:gluonc}
In this section we restore the lattice spacing $a$ and follow the notation of \myrefsname~\cite{Bali:2014fea,Bali:2014sja}: the gluon condensate is defined as the vacuum expectation value of the operator
\begin{equation}
	O_G=-\frac{2}{\beta_0} \frac{\beta(\alpha)}{\alpha}\sum_{a,\mu,\nu}G_{\mu\nu}^a G_{\mu\nu}^a\,,
\end{equation}
where the coupling $\alpha$ is related to the Wilson action coupling by $\alpha=\frac{N_c}{2\pi\beta}$ and the beta function is
\begin{equation}
	\label{eq:BetaDef}
	\beta(\alpha)=\frac{d\alpha}{d\ln\mu}=-2\alpha\left[\beta_0\frac{\alpha}{4\pi}+\beta_1\left(\frac{\alpha}{4\pi}\right)^2+\dots \right]\,,
\end{equation}
with the scheme-independent coefficients
\begin{subequations}
	\begin{align}
		\beta_0&=\frac{11}{3} N_c-\frac{2}{3}N_f\\
		\beta_1&=\frac{34}{3}N_c^2-N_f\left(\frac{13}{3}N_c-\frac{1}{N_c}\right)\,.
	\end{align}
\end{subequations}
It is useful to remember that, in the massless limit, $O_G$ is renormalisation group invariant and depends on the scheme only through the renormalisation condition used to define the composite operator.

It is easy to relate the gluon condensate and the plaquette in the naive continuum limit:
\begin{subequations}
	\begin{align}
		\label{eq:PlaqNaivLim}
		& a^{-4}P \xrightarrow{a\to0} \frac{\pi^2}{12N_c}O_G=\frac{\pi^2}{12N_c}\left(  \frac{\alpha}{\pi}G^2  \right)\,, \\
		\label{eq:OGNaivLim}
		& O_G= \frac{\alpha}{\pi}G^2 \left[1+O(\alpha) \right]\,.
	\end{align}
\end{subequations}
In the interacting theory mixing with operators of lower or equal dimension occurs. For the case of the plaquette, the mixing with the identity needs to be considered, yielding
\begin{equation}
	\label{eq:plaquetterenormalisation}
	a^{-4}P = a^{-4}Z(\beta)\mathbb{1}+\frac{\pi^2}{12N_c}C_G(\beta)O_G+O(a^2\Lambda_\text{QCD}^6)\, ,
\end{equation}
which shows explicitly the subtraction of the quartic power
divergence~\footnote{We mention that, in a theory with fermions, the operator
	$O_G$ must be combined with $m\bar\psi\psi$ to give a renormalisation group
	invariant quantity; moreover mixing with the operators $m\bar\psi\psi$ and
	$\bar\psi(i\slashed{D}-m)\psi$ should also be
	considered~\cite{Tarrach:1981bi,Grinstein:1988wz}. Clearly such complications
	are not present in the massless case and the operator $i\bar\psi\slashed{D}\psi$
	can be neglected in the following discussions since it vanishes when the
	equation of motion are used.}.

As a consequence
\begin{equation}
	\label{eq:PlaqMCValue}
	\braket{P}_\text{MC} = Z(\beta) + \frac{\pi^2}{12N_c} C_G(\beta)
	a^4 \braket{O_G} 
	+ O(a^6\Lambda_\text{QCD}^6)\,,
\end{equation}
where $\braket{P}_\text{MC}$ is the plaquette expectation value obtained from a
nonperturbative Monte Carlo simulation. As such, $\braket{P}_\text{MC}$ is
expected to depend on the cut-off scale $a$, and $\Lambda_\text{QCD}$. In the
limit $a^{-1}\gg\Lambda_\text{QCD}$, \eq~\eqref{eq:PlaqMCValue} can be seen as
an Operator Product Expansion (OPE)~\cite{Wilson:1969zs,Shifman:1978bx,Shifman:1978by}, which factorises the
dependence on the small scale $a$. In this framework~\footnote{
It is useful to keep in mind that other definitions of the gluon condensate are possible, see e.g. \myrefname~\cite{DelDebbio:2013sta}.
}, condensates like
$\braket{O_G}$ are process-independent parameters that encode the
nonperturbative dynamics, while the Wilson coefficients are defined in
perturbation theory,
\begin{equation}
	\label{eq:ZetaBetaPertExp}
	Z(\beta)=\sum_{n=0}p_n\beta^{-(n+1)}\,,
	\quad C_G(\beta)=1+\sum_{n=0}c_n\beta^{-(n+1)}\,.
\end{equation}
Note that both $Z$ and $C_G$ depend only on the bare coupling $\beta^{-1}$, and
do not depend on the renormalisation scale $\mu$, as expected for both
coefficients~\cite{DiGiacomo:1990gy,Testa:1998ez}. Nonperturbative contributions
to $Z$, or $C_G$, originating for example from instantons, would correspond to subleading terms in $\Lambda_\text{QCD}$. This procedure defines a renormalisation
scheme to subtract power divergences: condensates are chosen to vanish in
pertubation theory or, in other words, they are normal ordered in the
perturbative vacuum. This definition matches the one that is natural in
dimensional regularisation, where power divergences do not arise. Nevertheless,
it is well known that such a definition of the condensates might lead to
ambiguities, since the separation of scales in the OPE does not necessarily
correspond to a separation between perturbative and nonperturbative physics (see
the interesting discussions in \myrefsname~\cite{Novikov:1984rf,Shifman:1998rb}). For
example, the fermion condensate in a massless theory is well-defined since,
being the order parameter of chiral symmetry breaking, it must vanish in
perturbation theory. The same cannot be said for the gluon
condensate~\cite{David:1983gz}, and indeed the ambiguity in its definition is
reflected in the divergence of the perturbative expansion of the plaquette. For
this picture to be consistent, it must be possible to absorb in the definition
of the condensate the ambiguity in resumming the perturbative series.

In the following, we are going to study the asymptotic behaviour of the
coefficients $p_n$ determined in the previous section and discuss the
implications for the definition of the gluon condensate in massless QCD.

\subsection{Growth of the coefficients}
\begin{figure}[tbp]
	\centering
	\begin{subfigure}{.99\textwidth}
		\includegraphics[width=\textwidth,trim={0.8cm 0 1.9cm 0},clip]{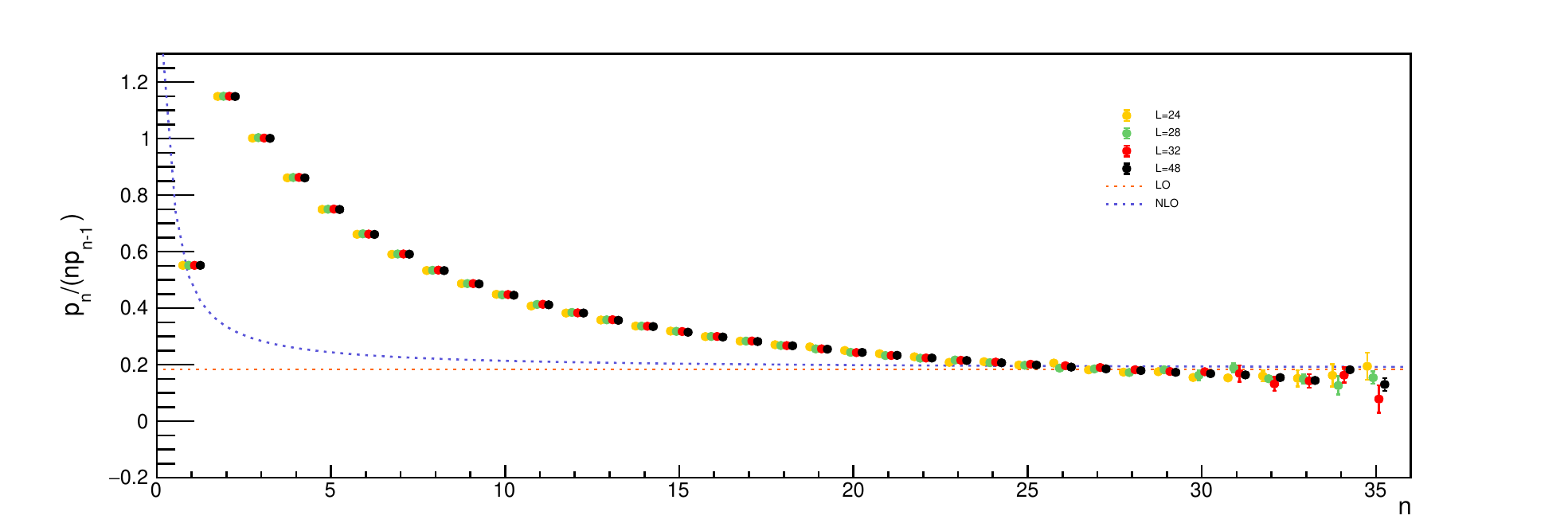}
	\end{subfigure}
	\caption{\label{fig:factgrowth}Ratio $p_n/(np_{n-1})$ extracted from our data at $L=24$, $28$, $32$, $48$. In order to be visible, points  referring to different volumes are placed side by side. The leading order (LO) prediction refers to the $n\to\infty$ limit, while the next-to-leading order (NLO) one includes the first $1/n$ correction.}
\end{figure}

\begin{figure}[h!]
	\centering
	\begin{subfigure}{.49\textwidth}
		\includegraphics[width=\textwidth,trim={0.8cm 0 1.9cm 0},clip]{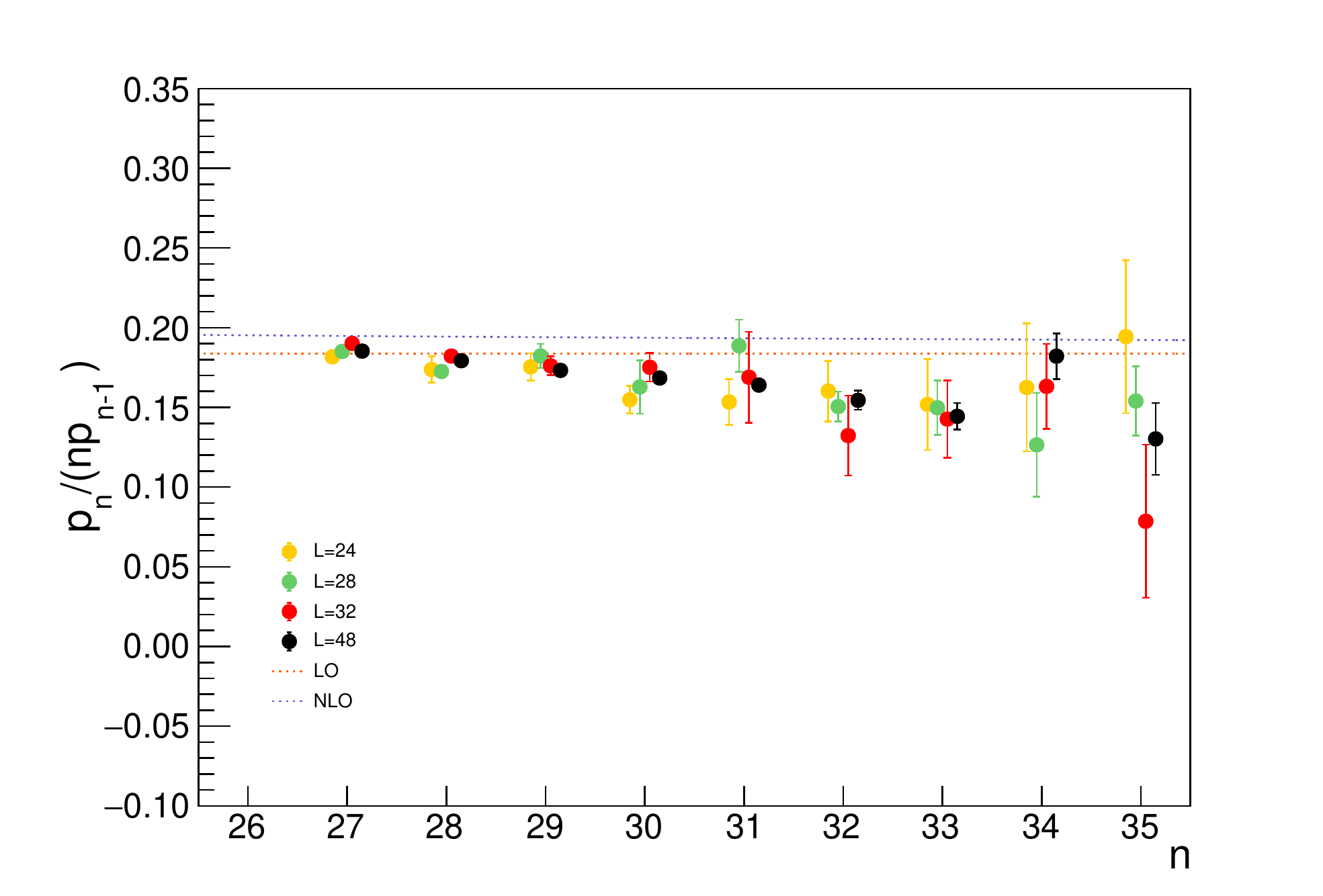}
	\end{subfigure}
	\caption{\label{fig:factgrowthzoom}Same as \myfigurename~\ref{fig:factgrowth}, but the region at large $n$ is enlarged.}
\end{figure}

From the analysis in \myrefsname~\cite{DiRenzo:1995qc,Bali:2014fea}, it is possible to predict the
asymptotic behaviour of the ratio
\begin{equation}
	\label{eq:asymptoticbehaviour}
	\frac{p_n}{np_{n-1}} = \frac{3\beta_0}{16\pi^2}
	\left[1+\frac{2\beta_1}{\beta_0^2}\frac{1}{n} + 
	O\left(\frac{1}{n^2}\right)\right]\,,
\end{equation}
where the use of the Wilson action with $N_c=3$ is assumed. This relation can be
derived under the hypothesis that the plaquette series has a fixed-sign
factorial divergence and the corresponding singularity in the Borel plane is the
source of an ambiguity that can be absorbed by redefining the condensate. It is
not possible to go further in the $1/n$ expansion since the $\beta_2$
coefficient is scheme-dependent and it is not known for staggered fermions. In
\myfigurename~\ref{fig:factgrowth} and \myfigurename~\ref{fig:factgrowthzoom}, the comparison between
\eq~\eqref{eq:asymptoticbehaviour} and our data at different volumes is shown.

\begin{figure}[h!]
	\centering
	\begin{subfigure}{.49\textwidth}
		\includegraphics[width=\textwidth,trim={0.5cm 0 1.4cm 0},clip]{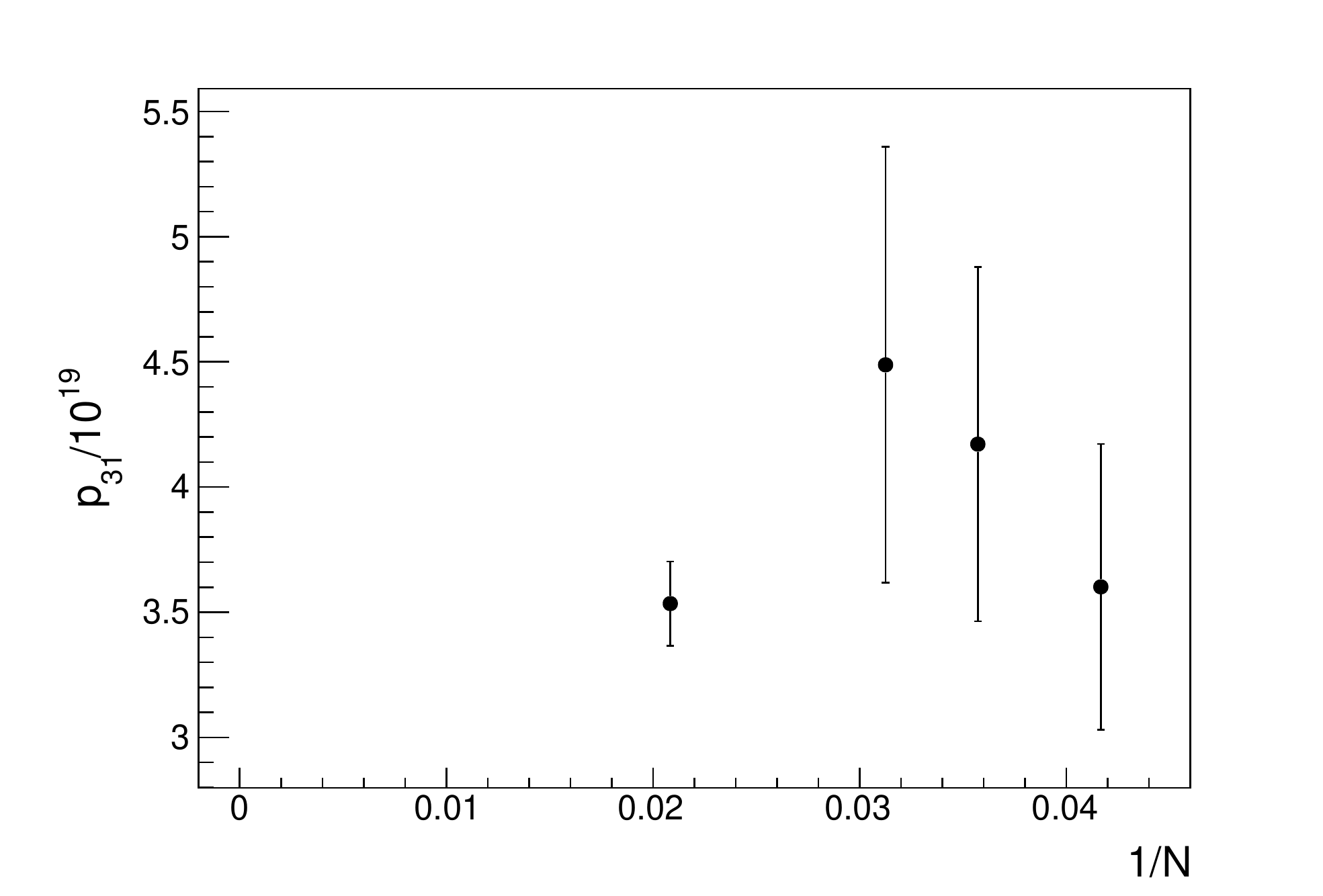}
	\end{subfigure}%
	\begin{subfigure}{.49\textwidth}
		\includegraphics[width=\textwidth,trim={0.5cm 0 1.4cm 0},clip]{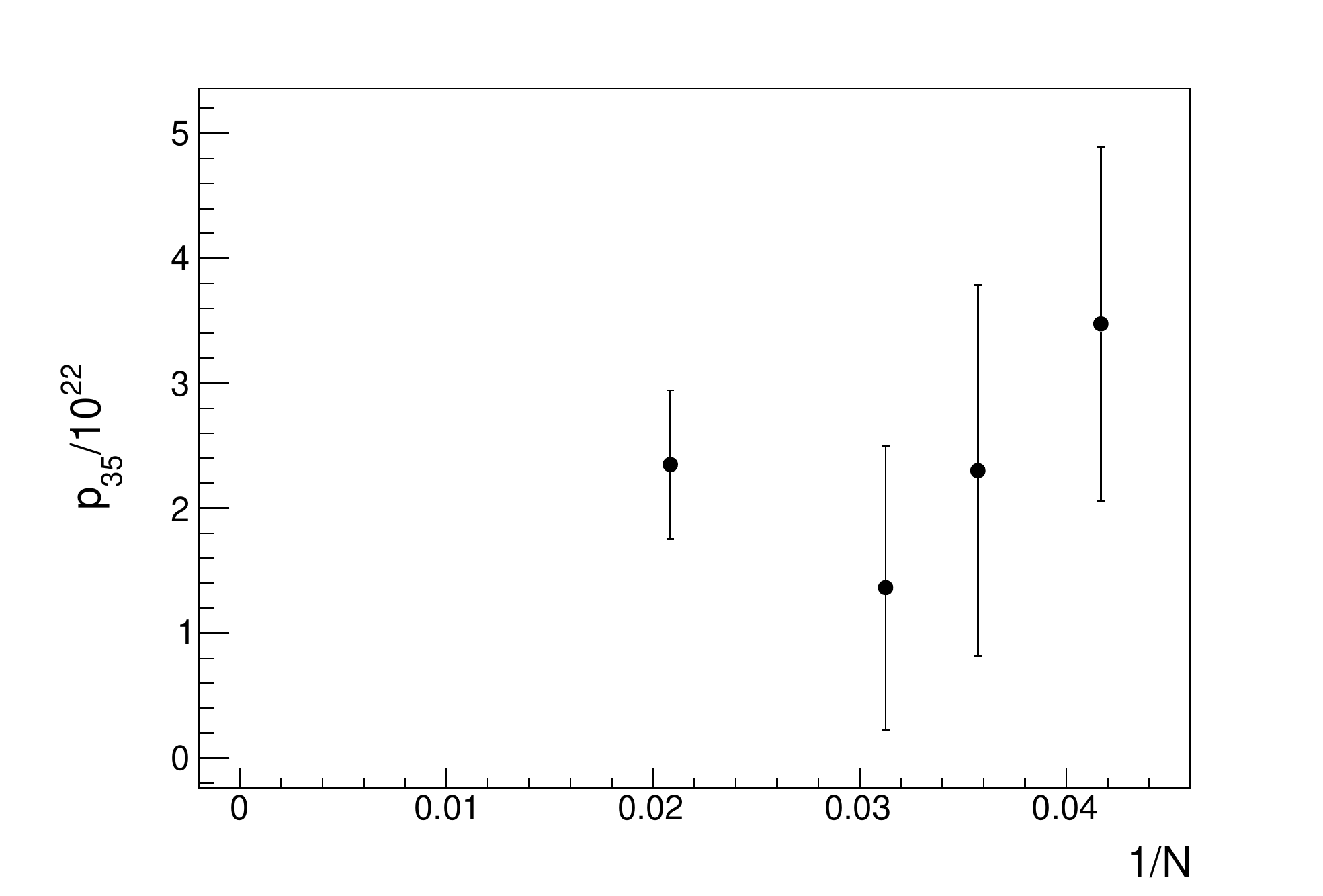}
	\end{subfigure}
	\caption{\label{fig:pnvolume}Coefficients $p_{31}$ and $p_{35}$ drawn as a function of the volume. No significant finite-volume effects are observed at our level of precision.}
\end{figure}

How finite-volume effects influence the values of the coefficients $p_n$ has
already been studied in the literature~\cite{DiRenzo:2000ua,Bali:2014fea}. From
a standard renormalon-based analysis, the value of the loop momenta that
contribute the most to $p_n$ decreases exponentially with $n$. Since the finite
size of the lattice provides a natural infrared cutoff, we expect finite-volume
effects to be larger at larger perturbative orders. The dependence of $p_n$ on
the lattice size $N$ can be modelled with a finite-volume OPE, exploiting the
separation of scales $a^{-1}\gg(Na)^{-1}$: the leading correction is~\cite{Bali:2014fea}
\begin{equation}
	\sum_{n=0}p_n(N)\beta^{-(n+1)}=\sum_{n=0}p_n\beta^{-(n+1)}-\frac{1}{N^4}\,C_G(\beta)
	\sum_{n=0}f_n\alpha((Na)^{-1})^{n+1}+O\left(\frac{1}{N^6}\right)\,,
\end{equation}
where $\alpha((Na)^{-1})$ must be expressed in terms of the coupling $\beta$ at
the scale $a^{-1}$ using \eq~\eqref{eq:BetaDef}. We do not attempt to take into
account $1/N^4$ effects, as our data do not allow to perform a reliable combined
fit. Apparently no significant finite-volume effects are visible where they
would be expected the most, i.e. at larger $n$. This is shown in the two
examples of \myfigurename~\ref{fig:pnvolume}. A similar behaviour has been observed in
\myrefname~\cite{Bali:2014fea}, where the data points computed on comparable volumes
show little dependence on the lattice size. In that study, a detailed analysis with a large number of volumes was
needed in order to be able to
fit the finite-volume corrections. The overall effect is found to be an increase
of the ratio $p_n/(n p_{n-1})$, see e.g. \myfigurename~6 in \myrefname~\cite{Bali:2014fea}.
In our case, data in finite volume do cross the theoretical expectation; still, considering 
the spread between points at different volumes in \myfigurename~\ref{fig:factgrowthzoom} as a source of systematic error, 
we could consider our measurements to be compatible with the asymptotic behaviour of \eq~\eqref{eq:asymptoticbehaviour}.
We also ascertain the existence of an inversion point when resumming the perturbative series, as explained in \mysectionname~\ref{sec:DetMinTerm}. Despite this encouraging behaviour, any definite conclusion about the existence of the expected renormalon can only be drawn after performing an appropriate infinite-volume study.
We emphasise that in this work the discrepancies in the determination of the $p_ n$
from different volumes must be interpreted as part of our systematic
uncertainty, being this an exploratory study. A precise assessment of the
finite-volume effects will be sought for a precise determination of the gluon
condensate; we are currently planning a set of dedicated simulations in the near
future to settle this issue. 


\subsection{Monte Carlo plaquette}
\begin{figure}[tbp]
	\centering
	\begin{subfigure}{.49\textwidth}
		\includegraphics[width=\textwidth,trim={0.1cm 0 1.5cm 0},clip]{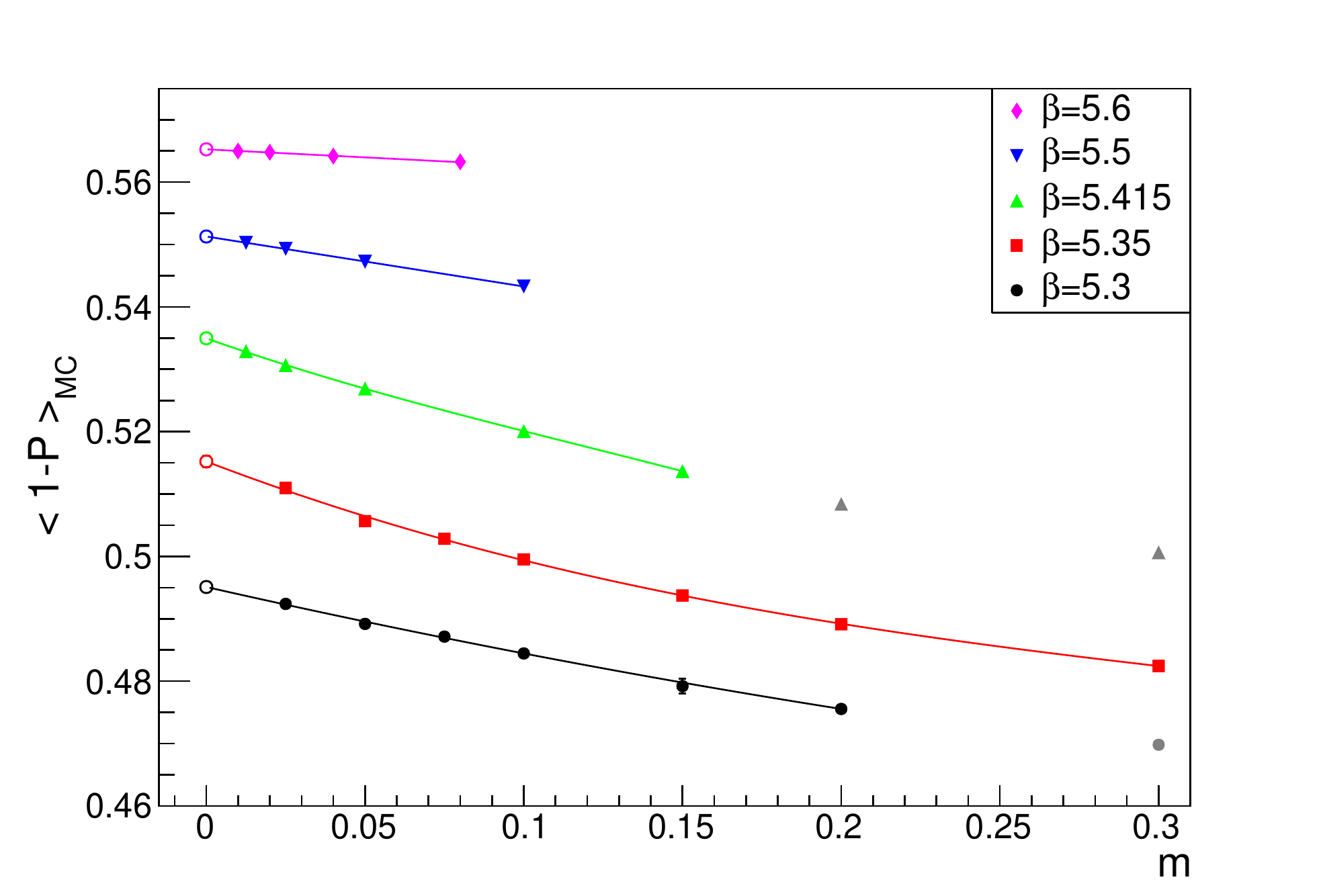}
	\end{subfigure}%
	\begin{subfigure}{.49\textwidth}
		\includegraphics[width=\textwidth,trim={0.1cm 0 1.5cm 0},clip]{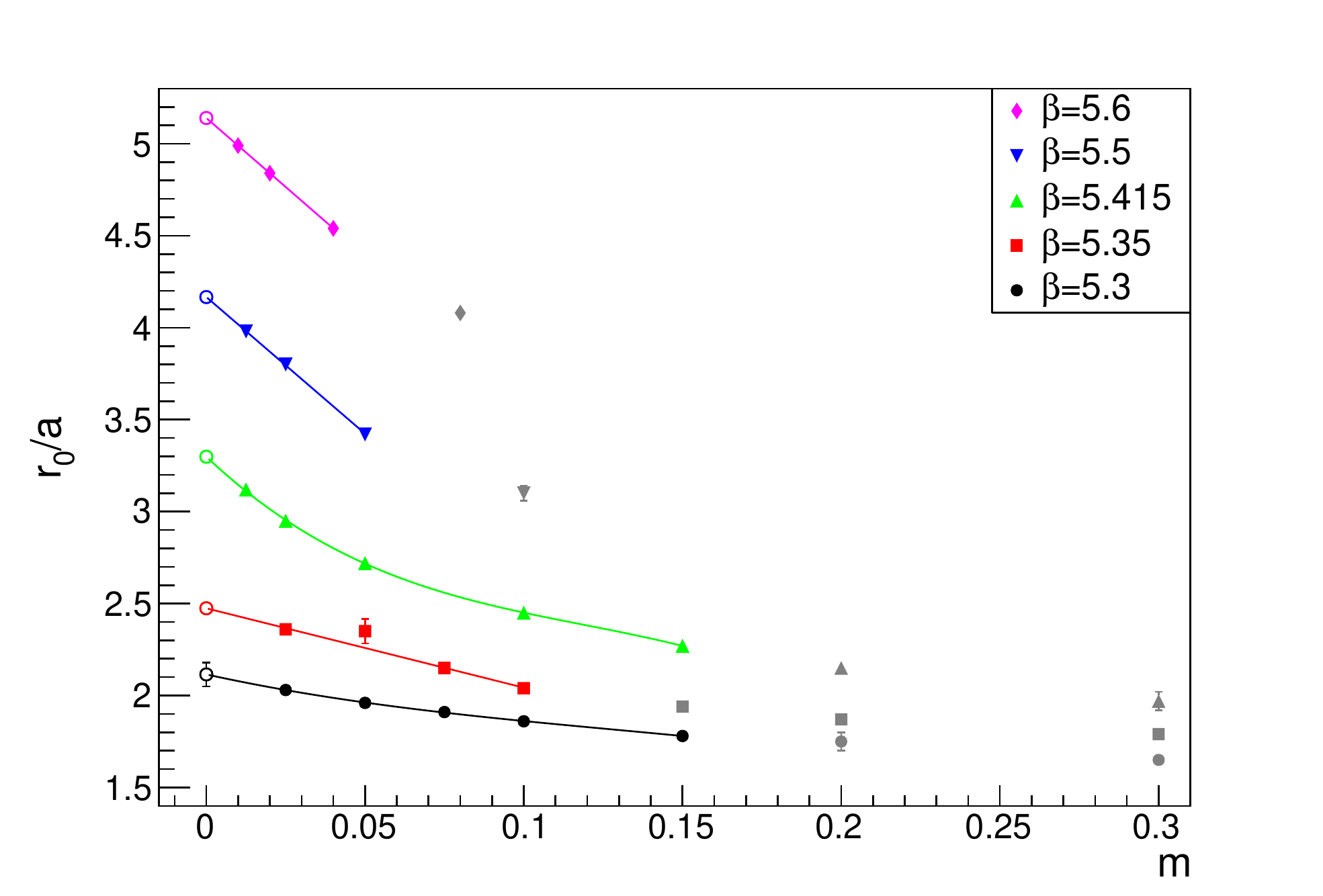}
	\end{subfigure}
	\caption{\label{fig:npplaq}Chiral extrapolation of the nonperturbative plaquette (left panel) and the ratio $r_0/a$ (right panel) at five different values of $\beta$. The grey points are available from \myrefname~\cite{Bali:2002wf} but are excluded because of our fit procedure. In most cases the error bar is smaller than the symbol. The orders of the polynomials used in the fits are in \mytablename~\ref{tab:npplaq}.}
\end{figure}

Nonperturbative values for the $\sun(3)$ plaquette with $N_f=2$ (rooted)
staggered fermions can be found in \myrefname~\cite{Bali:2002wf}, where data are
collected from \myrefsname~\cite{Tamhankar:1999ce,Heller:1994rz}. For each value of
the bare coupling, the physical scale is provided via the Sommer parameter
$r_0$~\cite{Sommer:1993ce}. The data are given for several values of the fermion
bare mass, and need to be extrapolated to the chiral limit for our purposes. A
reasonable assumption (for example adopted and verified also in
\myrefname~\cite{DellaMorte:2004bc} for the ratio $r_0/a$) is that the plaquette and
the ratio $r_0/a$ have a polynomial behaviour at small masses. We performed fits
with linear to cubic polynomials and varied the fit ranges to exclude points at
larger values of the masses, but in many cases the fits did not return a
satisfactory description of the data with sensible values of
$\chi^2/\text{dof}$.
Because we are using results from past simulations, it is difficult to track
accurately the systematic errors in the data. For this reason, we decided to
choose the fit with smaller $\chi^2/\text{dof}$ among those we tried and if
$\chi^2/\text{dof}>1$ the errors in the data were rescaled by a common
factor in order to have a reduced chi-squared equal to $1$. The fits resulting
from this approach are shown in \myfigurename~\ref{fig:npplaq}; the extrapolated values
for plaquettes and $r_0/a$ are in \mytablename~\ref{tab:npplaq}. Another approach
consists in considering the average between the largest and smallest
extrapolated values among all the different fits we tried (without rescaled
errors and with reduced chi-squared smaller than some reasonable threshold) and
assigning an error equal to the sum in quadrature between the largest error from
the fits and half the difference between the largest and smallest extrapolated
values. In this way we obtain results compatible (both for central values and
errors) with those in \mytablename~\ref{tab:npplaq}, confirming that the chiral
extrapolation is sound and the error bars conservative enough. Note that in this
paper we do not aim at a precise determination of the condensate, and therefore
we can be satisfied with an inflated error on the Monte Carlo data points. 

\begin{table}[tbp]
	\centering
	\begin{tabular}{|c|c c c c|}
		\hline
		$\beta$ & $\langle 1-P\rangle_\text{MC}$ & pol. ord. & $r_0/a$ & pol. ord. \\
		\hline
		$5.3$ & $0.4951(4)$ & $2$ & $2.11(7)$ & $3$ \\
		$5.35$ & $0.5152(9)$ & $3$ & $2.47(3)$ & $1$ \\
		$5.415$ & $0.5350(3)$ & $3$ & $3.30(3)$& $3$ \\
		$5.5$ & $0.55128(3)$ & $1$ & $4.17(2)$ & $1$ \\
		$5.6$ & $0.56526(5)$ & $1$ & $5.14(1)$ & $1$ \\
		\hline
	\end{tabular}
	\caption{\label{tab:npplaq} Results of the chiral extrapolation for the plaquette and the scale. The order of the polynomials used in the fits is indicated.}
\end{table}

\subsection{Determination of the minimal term}
\label{sec:DetMinTerm}
\begin{figure}[tbp]
	\centering
	\begin{subfigure}{.49\textwidth}
		\includegraphics[width=\textwidth,trim={0.1cm 0 1.2cm 0},clip]{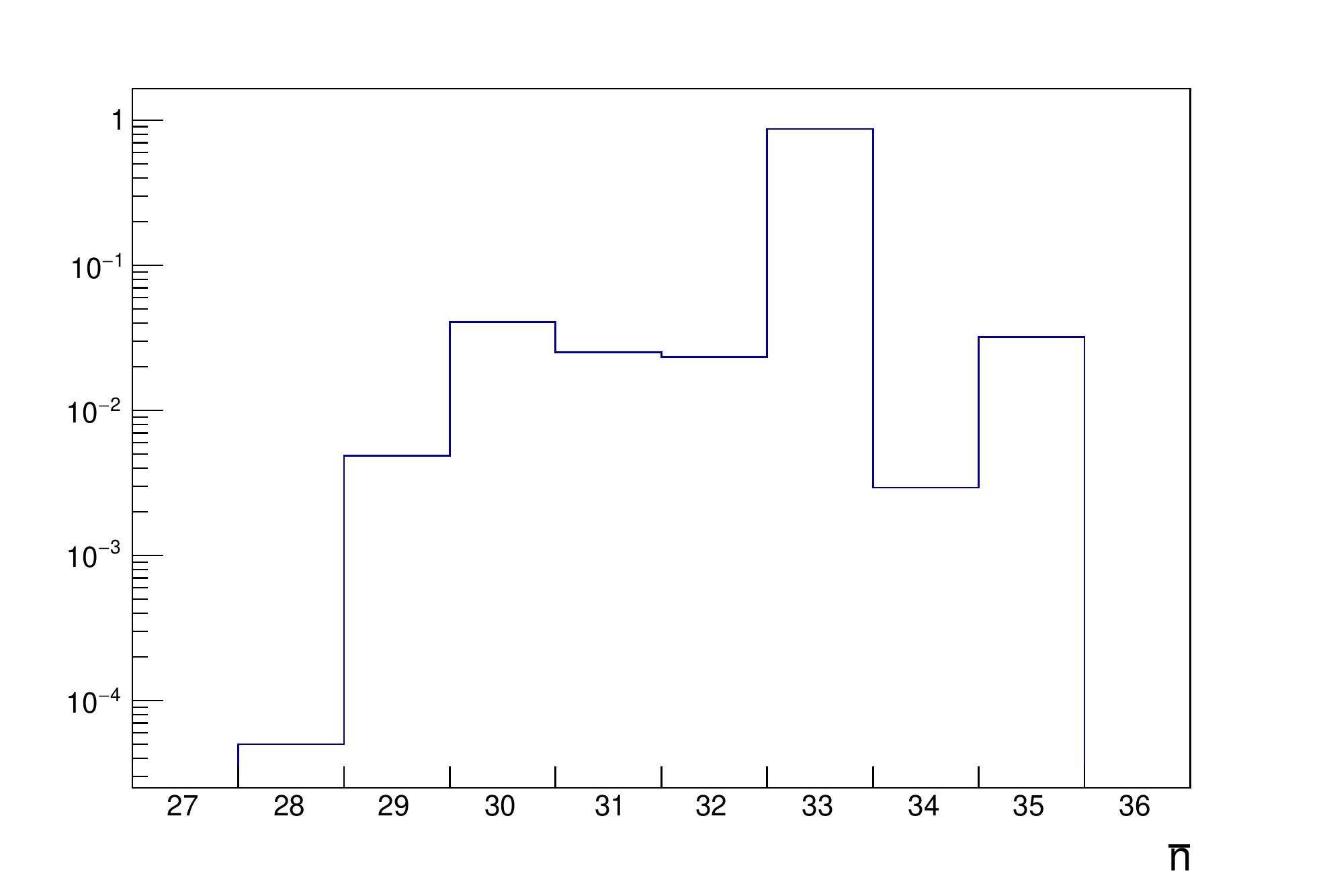}
	\end{subfigure}%
	\begin{subfigure}{.49\textwidth}
		\includegraphics[width=\textwidth,trim={0.1cm 0 1.2cm 0},clip]{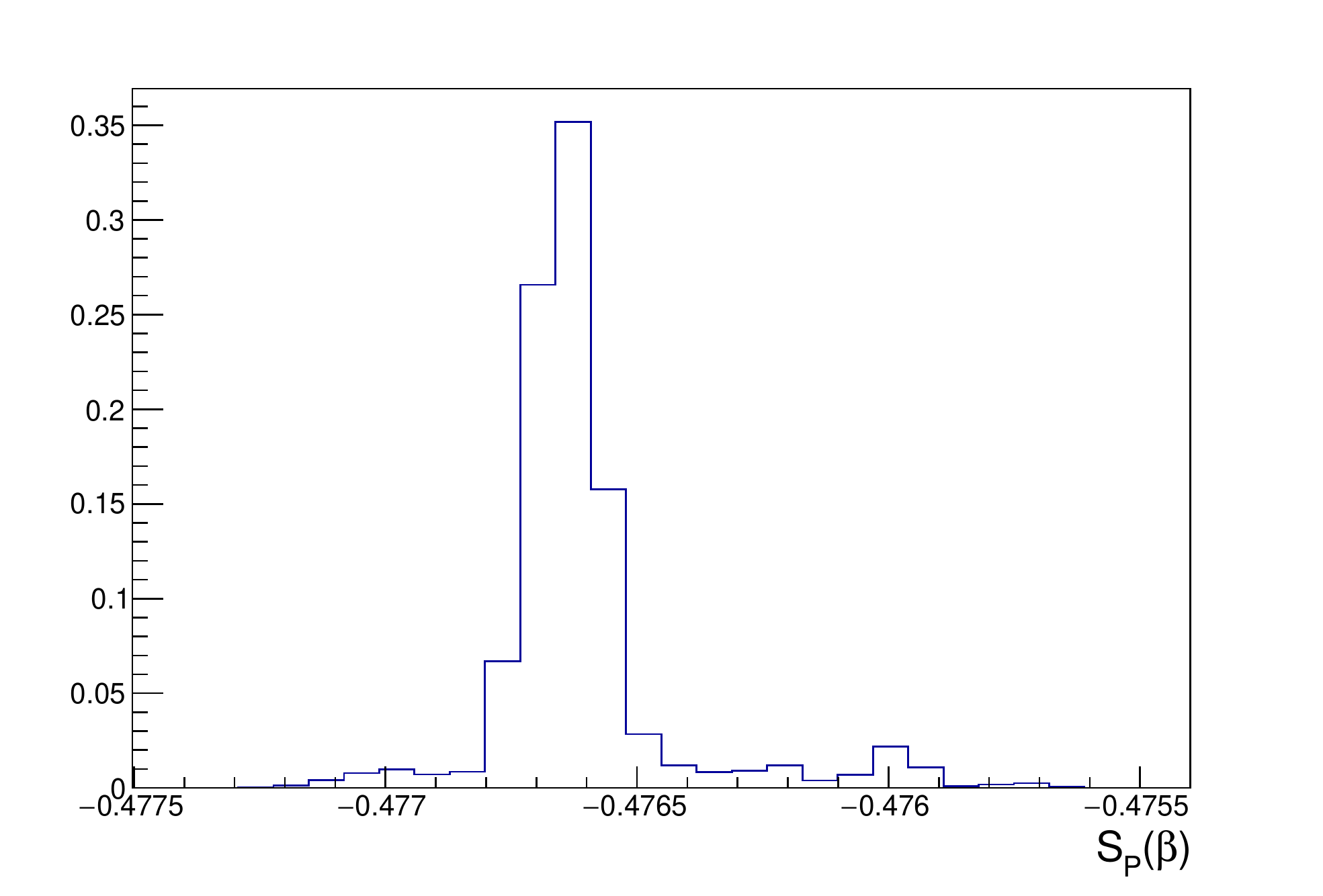}
	\end{subfigure}
	\caption{\label{fig:bootstrapnbar}Normalised distributions, over $10^5$ bootstrap samples, of $\bar{n}$ (left panel) and $S_P(\beta)$ (right panel) for $L=48$, $\beta=5.3$.}
\end{figure}

The perturbative contribution to the plaquette can be defined by the sum of the
series up to the minimal term. The determination of the minimal term, and the
summation of the series are performed separately for each volume. We choose the
prescription adopted in \myrefname~\cite{Bali:2014sja}, i.e. we define the minimal term to be the value $\bar{n}$ that minimises the product $p_n \beta^{-(n+1)}$ and resum the series,
\begin{equation}
	S(\beta)_P=\sum_{n=0}^{\bar n}p_n\beta^{-(n+1)}\,.
\end{equation}
Our results for all combinations of $L$ and $\beta$ are summarised in
\mytablename~\ref{tab:subtraction}. The order $\bar n$ at which the series starts to
diverge depends only on the central value of the coefficients $p_n$ and not on
their errors: in order to check that the inversion point determined by our
procedure is stable, we bootstrapped the procedure by generating an ensemble of
sets of coefficients $\left\{p_n\right\}$. For each set, the coefficients $p_n$
are drawn from a Gaussian probability, whose mean and covariance are taken from
the fit procedure described in \mysectionname~\ref{sec:plaquette}. We then determine $\bar n$ for each of these sets. The
inversion point turns out to be stable, as shown in
\myfigurename~\ref{fig:bootstrapnbar} for a the case $L=48$, and $\beta=5.3$. This
particular case is shown for illustration purposes, and the same features are
seen in all other combinations of $L$ and $\beta$.

The gluon condensate is then determined from
\begin{equation}
	\label{eq:subtraction}
	\braket{O_G}=\frac{36}{\pi^2} \, C^{-1}_G(\beta)\,a^{-4}\,[\braket{P}_\text{MC}(\beta)-S_P(\beta)]
\end{equation}
with
\begin{equation}
	C^{-1}_G(\beta)=1+\frac{3}{8\pi^2}\frac{\beta_1}{\beta_0} \frac{1}{\beta}+O\left(\frac{1}{\beta^2}\right)\,.
\end{equation}
The coefficient $\beta_2$ is not universal, and is actually unknown for the
discretisation used in this work. Not knowing $\beta_2$ prevents us from going
further in the expansion of $C_G$; since the correction due to the Wilson
coefficient falls between $5\%$ and $6\%$ for the values of $\beta$ considered,
a $6\%$ systematic uncertainty is added in quadrature after the subtraction.

\begin{figure}[tbp]
	\centering
	\begin{subfigure}{.49\textwidth}
		\includegraphics[width=\textwidth,trim={0.1cm 0 1.5cm 0},clip]{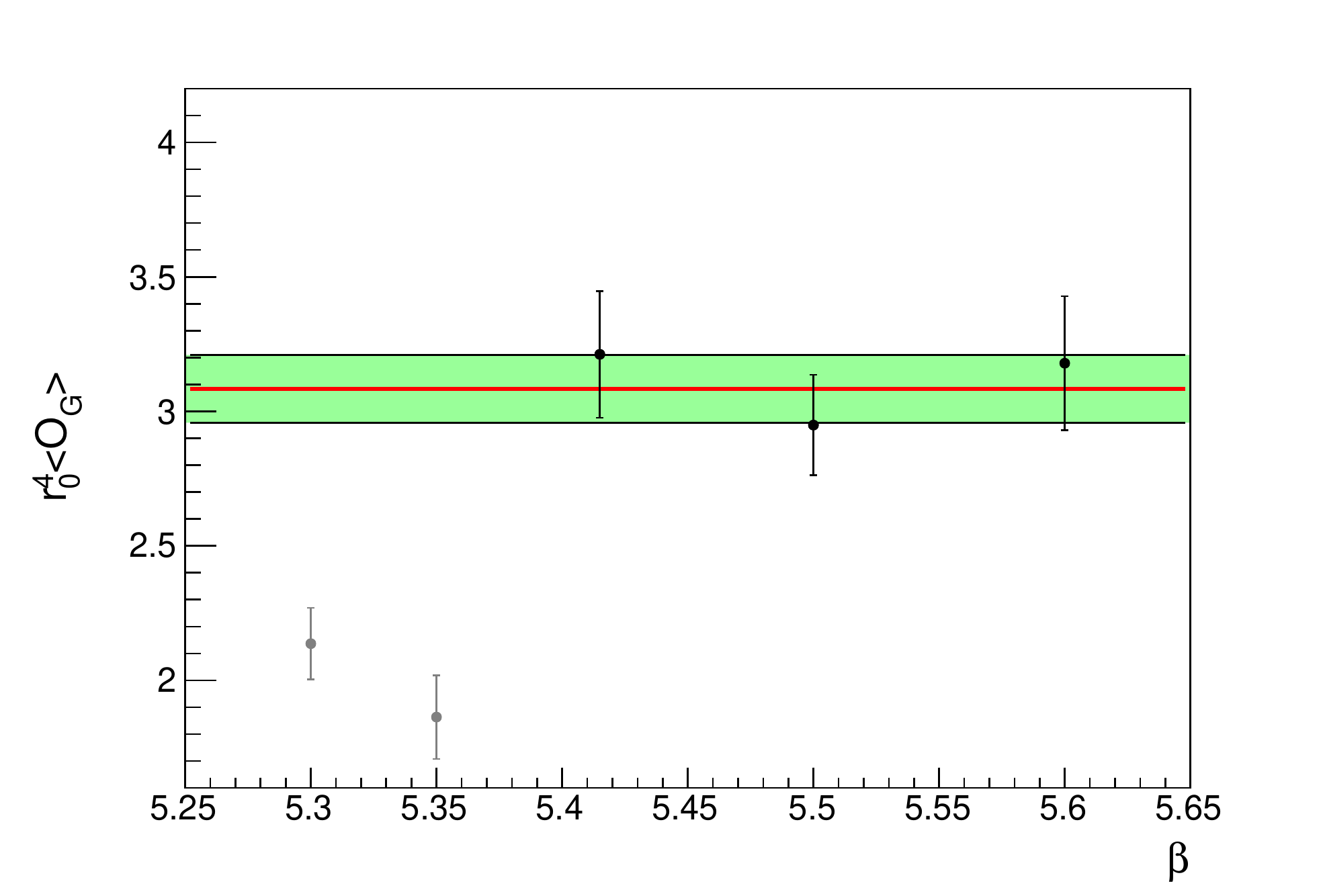}
	\end{subfigure}%
\begin{subfigure}{.49\textwidth}
		\includegraphics[width=\textwidth,trim={0.1cm 0 1.5cm 0},clip]{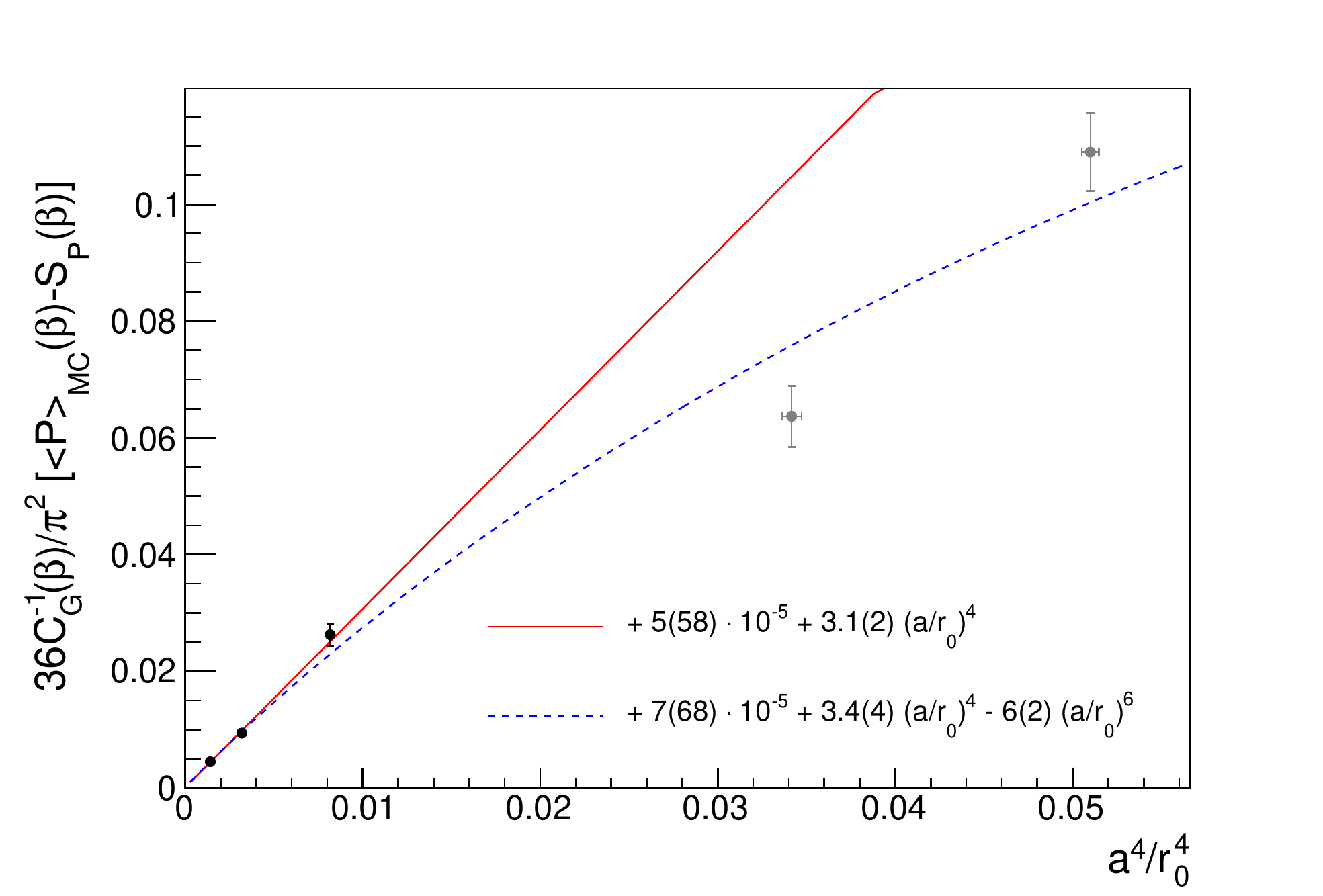}
	\end{subfigure}
	\caption{\label{fig:condensate}In the left panel, determination of the gluon condensate from \eq~\eqref{eq:subtraction}. The line corresponds to the weighted average of the three largest values of $\beta$. In the right panel, scaling of the condensate with $a^4$ (solid red line, grey points are excluded), with possibly a $a^6$ correction (dashed blue line, grey points are included). Both panels refer to $L=48$.}
\end{figure}

The result of the subtraction is shown in the left panel of
\myfigurename~\ref{fig:condensate}, for the largest volume. Since only a few values of $\beta$ is available,
it is hard to assess unambiguously the
presence of a plateau. We decided to discard from the analysis the two values of
the coupling corresponding to the coarser lattices, and define our best estimate
of the condensate as the weighted average of the values obtained at the
remaining $\beta$s. Our final results are summarised in the first column of
\mytablename~\ref{tab:condensatesummary}.

In order to put the choice of fit range on more solid ground, we studied the
scaling of $a^4\braket{O_G}$ as a function of $a^4$, as shown in \myfigurename~\ref{fig:condensate}.
The slope of a linear fit of the three finest lattice
spacings should give a determination of the condensate compatible with the value
extracted from the weighted average.
The spread between these two determinations and among the different volumes gives an idea of the magnitude of the systematic uncertainties involved.
We also tried to include in the analysis all the available values of $\beta$ and add a $a^6$ correction, in the attempt to model the deviations at large values of the coupling; this procedure gives again consistent results (despite
a larger $\chi^2$).

Truncating the sum up to the minimal term is one of the possible prescriptions to define the sum of a divergent series.
The intrinsic ambiguity associated to $S_P(\beta)$ can be defined as the imaginary part of the
Borel integral, which at leading order in $1/n$ is $\sqrt{\pi\bar n/2}\,p_{\bar
	n}\,\beta^{-\bar n -1}$~\cite{Beneke:1998ui}.  In
\mytablename~\ref{tab:ambiguitysummary}, the ambiguity associated to the
gluon condensate
\begin{equation}
	\label{eq:CondensateAmbiguity}
	\delta\braket{O_G}=\frac{36}{\pi^2} \, C^{-1}_G(\beta)\,a^{-4}\, \sqrt{\frac{\pi\bar n}{2}} \, p_{\bar n}\beta^{-\bar n -1}
\end{equation}
is summarised~\footnote{
	Our definition of the ambiguity differs from the one in \myrefname~\cite{Bali:2014fea} by a factor $\sqrt{\pi/2}$.
}.

\begin{table}[tbp]
	\centering
	\begin{tabular}{|c|c c c c|}
		\hline
		$\beta$ & $L$ & $S_P(\beta)$ & $\bar n$ & $p_{\bar{n}} \beta^{-(\bar{n}+1)}$ \\
		\hline 
		\multirow{4}{*}{$5.3$}
		& $24$ & $0.47515(9)$ & $25$ & $3.70 \cdot 10 ^{-4}$ \\
		& $28$ & $0.4767(1)$ & $30$ & $2.52 \cdot 10 ^{-4}$ \\
		& $32$ & $0.4775(4)$ & $35$ & $5.23 \cdot 10 ^{-5}$ \\
		& $48$ & $0.47665(7)$ & $33$ & $1.97 \cdot 10 ^{-4}$ \\
		\hline
		\multirow{4}{*}{$5.35$}
		& $24$ & $0.46718(8)$ & $25$ & $2.90 \cdot 10 ^{-4}$ \\
		& $28$ & $0.46843(9)$ & $30$ & $1.88 \cdot 10 ^{-4}$ \\
		& $32$ & $0.4690(3)$ & $35$ & $3.73 \cdot 10 ^{-5}$ \\
		& $48$ & $0.46826(5)$ & $33$ & $1.43 \cdot 10 ^{-4}$ \\
		\hline
		\multirow{4}{*}{$5.415$}
		& $24$ & $0.4587(1)$ & $33$ & $1.06 \cdot 10 ^{-4}$ \\
		& $28$ & $0.45844(7)$ & $30$ & $1.29 \cdot 10 ^{-4}$ \\
		& $32$ & $0.4588(2)$ & $35$ & $2.42 \cdot 10 ^{-5}$ \\
		& $48$ & $0.45822(4)$ & $33$ & $9.51 \cdot 10 ^{-5}$ \\
		\hline
		\multirow{4}{*}{$5.5$}
		& $24$ & $0.44663(9)$ & $33$ & $6.22 \cdot 10 ^{-5}$ \\
		& $28$ & $0.44651(6)$ & $30$ & $7.98 \cdot 10 ^{-5}$ \\
		& $32$ & $0.4466(1)$ & $35$ & $1.38 \cdot 10 ^{-5}$ \\
		& $48$ & $0.44627(4)$ & $33$ & $5.60  \cdot 10 ^{-5}$ \\
		\hline
		\multirow{4}{*}{$5.6$}
		& $24$ & $0.43384(6)$ & $34$ & $3.32 \cdot 10 ^{-5}$ \\
		& $28$ & $0.43380(5)$ & $30$ & $4.57 \cdot 10 ^{-5}$ \\
		& $32$ & $0.43383(6)$ & $35$ & $7.21 \cdot 10 ^{-6}$ \\
		& $48$ & $0.43357(3)$ & $33$ & $3.03 \cdot 10 ^{-5}$ \\
		\hline
	\end{tabular}
	\caption{\label{tab:subtraction} Summation up to the minimal term of the perturbative series of the plaquette.}
\end{table}

\begin{table}[tbp]
	\centering
	\begin{tabular}{|c|c c c|}
		\hline
		$L$ & $r_0^4\braket{O_G}_1$ & $r_0^4\braket{O_G}_2$ & $r_0^4\braket{O_G}_3$\\
		\hline
		$24$ & $2.6(1)$ & $2.9(2)$ & $3.1(4)$ \\
		$28$ & $2.8(1)$ & $3.1(2)$ & $3.4(4)$ \\
		$32$ & $2.4(1)$ & $2.9(2)$ & $3.2(4)$ \\
		$48$ & $3.1(1)$ & $3.1(2)$ & $3.4(4)$ \\
		\hline
	\end{tabular}
	\caption{\label{tab:condensatesummary}Determination of the gluon condensate at different volumes. The determination labelled with $1$ is obtained from the weighted average of the values at the three largest values of $\beta$. The determinations labelled with $2$ and $3$ are obtained by studying the scaling of $a^4\braket{O_G}$ with $a^4$, as in the right panel of \myfigurename~\ref{fig:condensate}; they correspond respectively to the fit without and with $a^6$ correction (see text for the details).}
\end{table}

\begin{table}[tbp]
	\centering
	\begin{tabular}{|c|c c c|}
		\hline
		\multirow{2}{*}{$L$}
		& \multicolumn{3}{c|}{$r_0^4\delta\braket{O_G}$}\\
		& $\beta=5.415$ & $\beta=5.5$ & $\beta=5.6$\\
		\hline
		$24$ & $0.4(2)$ & $0.5(4)$ & $0.7(5)$ \\
		$28$ & $0.4(3)$ & $0.7(4)$ & $0.9(5)$ \\
		$32$ & $0.3(2)$ & $0.5(3)$ & $0.3(3)$ \\
		$48$ & $0.3(2)$ & $0.5(3)$ & $0.6(4)$ \\
		\hline
	\end{tabular}
	\caption{\label{tab:ambiguitysummary}Ambiguity of the gluon condensate determined from \eq~\eqref{eq:CondensateAmbiguity} at the three largest values of $\beta$.}
\end{table}

\section{Conclusions}
\label{sec:conclusions}
We used NSPT to perform for the first time large-order computations in lattice
gauge theories coupled to massless fermions. We adopted twisted boundary
conditions for the gauge fields to remove the zero-momentum mode. Since our
fermions are in the fundamental representation, we consistently provided them
with a smell degree of freedom. Both Wilson and (for the first time in NSPT)
staggered fermions have been implemented. While for the former we performed an
exploratory study of the critical mass up to order $O(\beta^{-7})$, the latter
are ultimately the best choice to reach very high orders, due to their residual
chiral symmetry that bypasses the need of an additive mass renormalisation.

Numerical instabilities were noticed in the study of simple models in NSPT since
the early days of the method, but gauge theories have always been reported to
stay on a safe side in      this respect, even at orders as high as the ones we
investigated in this work. With fermions in place, we now found that numerical
instabilities arise for lattice gauge theories at high orders. While we plan to
investigate the causes and develop a solution to this, the problem did not
prevent us to reach order $O(\beta^{-35})$ in the expansion of the basic
plaquette for $N_c=3$ and $N_f=2$.

The plaquette has been for a long time the stage for the determination of the gluon condensate, to which is connected in the continuum limit.
The perturbative expansion of the plaquette, which corresponds to the power divergent contribution associated to the identity operator in the relevant OPE, must be subtracted from nonperturbative Monte Carlo lattice computations.
This long-standing and tough problem was eventually solved a few years ago in pure gauge~\cite{Bali:2014fea,Bali:2014sja}, thanks to NSPT.
Equipped with our high-orders expansions, we tackled once again the problem in the lattice regularisation of full QCD.
We computed the perturbative expansion of the plaquette, and subtracted it from Monte Carlo measurements. In this context, NSPT is crucial: it is actually the only tool enabling this procedure, which asks for having the asymptotic behaviour of such series under control. This happens since the perturbative expansion of the plaquette is expected to be plagued by renormalon ambiguities. 
Under the assumption of considering finite-volume effects as a source of systematic errors, the observed growth of the coefficients in the expansion could be compatible with the leading IR renormalon; nevertheless, the large uncertainties and the lack of a study of finite-volume effects prevent us from drawing any definite conclusion. The IR renormalon forces to absorb the ambiguities attached to the perturbative series into the definition of the condensate itself. All in all, this implies that we needed a prescription to perform the computation. The one we chose amounts to summing the perturbative series up to its minimal term (which means computing the series up to orders that only NSPT can aim at).

We regard this project as a first exploratory study.
We could confirm both that the IR renormalon can be directly inspected, and that the series can be computed up to orders where the inversion point beyond which the expansion starts to diverge (at values of the coupling which are the typical ones in lattice simulations) is clearly visible.
We performed our simulations at different lattice extents, in order to have a first estimate of finite-size effects (again, in both the study of renormalon behaviour and in the truncation of the series).
This is the point which has to be better investigated in a following study.
At the moment, finite-size effects are still to be considered as a systematic source of errors in our procedure.

On top of the follow-ups we have already discussed, we plan to extend our study to different number of colours, number of flavours and fermionic representations.
It would be of the utmost importance to assess the high-order behaviour of perturbative coefficients in gauge theories different from QCD, to probe regions in the space of theories in which a (quasi-)conformal window can be present.
This could be a powerful, alternative method to look for candidate theories for physics beyond the Standard Model.

\section*{Acknowledgements}
We would like to thank Gunnar Bali and Antonio Pineda for useful comments on the
manuscript. A special thought goes to the organiser of the workshop
"High-precision QCD at low energies" in Benasque, where these ideas were first
discussed. LDD is supported by an STFC Consolidated Grant, ST/P0000630/1, and a
Royal Society Wolfson Research Merit Award, WM140078. FDR acknowledges support
from INFN under the research project {\sl i.s. QCDLAT}. Access to MARCONI was
obtained through a CINECA-INFN agreement. Most of this work was performed using
the Cambridge Service for Data Driven Discovery (CSD3), part of which is
operated by the University of Cambridge Research Computing on behalf of the STFC
DiRAC HPC Facility (\url{www.dirac.ac.uk}). The DiRAC component of CSD3 was
funded by BEIS capital funding via STFC capital grants ST/P002307/1 and
ST/R002452/1 and STFC operations grant ST/R00689X/1. DiRAC is part of the
National e-Infrastructure.

\appendix
\section{Group theory conventions}
\label{sec:group-theory-conv}
The conventions used for group theoretical manipulations are summarised here. We consider the gauge group $\sun(N_c)$.

The generators of the group are denoted by $T^a$; the indices $a,b,c=1, \ldots, N_c^2-1$ are assumed to be indices in the adjoint representation. The generators are defined to be Hermitian, and satisfy the commutation relations
\begin{equation}
	\label{eq:StructConst}
	\left[T^a,T^b\right] = \sum_c i f^{abc} T^c\, ,
\end{equation}
where $f^{abc}$ are the group structure constants. The normalisation of the generators is chosen to be such that
\begin{equation}
	\label{eq:GenNorm}
	\mathrm{Tr}\left(T^a T^b\right) = \frac12 \delta^{ab}\, . 
\end{equation}
The left derivative on the group is defined as $\nabla_{x\mu}=\sum_a T^a \nabla^a_{U_{\mu}(x)}$, where the Lie derivative is given by
\begin{equation}
	\label{eq:LieDeriv}
	\nabla^a_V f(V) = \lim_{\alpha\to 0} \frac{1}{\alpha} 
	\left[
	f\left(e^{i \alpha T^a} V\right) - f(V)
	\right]\, .
\end{equation}
We define an operator, $\poa$, that projects on the algebra $\mathfrak{g}$ of the group: 
\begin{equation}
	\label{eq:PoADef}
	\poa(X) = \frac12 \left(
	X - X^\dagger - \frac{1}{N_c} \Tr\left(X-X^\dagger\right)
	\right)\,.
\end{equation}
The indices $i,j=1,\ldots,N_c$ will be used as indices in the fundamental representation, $r,s=1,\ldots,N_c$ as indices in the antifundamental representation.

\section{Optimisation of the fermion drift}
\label{sec:driftoptimisation}
A useful optimisation consists in improving on \eqs~\eqref{eq:wilsondrift} and~\eqref{eq:staggereddrift} so that it becomes numerically cheaper to evaluate the fermion drift. Considering for example Wilson fermions, we notice that it is possible to simplify the trace
\begin{align}
	\label{eq:franztrick}
	\Tr(\nabla^a_{x\mu} M) M^{-1}&=i\tilde\Tr \left[\left(T^aD(\mu)M^{-1}\right)_{x,x}-\left(\gamma_5D(\mu)^\dagger\gamma_5T^aM^{-1}\right)_{x,x}\right]=\notag\\
	&=i\sum_{y,\beta,i,r} \left(\delta_{x,y}[T^aD(\mu)M^{-1}]_{y\beta i r, y\beta i r}-\text{h.c.}\right)\,,
\end{align}
where $\tilde\Tr$ is tracing all indices but the position one, and we used the fact that the inverse Wilson operator is $\gamma_5$-Hermitian. For staggered fermions the simplification is analogous because the inverse staggered operator is antihermitian.
The step must be done before the stochastic evaluation of the trace: once the random sources are introduced, cyclic invariance gets broken and will be restored only on average.
Using \eq~\eqref{eq:franztrick} as a starting point, we obtain a drift which is already in the algebra (no need of taking the real part) and reads
\begin{subequations}
	\begin{align}
		F^{f}_\mu(x)_{ij}&=\frac{N_f}{N_c}\frac{\tau}{\beta}\poa\left(\sum_\beta \varphi^{(\mu)}(x)_\beta\,\xi(x)^\dag_\beta\right)_{ij}&\qquad&\text{(Wilson fermions)}\\
		F^{f}_\mu(x)_{ij}&=\frac{N_f}{4N_c}\frac{\tau}{\beta}\poa\left(\varphi^{(\mu)}(x)\,\xi(x)^\dag\right)_{ij}&\qquad&\text{(staggered fermions)}\,.
	\end{align}
\end{subequations}
In a similar fashion, it could be possible to show that also
\begin{subequations}
	\begin{align}
		F^{f}_\mu(x)_{ij}&=\frac{N_f}{N_c}\frac{\tau}{\beta}\poa\left(\sum_\beta \tilde\varphi^{(\mu)}(x)_\beta\,\psi(x)^\dag_\beta\right)_{ij}&\qquad&\text{(Wilson fermions)}\\
		F^{f}_\mu(x)_{ij}&=\frac{N_f}{4N_c}\frac{\tau}{\beta}\poa\left(\tilde\varphi^{(\mu)}(x)\,\psi(x)^\dag\right)_{ij}&\qquad&\text{(staggered fermions)}
	\end{align}
\end{subequations}
are legitimate expressions for the drift.
All these new formulae are numerically different from those in \eqs~\eqref{eq:wilsondrift} and~\eqref{eq:staggereddrift} but lead to the same results on average; clearly the advantage is that only half of the Lie derivative has to be computed.

\section{Fourier transforms with twisted boundary conditions}
\label{sec:appendixft}
If $f(x)$ is a periodic function defined on the $L^4$ lattice, its Fourier transform and inverse are
\begin{equation}
	\label{eq:fourierexp}
	f(x) = \frac{1}{L^4}\sum_{p_\parallel} e^{ip_\parallel x}\tilde f(p_\parallel)\,,
	\qquad\tilde f(p_\parallel) = \sum_x e^{-ip_\parallel x} f(x)\,,
\end{equation}
where $p_\parallel$ is the quantised vector $p_\parallel=\frac{2\pi}{L}(n_1,n_2,n_3,n_4)$ and the sum is to be read $\sum_{p_\parallel} = \sum_{n_1,n_2,n_3,n_4=0}^{L-1}$. Antiperiodicity in the direction $\hat\nu$ would lead again to \eq~\eqref{eq:fourierexp} but with a quantised momentum $(p_\parallel)_\mu=\frac{2\pi}{L}n_\mu+\frac{\pi}{L}\delta_{\mu\nu}$.

\paragraph{Twisted boundary conditions on a plane}
Let us consider some $N_c\times N_c$ matrix $M(x)$ (which for example can be a gauge link or the vector potential seen as matrices in colour space, or a fundamental fermion field seen as a matrix in colour-smell).
We impose twisted boundary condition in the $\hat 1,\hat 2$ plane so that
\begin{equation}
	\label{eq:planetwist}
	M(x+L\hat1)=\Omega_1 M(x)\Omega_1^\dag\,,\qquad
	M(x+L\hat2)=\Omega_2 M(x)\Omega_2^\dag\,,
\end{equation}
with $\Omega_2\Omega_1=z\Omega_1\Omega_2$, $z=z_{12}\in Z_N$.
If we had just (anti)periodic boundary conditions, we would treat the matrix as $N_c^2$ independent scalar functions; twisted boundary conditions actually couple the different components, therefore in order to expand $M(x)$ in plane waves we need to find a good basis for the matrix space: it can be proved (see \myrefsname~\cite{GonzalezArroyo:1982hz,Luscher:1985wf}) that the Fourier transform and its inverse are
\begin{equation}
	\label{eq:ffttwistexpansion}
	M(x)=\frac{1}{N_cL^4}\sum_{p_\parallel,p_\perp} e^{ipx}\,\Gamma_{p_\perp} \tilde M(p_\parallel)_{p_\perp}\,,\qquad
	\tilde M(p_\parallel)_{p_\perp}=\sum_x e^{-ip x}\,\Tr\Gamma_{p_\perp}^\dag M(x)\,,
\end{equation}
where $p=p_\parallel+p_\perp$, $p_\perp$ is the quantised vector $p_\perp=\frac{2\pi}{N_cL}(\tilde n_1,\tilde n_2,0,0)$ and the sum is to be read $\sum_{p_\perp} = \sum_{\tilde n_1,\tilde n_2=0}^{N_c-1}$.
The matrices $\Gamma_{p_\perp}$ form the sought basis in the matrix space: assuming a twist with $z=\exp(2\pi i/N_c )$, we can choose for example
\begin{equation}
	\label{eq:twistedbasis}
	\Gamma_{p_\perp}=\Omega_1^{\tilde n_2}\Omega_2^{-\tilde n_1}\,.
\end{equation}
A different choice for $z$ would have  somehow reshuffled the exponents in \eq~\eqref{eq:twistedbasis}. We see that the Fourier transform of $M(x)$ is a scalar function $\tilde M(p_\parallel)_{p_\perp}$, but momentum has a finer resolution compared to (anti)periodic boundary conditions: spatial and colour degrees of freedom mix in momentum space. Moreover, traceless matrices naturally do not have a zero momentum component, because $\tilde M(p_\parallel)_0=\sum_x e^{-ip_\parallel x}\Tr M(x)$.

\paragraph{Twisted boundary conditions in three directions}
The conditions in \eq~\eqref{eq:planetwist} are supplemented by
\begin{equation}
	M(x+L\hat3)=\Omega_3 M(x)\Omega_3^\dag\,,
\end{equation}
with $\Omega_3=\Omega_1^\rho\Omega_2^\sigma$ and $\rho,\sigma$ span all the possible twist choices. It can be shown that \eq~\eqref{eq:ffttwistexpansion} still holds but with a fine momentum $p_\perp=\frac{2\pi}{N_cL}(\tilde n_1,\tilde n_2,\tilde n_3,0)$. The component $\tilde n_3$ is not a new degree of freedom but depends on the values of $\tilde n_1,\tilde n_2$. For example, in the case $z=\exp(2\pi i/N_c ),\rho=\sigma=1$, then $\tilde n_3=(\tilde n_1+\tilde n_2)\mod N_c$. Other choices of $z,\rho,\sigma$ just give a new relation between $\tilde n_3$ and $z,\tilde n_1,\tilde n_2$.

\paragraph{Numerical implementation}
The Fast Fourier Transform (FFT) algorithm encodes \eq~\eqref{eq:fourierexp}, $\text{FFT}[f(x)]=\tilde f(p)$.
We cannot apply directly the FFT to each matrix element of $M(x)$, because the Fourier expansion has a dependence on $p_\perp x$. First, we need to project onto one of the $p_\perp$,
\begin{equation}
	\hat M(x)_{p_\perp}=e^{-ip_\perp x}\Tr\Gamma_{p_\perp}^\dag M(x)= \frac{1}{L^4} \sum_{p_\parallel} e^{ip_\parallel x}
	\tilde M(p_\parallel)_{p_\perp}\,,
\end{equation}
and then to each of these we apply the FFT,
\begin{equation}
	\tilde M(p_\parallel)_{p_\perp}=\text{FFT}[\hat M(x)_{p_\perp}]\,.
\end{equation}
At the end, $N_c^2$ projections and $N_c^2$ FFTs have been performed.
The inverse transform will be simply
\begin{equation}
	\hat M(x)_{p_\perp}=\text{FFT}^{-1}[\tilde M(p_\parallel)_{p_\perp}]\,
\end{equation}
followed by
\begin{equation}
	M(x) = \frac{1}{N_c} \sum_{p_\perp} e^{ip_\perp x}\,
	\Gamma_{p_\perp} \hat M(x)_{p_\perp}\,.
\end{equation}
Note that $\tilde M(p_\parallel)_{p_\perp}$ is a scalar function but the dependence on $p_\perp$ is through $\tilde n_1,\tilde n_2$, where each integer runs from $0$ to $N_c-1$: this allows a representation of the Fourier transform again with a $N_c\times N_c$ matrix field, $\left(M(p_\parallel)\right)_{\tilde n_1\tilde n_2}$. Of course this has to be understood only as a useful representation of the momentum degrees of freedom, not as a matrix in colour space.

\section{Autocorrelations and cross-correlations}
\label{sec:autocorrelation}
We consider a sample $\{a_i,b_i\}_{i=1}^N$ of measures of two observables $A,B$ taken from the stochastic process at equilibrium. Let $\braket{A}=a,\braket{B}=b$ be the expectation values respectively of the observables $A,B$.The \emph{cross-correlation function} is defined as
\begin{equation}
	\Gamma_{AB}(t)=\braket{(a_i-a)(b_{i+t}-a)}=\braket{a_ib_{i+t}}-ab\,,
\end{equation}
where we used the fact that the expectation value is not dependent on $i$ because the equilibrium distribution is time-independent.
The cross-correlation function is not an even function, $\Gamma_{AB}(-t)=\Gamma_{BA}(t)$.
In particular, $\Gamma_{AB}(0)=\text{Cov(A,B)}$ is the covariance between $A$ and $B$.
The average $\bar a=\frac{1}{N}\sum_{i=1}^Na_i$ is a stochastic variable that satisfies $\braket{\bar a}=a$.
The covariance between the estimators $\bar a$ and $\bar b$ is
\begin{align}
	\text{Cov}(\bar a,\bar b)&=\braket{(\bar a-a)(\bar b-b)}=\frac{1}{N^2}\sum_{i,j=1}^N\Gamma_{AB}(i-j)=\notag\\
	&=\frac{\text{Cov(A,B)}}{N}\left[1+\sum_{r=1}^{N-1}\left(1-\frac{r}{N}\right)\frac{\Gamma_{AB}(r)}{\Gamma_{AB}(0)}+\sum_{r=1}^{N-1}\left(1-\frac{r}{N}\right)\frac{\Gamma_{AB}(-r)}{\Gamma_{AB}(0)}\right]
\end{align}
but since the cross-correlation function is expected to drop exponentially at large times, it is possible to approximate
\begin{equation}
	\label{eq:covaverage}
	\text{Cov}(\bar a,\bar b)\simeq\frac{\text{Cov(A,B)}}{N}\,(\tau_{AB}^\text{int}+\tau_{BA}^\text{int})
\end{equation}
with the \emph{integrated cross-correlation time}
\begin{equation}
	\tau_{AB}^\text{int}=\frac{1}{2}+\sum_{r=1}^\infty\frac{\Gamma_{AB}(r)}{\Gamma_{AB}(0)}\,.
\end{equation}
We expect $\tau_{AB}^\text{int}\neq\frac{1}{2}$ when the observable $B$ has some dependence on $A$.
If $B$ is independent of $A$, we can assume $\tau_{AB}^\text{int}=\frac{1}{2}$.
An estimator for the cross-correlation function is
\begin{equation}
	\bar\Gamma_{AB}(t)=\frac{1}{N-t}\sum_{i=1}^{N-t}(a_i-\bar a)(b_{i+t}-\bar b)\,.
\end{equation}
and the integrated cross-correlation time can be extracted in the Madras-Sokal approximation~\cite{Madras:1988ei,Luscher:2005rx}.
Note that when $A=B$ then $\Gamma_{AA}(t)$ is the \emph{autocorrelation function} and \eq~\eqref{eq:covaverage} becomes $\text{Var}(\bar a)=2\tau_{AA}^\text{int}\text{Var}(\bar a)/N$, where $\tau_{AA}^\text{int}$ is the \emph{integrated autocorrelation time}.

\section{Twisted lattice perturbation theory}
\label{sec:twistedpt}
Twisted lattice perturbation theory for the a pure gauge theory was introduced in \myrefname~\cite{GonzalezArroyo:1982hz} (see also \myrefname~\cite{Snippe:1997ru}). Recently, the computation of Wilson loops has been treated in detail in \myrefname~\cite{Perez:2017jyq}. Here we focus on two vertices, introducing Wilson and staggered fermions with smell in the fundamental representation.
Feynman rules are fairly similar to those of lattice perturbation theory, apart from phases in propagators and vertices; all phases cancel in the first-order computations we considered. We recall also that the sum over momenta is inherited from the Fourier transform in \myappendixname~\ref{sec:appendixft},
\begin{equation}
	\cancel{\sum_k}=\frac{1}{N_cL^4}\sum_{k_\parallel,k_\perp}\,,
\end{equation}
and each fermion loop has to be divided by $N_c$, i.e. by the numbers of smells running in the loop. The function $f(p_\perp,p_\perp^\prime)=z^{-\tilde n_1\tilde n_2^\prime}$ is introduced for convenience. The gluon propagator is
\begin{equation}
	\label{eq:gluonpropagator}
	\braket{\tilde A_\mu(p)\tilde A_\nu(q)}=\delta_{p,q}\frac{(1-\delta_{p_\perp,0})}{2}f(p_\perp,p_\perp)\frac{1}{4\sum_{\rho}\sin^2\left(\frac{p_\rho}{2}\right)}\left[\delta_{\mu\nu}-(1-\xi)\frac{\sin\left(\frac{p_\mu}{2}\right)\sin\left(\frac{p_\nu}{2}\right)}{\sum_{\sigma}\sin^2\left(\frac{p_\sigma}{2}\right)}\right]\,,
\end{equation}
where $\xi$ is the gauge fixing parameter; note that the traceless property of the gauge field forces the propagator to vanish for $p_\perp=0$.  The Wilson and staggered propagators are defined respectively in \eqs~\eqref{eq:wilsonpropagator} and~\eqref{eq:staggeredpropagator}.  Below we write the fermion-fermion-gluon and fermion-fermion-gluon-gluon vertices in the Wilson and staggered case; $p_1, p_2$ are respectively the incoming and outgoing momenta of the fermions, $k_1, k_2$ are the outgoing momenta of the gluons.

\paragraph{Wilson fermions}
\begin{subequations}
	\begin{align}
		V_{ffg}(p_1,p_2,k_\perp)_{\mu}&=-g\,f(k_\perp,p_{2\perp})\left[
		i\gamma_\mu\cos\frac{1}{2}\left(p_1^A+p_2^A\right)_\mu+\sin\frac{1}{2}\left(p_1^A+p_2^A\right)_\mu
		\right]\\
		V_{ffgg}(p_1,p_2,k_{1\perp},k_{2\perp})_{\mu\nu}&=-g^2\delta_{\mu\nu}f(k_{1\perp}+k_{2\perp},p_{2\perp})\,\frac{1}{2}[f(k_{1\perp},k_{2\perp})+f(k_{2\perp},k_{1\perp})]\,\cdot\notag\\
		&\qquad\cdot\,\left[
		\cos\frac{1}{2}\left(p_1^A+p_2^A\right)_\mu-i\gamma_\mu\sin\frac{1}{2}\left(p_1^A+p_2^A\right)_\mu
		\right]
	\end{align}
\end{subequations}
\paragraph{Staggered fermions}
Here momentum conservation is made explicit, because the vertices are not diagonal in momentum space.
\begin{subequations}
	\begin{align}
		V_{ffg}(p_1,p_2,k_1)_\mu&=-ig\,f(k_{1\perp},p_{2\perp})\,\cos\left(p_2+\frac{k_1}{2}\right)_\mu\,\cdot\notag\\
		&\qquad\cdot\,\bar\delta(-p_{1\parallel}+k_{1\parallel}+p_{2\parallel}+\pi\bar\mu)\delta_{-p_{1\perp}+k_{1\perp}+p_{2\perp},0}\\
		V_{ffgg}(p_1,p_2,k_1,k_2)_{\mu\nu}&=ig^2\,f(k_{1\perp}+k_{2\perp},p_{2\perp})\frac{1}{2}\left[f(k_{1\perp},k_{2\perp})+f(k_{2\perp},k_{1\perp})\right]\,\cdot\notag\\
		&\qquad\cdot\,\sin\left(p_2+\frac{k_1}{2}+\frac{k_2}{2}\right)_\mu\,\cdot\notag\\
		&\qquad\cdot\,\delta_{\mu\nu}\,\bar\delta^{(4)}(-p_1+k_1+k_2+p_2+\pi\bar\mu)\delta_{k_{1\perp}-p_{1\perp}+k_{2\perp}+p_{2\perp},0}
	\end{align}
\end{subequations}

\section{Code development for NSPT}
\label{sec:codenspt}
We developed two independent NSPT codes in order to cross-check and improve our implementation.

\begin{figure}[tbp]
	\centering
	\begin{subfigure}{.49\textwidth}
		\includegraphics[width=\textwidth]{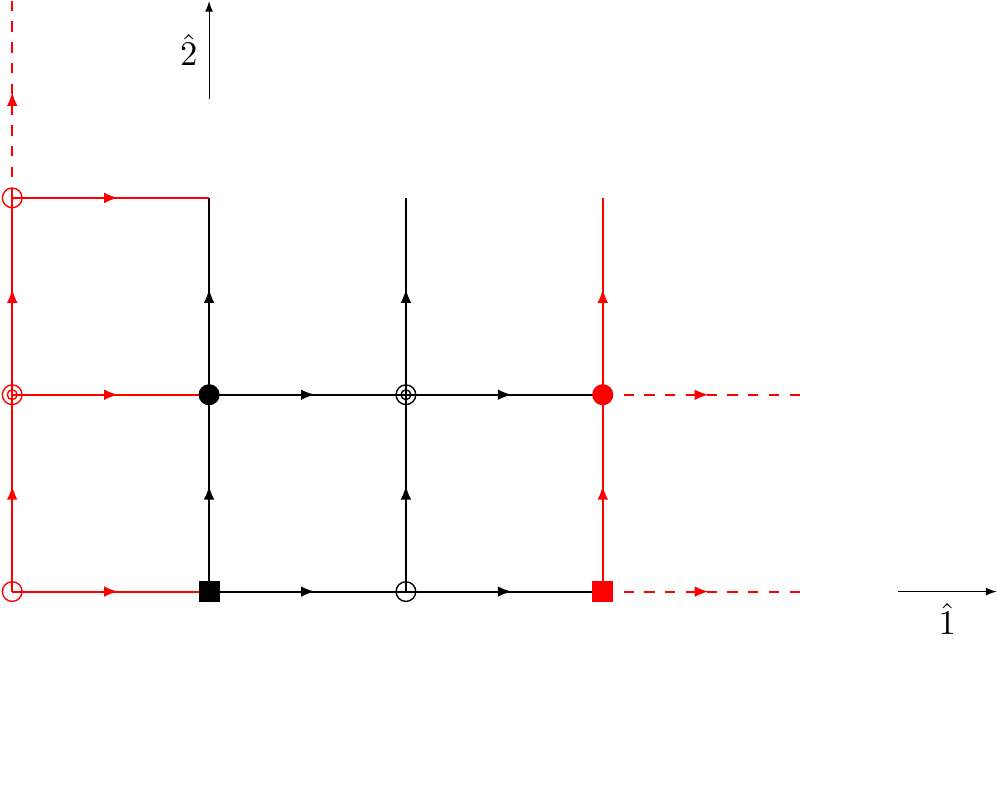}
	\end{subfigure}
	\begin{subfigure}{.49\textwidth}
		\includegraphics[width=\textwidth]{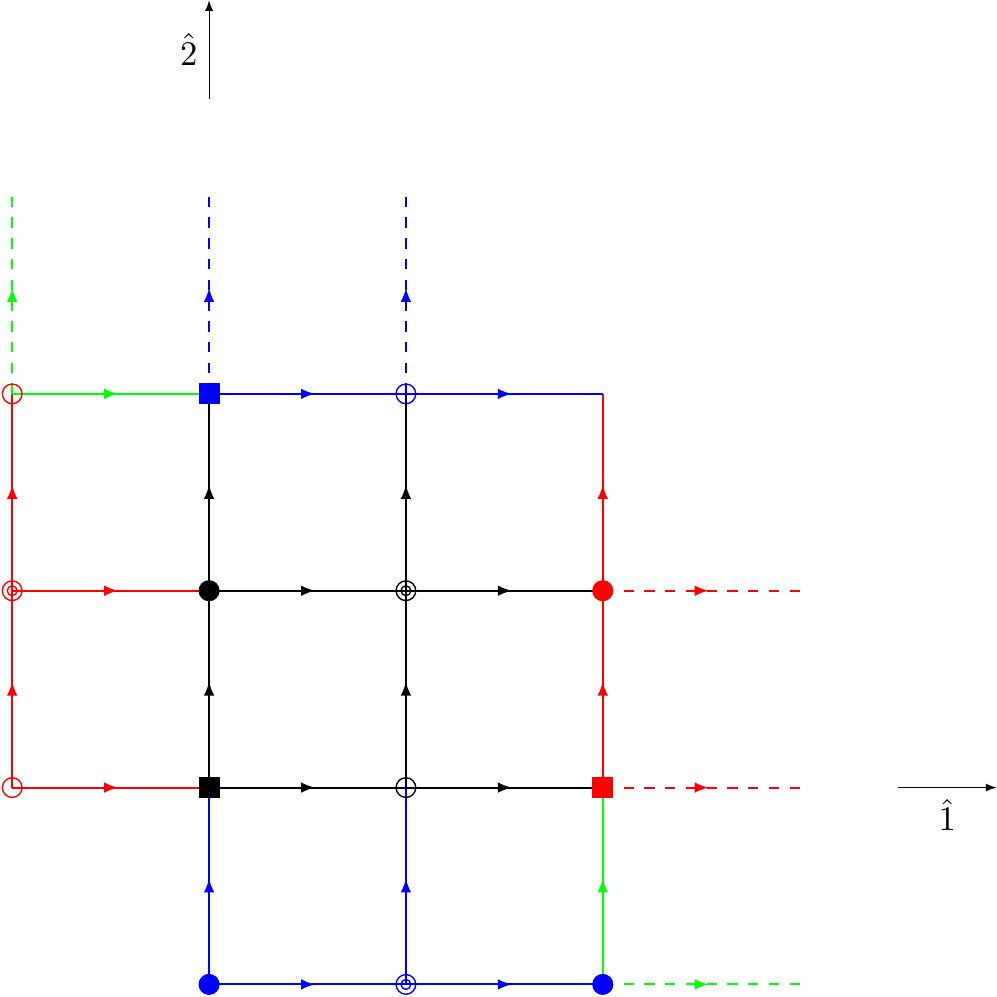}
	\end{subfigure}
	\caption{\label{fig:twistimplementation}Sketch of the PRlgt auxiliary links in a plane of a $2\times2$ lattice. Physical links are in black and sites identified by the same symbol represent the same physical site. Dashed lines highlight links that are allocated but do not participate in the update. In the left panel direction $\hat 1$ is twisted, $\hat 2$ is not: in red there are the auxiliary sites (two forward, three backward) and the red links beginning there correspond to physical links twisted according to the matrix $\Omega_1$. In the right panel both $\hat 1$ and $\hat 2$ are twisted directions: in blue there are auxiliary sites whose links are twisted according to the matrix $\Omega_2$. In the latter case there are sites which pass the boundary in two twisted directions: the green links undergo both the $\Omega_1$ and the $\Omega_2$ twist (the two operations commute by definition).}
\end{figure}

PRlgt~\footnote{
	For recent developments on the code see \myrefname~\cite{Brambilla:2014bta}.
} stems from the first NSPT codes developed by the Parma lattice gauge theory group, allowing for $\sun(3)$ simulations with Wilson fermions.
We implemented twisted boundary conditions, smell for Wilson fermions and added support for $\sun(2)$ simulations.
The code is tailored for perturbation theory. The underlying idea is to have base classes \texttt{ptSU2} and \texttt{ptSU3} that describe perturbative matrices. The operator \texttt{*} is overloaded with the Cauchy product, so that it is possible to write the product of two series in a natural way. This is one of the operations that, especially at high orders, becomes very time-consuming: thus, having perturbative matrices as base classes allows to keep the perturbative orders close in memory and to speed up the multiplication of series. In particular, the perturbative expansion is hardcoded to start from $1$ for an element of the group and from $0$ for an element of the algebra, in order to avoid multiplying by the identity or zero matrix; this choice also improves numerical stability in keeping the series within the group or algebra. All the other structures are built from the base classes by adding Lorentz, Dirac or lattice degrees of freedom. The fermion field too is described by matrices in colour-smell space.
The update of the configuration is done one link at a time: this is possible, faster and less memory consuming for the first order integrator we are using; indeed the staples around a link can be computed also if the neighbour links have already been updated, since the effect of doing so gives higher-order effects in the time step.
Twisted boundary conditions are implemented ad hoc for the Wilson action, as shown in \myfigurename~\ref{fig:twistimplementation}: a system of twisted copies of the links on the boundary is updated at each Langevin step.
The code makes heavy use of multithreading in all loops over lattice sites.
Even though the performance of PRlgt is extremely good for small lattices, it is hard to scale to large volumes due to the scalar nature of the code.

We have also developed the GridNSPT code~\footnote{Available at \url{https://github.com/gfilaci/GridNSPT}.}, based on the Grid library~\cite{Boyle:2016lbp}.
GridNSPT has been debugged against PRlgt, and we are able to obtain the very same outputs from these two completely different implementations (but staggered fermions have been implemented in GridNSPT only).
The Grid library provides an environment where message passing, multithreading and vector parallelism are fully exploited: the lattice is geometrically decomposed into MPI domain, each one mapped to a set of processors; it is also overdecomposed over virtual nodes in order to fill a SIMD vector, assuring very high vectorisation efficiency. For example, on KNL and Skylake machines we can exploit the AVX-512 instruction set and a SIMD vector has room for $4$ complex numbers in double precision; the virtual node decomposition results in the layout 1.1.2.2, where we are referring respectively to the coordinates $x.y.z.t$.
Within the MPI task, multithreading is automatic because it is included in the closure of Grid lattice object expression templates.
Grid incorporates \CCel internal template classes representing scalars, vector or matrices. We introduced a new template class representing a perturbative series, that embeds the overloading of the \texttt{*} operator. 
\begin{Verbatim}
template<class vtype, int Np> class iPert
{
vtype _internal[Np];
};
\end{Verbatim}
All the structures are tensors built from these templates: for example, the gauge field is \texttt{Lattice<iVector<iScalar<iPert<iMatrix<vComplexD,Nc>,Np>{>},Nd>{>}}, where (starting from the outer template) we have the lattice, Lorentz, spin, perturbative, colour structure and the base type is a vectorised complex number in double precision.
With this in place, every operation in Grid is performed consistently with almost no modification.
We rely on the Grid library for the optimal implementation of the gauge action and for the Wilson and staggered fermion kernel. Twisted boundary conditions have been implemented modifying the covariant circular shifts.
Even though GridNSPT lacks of many optimisations compared to PRlgt (for example the Langevin update is not performed one link at a time, but all operations and shifts are performed on the lattice as whole), it allows to have a more flexible environment and to scale easily end very efficiently to large volumes.

\section{The nearest covariance matrix}
\label{sec:higham}
If $C$ is a covariance matrix, the corresponding correlation matrix is defined as
\begin{equation}
	\hat C=S^{-1/2} \,C\, S^{-1/2}\,,
\end{equation}
where $S$ is the matrix which is equal to $C$ on the diagonal and vanishes everywhere else. $\hat C$ has $1$ on the diagonal by construction; it might have some negative or zero eigenvalue if the estimator used in the determination of the covariance does not guarantee positive definiteness.
Given $\hat C$, Higham's algorithm~\cite{doi:10.1093/imanum/22.3.329} allows to find the nearest (in a weighted Frobenius norm) positive semidefinite matrix with unit diagonal.
The core of the procedure is alternating a projection $P_S$ onto the space of positive semidefinite matrices and a projection $P_U$ onto the matrices with unit diagonal.
The projection $P_S(X)=Y$ consists in
\begin{itemize}
	\item diagonalising $X=U^T \,\Lambda \,U$, where $U$ is an orthogonal matrix and $\Lambda$ is a diagonal matrix with the eigenvalues of $X$ on the diagonal
	\item setting to zero all the negative elements in $\Lambda$, obtaining $\tilde\Lambda$
	\item returning $Y=U^T \,\tilde\Lambda \,U$.
\end{itemize}
The projection $P_U(X)$ consists simply in putting $1$ on the diagonal of $X$.
We refer to the original work for the presentation and proof of the complete algorithm: after some iterations, the algorithm converges and returns a matrix $\hat C_H$ which is positive semidefinite and has $1$ on the diagonal.

However, the algorithm allows $\hat C_H$ to have some zero (within machine precision)  eigenvalue, preventing the inversion of the covariance matrix. If this is the case, we additionally project $\hat C_H$ onto the space of positive definite matrices. This projection consists in
\begin{itemize}
	\item diagonalising $\hat C_H=V^T \,\Gamma \,V$, where $V$ is an orthogonal matrix and $\Gamma$ is a diagonal matrix with the eigenvalues of $\hat C_H$ on the diagonal
	\item identifying $\epsilon=\delta \lambda_{max}$, where $\lambda_{max}$ is the maximum eigenvalue and $\delta$ is the tolerance of the projection
	\item setting to $\epsilon$ all the diagonal elements of $\Gamma$ whose absolute value is smaller than $\epsilon$, obtaining $\tilde\Gamma$
	\item returning $\hat C_P=V^T \,\tilde\Gamma \,V$.
\end{itemize}
In conclusion, the nearest covariance matrix is
\begin{equation}
	C_P=S^{1/2} \,\hat C_P\, S^{1/2}\,.
\end{equation}

\clearpage
\bibliographystyle{JHEP}
\bibliography{gluonc.bib}

\end{document}